\documentclass{ws-ijmpb}
\usepackage{amsmath,braket}
\usepackage{bbm}
\usepackage[dvipsnames]{xcolor}

\newcommand{\av}[1]{\overline{#1}}
\newcommand{\mc}[1]{\mathcal{#1}}
\newcommand{\mr}[1]{\mathrm{#1}}
\newcommand{\mbb}[1]{\mathbb{#1}}
\newcommand{\mbf}[1]{\mathbf{#1}}

\newcommand{\lrs}[1]{\left( #1 \right)}
\newcommand{\lrm}[1]{{\left\{ #1 \right\}}}
\newcommand{\lrl}[1]{\left[ #1 \right]}
\newcommand{\lrv}[1]{\left| #1 \right|}
\newcommand{\braketL}[1]{\left\langle #1 \right\rangle}

\newcommand{\fracd}[2]{\frac{\mathrm{d} #1 }{\mathrm{d} #2 }}

\newcommand{\fracpd}[2]{\frac{\partial #1 }{\partial #2 }}

\newcommand{\rhcomment}[1]{\textcolor{black}{#1}}

\newcommand{\aln}[1]{
\begin{align}
#1
\end{align}
}

\newcommand{\ra}{\rightarrow}

\newcommand{\Tr}{\mr{Tr}}

\begin{document}

\markboth{Zongping Gong and Ryusuke Hamazaki}
{Bounds in Quantum Dynamics}

%
\catchline{}{}{}{}{}
%

\title{BOUNDS IN NONEQUILIBRIUM QUANTUM DYNAMICS} 

\author{ZONGPING GONG} 

\address{Max-Planck-Institut f\"ur Quantenoptik, Hans-Kopfermann-Stra{\ss}e 1\\
Garching 85748, Germany 
\\
zongping.gong@mpq.mpg.de} 

\author{RYUSUKE HAMAZAKI}

\address{Nonequilibrium Quantum Statistical Mechanics RIKEN Hakubi Research Team, RIKEN Cluster for Pioneering Research (CPR), RIKEN iTHEMS, \\
Wako, Saitama 351-0198, Japan\\
ryusuke.hamazaki@riken.jp}

\maketitle

\begin{history}
\received{Day Month Year}
\revised{Day Month Year}
\end{history}

\begin{abstract}
We review various bounds concerning out-of-equilibrium dynamics in few-level and many-body quantum systems. 
We primarily focus on closed quantum systems but will
also mention some related results for open quantum systems and classical stochastic systems. We start from the speed limits, the universal bounds on the speeds of (either quantum or classical) dynamical evolutions. We then turn to review the bounds that address how good and how long would a quantum system equilibrate or thermalize. Afterward, we focus on the stringent constraint set by locality in many-body systems, rigorously formalized as the Lieb-Robinson bound. We also review the bounds related to the dynamics of entanglement, a genuine quantum property. Apart from some other miscellaneous topics, several notable error bounds for approximated quantum dynamics are discussed. While far from comprehensive, this topical review covers a considerable amount of recent progress and thus could hopefully serve as a convenient starting point and up-to-date guidance for interested readers.
\end{abstract}

\keywords{quantum speed limit; quantum thermalization; Lieb-Robinson bound; entanglement dynamics; error bound}


\section{Introduction}
While physics laws are typically equalities, there are also well-known inequalities represented by the second law of thermodynamics and Heisenberg's uncertainty principle. The former states that the entire entropy change during a thermodynamic process can never be negative,\cite{GEU63} while the latter states that the product of the standard deviations of position and momentum is always lower bounded by $\hbar/2$.\cite{WH27} Ultimately, these inequalities or bounds can be derivable from some underlying equality relations, which are the entropy fluctuation theorem\cite{US05} and the canonical commutation relation\cite{EHK27} for the two specific examples mentioned above. Historically, however, these inequalities are of fundamental importance in constraining the consistent theories (e.g., stochastic thermodynamics\cite{KS10} and quantum mechanics\cite{JvN55}) based on equalities. 


In this review, we focus on inequalities or bounds concerning nonequilibrium quantum dynamics, which has become an extremely active interdisciplinary field at the interface of quantum information, quantum optics, condensed matter, statistical mechanics, and even high energy physics. Primarily, we consider closed quantum systems governed by the Sch\"odinger equation
\begin{equation}
i\hbar \frac{d}{dt} |\psi(t)\rangle = \hat H |\psi(t)\rangle,
\end{equation}
where the Hamiltonian $\hat H$ may be time-dependent and may contain a few or a macroscopically large number of energy levels. 
This is our starting point for obtaining all the bounds, ranging from the celebrated time-energy uncertainty relation (cf. Sec.~\ref{Sec:SLUQD}) to the recently uncovered bound on chaos (cf. Sec.~\ref{Sec:MSSB}), which may further rely on additional properties of the Hamiltonian such as locality in quantum many-body systems (cf. Sec. ~\ref{Sec:LRbound}). For convenience, we may also work with the Heisenberg picture $i\hbar  \frac{d}{dt}\hat O(t) =[\hat O(t), \hat H]$ for an operator $\hat O$ or/and the discrete version $|\psi(t+1)\rangle= \hat U|\psi(t)\rangle$ with $\hat U$ being a unitary. Sometimes the essential mathematical structure used for deriving the bounds applies also to open quantum systems and even classical stochastic systems, the latter of which can be relevant to, e.g., biophysics and network science. These natural generalizations will also be mentioned briefly.     


The bounds we will discuss can be important in several aspects. First, they are of fundamental significance on the theoretical side, in the sense that they unveil the ultimate limitations set by the basic laws of (quantum) physics. For example, the Mandelstam-Tamm bound (\ref{MTB}) on the quantum speed limit makes it clear that one cannot simultaneously suppress the evolution rate and energy fluctuation. Second, they can be testified in state-of-the-art experiments. For example, the emergent light cone for correlation propagation, which is rigorously formalized as the Lieb-Robinson bound (\ref{LRB}), has been observed in locally interacting many-body systems emulated by, e.g., ultracold atoms in optical lattices and trapped ions. Third, they may have crucial application implications for quantum science and technology. For example, the Trotter-Suzuki error bound justifies the faithfulness of digital quantum simulations based on Trotterization (cf. Sec.~\ref{Sec:Trot}). 


The purpose of this introductory review is to provide a brief overview on some notable topics in nonequilibrium quantum dynamics to newcomers in the field and to remind more professional researchers of some recent progress scattered in the literature. In each section, we provide some more specialized and comprehensive reviews for those readers with particular interests. Some open problems will also be mentioned, either in the relevant sections or the outlook part. Due to a large amount of references, it is inevitable that our selection of topics and materials is highly biased and far from complete. Nevertheless, we hope this review could provide strong evidence that the known results are no more than the tip of the iceberg, and thus stimulate further studies on deriving bounds or other rigorous results concerning nonequilibrium quantum dynamics.


The rest of the review is structured as follows. In Sec.~\ref{Sec:SL}, we focus on the speed limits in both few-level and macroscopic quantum systems, as well as their natural generalizations to classical systems. In Sec.~\ref{Sec:QT}, we talk about the bounds related to quantum thermalization in both isolated and open quantum systems. So far, the results do not rely heavily on the Lieb-Robinson bound, which is reviewed in Sec.~\ref{Sec:LRbound} and will be frequently used afterward. In particular, we will discuss the Lieb-Robinson bounds for both short-range and long-range many-body systems, as well as their implications for equilibrium properties. Entanglement generation is the main topic of Sec.~\ref{Sec:EG}, where the Lieb-Robinson bound is exploited for proving general upper bounds for many-body systems. Section~\ref{Sec:EB} is devoted to covering a few remarkable error bounds for approximated quantum dynamics. Some miscellaneous topics with particular recent interest are given in Sec.~\ref{Sec:MT}. Finally, in Sec.~\ref{Sec:CO}, we conclude the review and provide some outlooks for future studies. 


\section{Speed limit}
\label{Sec:SL}
As a first topic, we discuss fundamental bounds on how fast the transition of a state occurs, which are known as  speed limits.
Speed limits are originally developed for unitary quantum dynamics.
It dates back to the seminal work by Mandelstam and Tamm\cite{LM45}, who showed that the transition time and the energy fluctuation of the system cannot be reduced simultaneously.
Later, speed limits {were} 
extended to open quantum systems, mixed states, and classical systems.
Here, we overview the basic concepts and several recent topics.
We refer interested readers to a more comprehensive review on quantum speed limits.\cite{SD17}

\subsection{Few-body systems}
\subsubsection{Bounds for unitary quantum dynamics}
\label{Sec:SLUQD}
Let us first review the most fundamental quantum speed limit, i.e., the Mandelstam-Tamm bound.\cite{LM45}
To begin with, let us consider an arbitrary {(time-independent)} observable $\hat{A}$ and its expectation value $\braket{\hat{A}(t)}{=\Tr[\hat A\hat\rho(t)]}$ with respect to {state $\hat{\rho}(t)$ at time $t$}.
{Suppose that the system evolves under a time-independent Hamiltonian $\hat H$. Then, the  speed of the observable,} 
$|d\braket{\hat{A}(t)}/dt|=|\braket{[\hat{A}(t),\hat{H}]}|/\hbar=|\braket{[\hat{A}(t)-\braket{\hat{A}(t)},\hat{H}-\braket{\hat{H}}]}|/\hbar$ \rhcomment{($\hat{A}(t)=e^{i\hat{H}t}\hat{A}e^{-i\hat{H}t}$ is the Heisenberg representation of $\hat{A}$)}, is bounded as 
\aln{\label{Eq:UR}
\lrv{\fracd{\braket{\hat{A}(t)}}{t}}\leq \frac{2}{\hbar}\sigma_A(t)\cdot\sigma_H,
}
where 
\aln{
\sigma_A(t)=\sqrt{\braket{(\hat{A}(t)-\braket{\hat{A}(t)})^2}}
}
is the quantum fluctuation of an observable $\hat{A}$ at time $t$ {($\sigma_H$ is defined similarly)}.
\rhcomment{Here, we have used an inequality $|\braket{[\hat{B},\hat{C}}]|\leq |\braket{\hat{B}\hat{C}}|+|\braket{\hat{C}\hat{B}}|\leq 2\sqrt{\braket{\hat{B}^2}\braket{\hat{C}^2}}$ for Hermitian operators $\hat{B}$ and $\hat{C}$.}
The inequality in \eqref{Eq:UR} means that the instantaneous speed of the observable is upper bounded by its fluctuation and the energy fluctuation $\sigma_H=\Delta E$.
Let us assume a pure state $\hat{\rho}(t)=\ket{\psi(t)}\bra{\psi(t)}$ and take $\hat{A}=\ket{\psi(0)}\bra{\psi(0)}$ {to be the projector onto the initial state. 
}  
Then, the quantum fidelity $F(t)=|\braket{\psi(t)|\psi(0)}|^2$ satisfies 
$|dF/dt|\leq 2\sqrt{F-F^2}\Delta E/\hbar$ and thus 
\aln{
\fracd{\mc{L}(t)}{t}\leq \frac{\Delta E}{\hbar},
}
where $\mc{L}(t)=\arccos\sqrt{F(t)}$ is the Bures angle between the initial and time-evolved states.
Integrating over time, we find that the time $T$ for the  state satisfying $F(0)=1$ becomes orthogonal to the
initial state, i.e., $F(T)=0$, is bounded as
\aln{
T\geq T_\mr{MT}=\frac{\hbar\pi}{2{\Delta E}}.
\label{MTB}
}
This inequality demonstrates that the energy fluctuation should be large for the fast transition of the state.
\rhcomment{This is regarded as an uncertainty relation between energy and time, as discussed by Mandelstam and Tamm.
Note that this uncertainty relation between time and energy is quite different from that between position and momentum\cite{WH27} formulated by Kennard\cite{EHK27} and Robertson.\cite{HPR29}
In fact, while the uncertainty for the latter is described by the standard deviation associated with measuring observables, i.e., $\sigma_x\sigma_p\geq\hbar/2$, we cannot make a similar argument for the former since time is usually not regarded a standard observable in quantum mechanics.
Instead, inequality \eqref{MTB} indicates that $T_\mr{MT}$ sets an intrinsic time scale for quantum dynamics.
}

While Mandelstam-Tamm bound relies on the energy fluctuation, \rhcomment{which were occasionally studied and  re-derived afterwards,\cite{GNF73,KB83,JA90}} Margolus and Levitin realized that the energy expectation value is also useful for bounding the transition time.\cite{NM98}
They showed
\aln{\label{Eq:ML}
T\geq T_\mr{ML}=\frac{\hbar\pi}{2(E-E_0)},
}
where $E=\braket{\hat{H}}$ and $E_0$ is the ground state of $\hat{H}$.
\rhcomment{
From the Mandelstam-Tamm bound and Margolus-Levitin bound, we have $T\geq \max({T_\mr{MT}, T_\mr{ML}})$.
Reference \refcite{LBL09} showed that 
the equality condition is attainable only for a two-level state $\ket{\psi}=(\ket{E_0}+\ket{E_1})/\sqrt{2}$, up to a degeneracy of the energy level $E_1$ and phase factors for the eigenstates. In this case, we have an evolution time $T =\hbar\pi/(E_1-E_0)$.
}

There are several generalizations of the bounds above.
For example, the Mandelstam-Tamm bound can be generalized to time-dependent Hamiltonian{s}, mixed states, and arbitrary angles.
Indeed, we can show that
\aln{\label{Eq:general_MT}
\tau\geq \frac{\hbar \mc{L}(\hat{\rho}(0),\hat{\rho}(\tau))}{\av{\Delta E}}
}
with the Bures distance for mixed states
\aln{
\mc{L}(\hat{\rho}(0),\hat{\rho}(\tau))=\arccos\sqrt{F(\hat{\rho}(0),\hat{\rho}(\tau))},
}
where $\av{\Delta E}=\frac{1}{\tau}\int_0^\tau\sigma_H(t)dt$ is the average energy fluctuation over time,
and $F(\hat{\rho}(0),\hat{\rho}(\tau))=\lrs{\Tr\lrm{\sqrt{\sqrt{\hat{\rho}(0)}{\hat{\rho}(\tau)}\sqrt{\hat{\rho}(0)}}}}^2$ is the fidelity for the mixed state{s}.
Inequality~\eqref{Eq:general_MT} was obtained by Ref.~\refcite{PP93} for the case with pure states, and by Ref.~\refcite{AU92} for mixed states, which were re-derived by a geometrical approach.\cite{SD13E}

\begin{figure}[bt]
\centerline{\psfig{file=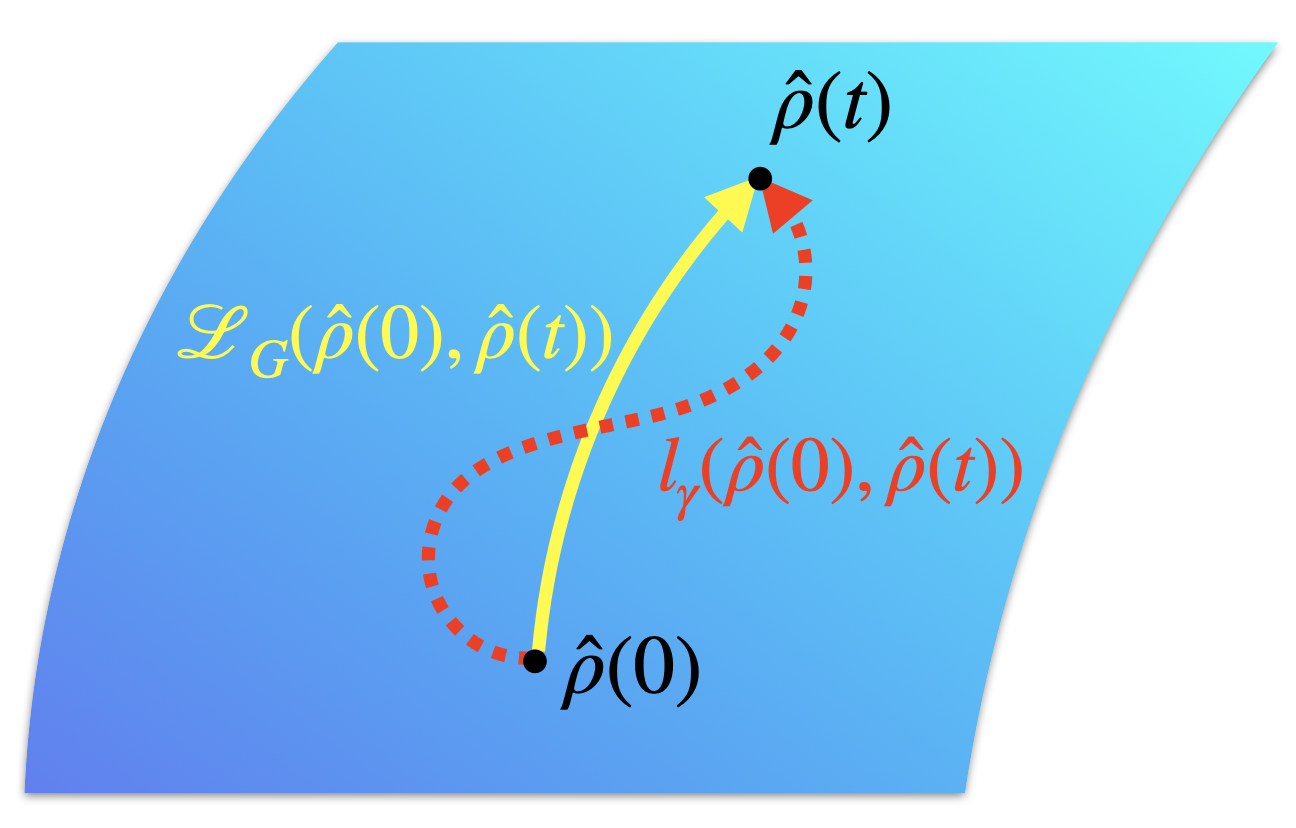,width=3in}}
\vspace*{8pt}
\caption{Geometry of a time-evolving quantum state. The length $l_\gamma(\hat{\rho}(0),\hat{\rho}(t))$ of the path $\gamma$ for the state to evolve from $\hat{\rho}(0)$ to $\hat{\rho}(t)$ (red) is always lower bounded by the geodesic distance $\mc{L}_G(\hat{\rho}(0),\hat{\rho}(t))$ (yellow).}
\label{fig:geometry}
\end{figure}

\rhcomment{
Speed limits turn out to be deeply related to a geometry of quantum states.\cite{JA90,PJJ10}
Let us consider an evolution generated by a unitary $\hat U_{\boldsymbol{\lambda}}$, where 
$\boldsymbol{\lambda}=\{\lambda_\mu\}_{\mu=1}^M$ are  parameters.
When we change the parameters from $\boldsymbol{\lambda}=\boldsymbol{\lambda}_i$ to $\boldsymbol{\lambda}=\boldsymbol{\lambda}_f$, the evolved state $\hat{\rho}_{\boldsymbol{\lambda}}=\hat U_{\boldsymbol{\lambda}} \hat{\rho}_0 \hat U_{\boldsymbol{\lambda}}^\dag$ draws a path $\gamma$ in the space of quantum states, which connects $\hat{\rho}_{\boldsymbol{\lambda}_i}$ and $\hat{\rho}_{\boldsymbol{\lambda}_f}$.
When a metric $ds^2=\sum_{\mu\nu}g_{\mu\nu}d\lambda_\mu d\lambda_\nu$ on the quantum-state space  is defined, the length of $\gamma$ can be calculated as 
$l_\gamma(\hat{\rho}_{\boldsymbol{\lambda}_i},\hat{\rho}_{\boldsymbol{\lambda}_f})=\int_\gamma ds$.
Now, let us assume that $\boldsymbol{\lambda}$ is time-dependent and that
$\boldsymbol{\lambda}(0)=\boldsymbol{\lambda}_i$ and
$\boldsymbol{\lambda}(\tau)=\boldsymbol{\lambda}_f$.
Then, the length of $\gamma$ can be expressed as
\aln{
l_\gamma(\hat{\rho}(0),\hat{\rho}(\tau))=\int_0^\tau dt\sqrt{\sum_{\mu\nu}g_{\mu\nu}\dot{\lambda}_\mu\dot{\lambda}_\nu},
}
where $\hat{\rho}(t)=\hat{\rho}_{\boldsymbol{\lambda}(t)}$.
On the other hand, this length is always lower bounded by the geodesic distance $\mc{L}_G(\hat{\rho}(0),\hat{\rho}(\tau))$ (subscript $G$ denotes ``geodesic"), i.e., $\mc{L}_G(\hat{\rho}(0),\hat{\rho}(\tau))\leq l_\gamma(\hat{\rho}(0),\hat{\rho}(\tau))$ (see Fig.~\ref{fig:geometry}).
In particular, it is known that the Bures distance $\mc{L}(\hat{\rho}(0),\hat{\rho}(\tau))$ is associated with the quantum Fisher information metric.
For simplicity, let us consider the case for a single parameter ($M=1$) and set  $\lambda(t)=t$ (we omit $\nu$ and $\mu$) with $\hat{H}(t)=-i\hbar \hat U_t\partial_t \hat U_t^\dag$. We then obtain}
\aln{\label{Eq:Fis_MT}
\mc{L}(\hat{\rho}(0),\hat{\rho}(\tau))\leq l_\gamma(\hat{\rho}(0),\hat{\rho}(\tau))=\frac{\tau}{2} \av{\sqrt{\mc{F}_Q}}.
}
Here 
$\mc{F}_Q$ is called the {(symmetric logarithmic derivative)} 
quantum Fisher information, \rhcomment{which was first introduced in quantum estimation theory\cite{CH67} (see Ref.~\refcite{JL19} for a recent review)},
\aln{
\mc{F}_Q=4g=\frac{2}{\hbar^2}\sum_{ij}\frac{(q_i-q_j)^2}{q_i+q_j}|\braket{i|\hat{H}|j}|^2=
\sum_{ij}\frac{2}{q_i+q_j}\lrv{\braketL{i\lrv{\fracd{\hat{\rho}}{t}}j}}^2
}
with the state being diagonalized as $\hat{\rho}=\sum_{i}q_i\ket{i}\bra{i}$ {(time dependence will be omitted for simplicity hereafter from time to time)}.
Since $\mc{F}_Q\leq 4\Delta E^2/\hbar^2$ (the equality condition is achieved \rhcomment{if and only if  $q_iq_j|\braket{i|\hat{H}-\braket{\hat{H}}|j}|^2=0$ for all $i,j$,\cite{SLB94} which is satisfied} for, e.g., pure states), inequality \eqref{Eq:Fis_MT}\cite{SD13E} gives a better bound than 
\eqref{Eq:general_MT} for a generic mixed state.
We also note that inequality \eqref{Eq:UR} is tightened by the quantum Fisher information as
\aln{\label{Eq:URfis}
\lrv{\fracd{\braket{\hat{A}(t)}}{t}}\leq \sigma_A(t)\cdot\sqrt{\mc{F}_Q(t)}.
}
This is known as a (generalized) quantum Cram\'{e}r-Rao inequality, which is a fundamental bound in quantum estimation theory.

\rhcomment{
We note that a different metric leads to a different inequality.
For example,  if we consider the Wigner-Yanase information metric $g'$, we have\cite{DPP15}
\aln{
\mc{L}'(\hat{\rho}(0),\hat{\rho}(\tau))\leq \frac{\tau}{\sqrt{2}\hbar}\av{\sqrt{\mc{I}}},
}
where $\mc{L}'(\hat{\rho}(0),\hat{\rho}(\tau))=\arccos[\Tr[\sqrt{\hat{\rho}(0)}\sqrt{\hat{\rho}(\tau)}]]$ is the distance based on the quantum affinity, and $\mc{I}=-(1/2)\Tr[[\sqrt{\hat{\rho}},\hat{H}]^2]=\hbar^2g'/2$ is the Wigner-Yanase skew information.
Generalizing these ideas, Ref.~\refcite{DPP16} derived an infinite family of geometric quantum speed limits, on the basis of  contractive Riemannian metrics under completely positive trace-preserving map.
}

\subsubsection{Bounds for classical dynamics}
While we have discussed quantum speed limits, fundamental dynamical constraints exist even for classical systems.
Shanahan {\it et al.}\cite{BS18} formulated the quantum speed limit for the Wigner representation and obtained the corresponding classical speed limit by its semiclassical expansion.
Okuyama and Ohzeki\cite{MO18} formally derived a speed limit for a large class of classical dynamics, including the classical Liouville equation, the Fokker-Planck equation, and the Markov equation.

Later, Shiraishi {\it et al.}\cite{NS18} realized that there is a tradeoff relation between the transition time and the entropy production rate in classical stochastic thermodynamics, as for the 
tradeoff relation between  time and energy (fluctuation) in isolated quantum systems.
They considered a stochastic system whose probability distribution obeys the Markov equation,
\aln{
\fracd{p_i}{t}=\sum_jW_{ij}p_j.
}
\rhcomment{Here, we assume the detailed-balance condition for simplicity, i.e.,  $W_{ij}e^{-\beta E_j}=W_{ji}e^{-\beta E_i}$ is satisfied for any $i$ and $j$, where $E_i$ is the energy of the $i$th state, and $\beta$ is the inverse temperature of the heat bath, although they also extended their results for systems without this condition.}
Let us consider the statistical distance between initial and final states as 
$d(p(0),p(t))=\sum_i|p_i(0)-p_i(t)|/2$, which is called the total variation distance.
The authors evaluated the instantaneous speed $|dp_i(t)/dt|$ as
\aln{
\sum_i\lrv{\fracd{p_i(t)}{t}}\leq \sqrt{2\mc{A}\dot{\Sigma}},
}
where $\mc{A}=\sum_{i\neq j}W_{ij}p_j$ quantifies the rate of transitions and is called the dynamical activity, and
$\dot{\Sigma}={(1/2)\sum_{i\neq j}(W_{ij}p_j-W_{ji}p_i)}\ln(W_{ij}p_j/W_{ji}p_i)$ is regarded as the entropy production rate (see E.g., Ref.~\refcite{US12} for a review on stochastic thermodynamics).
\rhcomment{This inequality states that the large entropy production is inevitable to enhance the transition speed, indicating the tradeoff relation between time and entropy production.}
In addition, from this inequality, the transition speed is evaluated as (cf. inequality~\eqref{Eq:general_MT}) 
\aln{\label{Eq:speed_shiraishi}
\tau{\geq \frac{d(p(0),p(\tau))}{\av{\sqrt{\frac{\mc{A}\dot{\Sigma}}{2}}}}}\geq \frac{d(p(0),p(\tau))}{\sqrt{\frac{\av{\mc{A}}\av{\dot{\Sigma}}}{2}}},
}
{where the rightmost result above arises simply from the Cauchy-Schwarz inequality.} 
\rhcomment{The results in Ref.~\refcite{NS18} are recently improved in several ways;
for example, while Ref.~\refcite{NS18} also  discussed the case without the the detailed-balance condition (using the so-called Hatano-Sasa entropy production\cite{TH01}), Ref.~\refcite{VTV20} improved it by a constant factor; Reference~\refcite{TVV21} derived a tighter bound using a geometric approach.
}
We note that the result in \eqref{Eq:speed_shiraishi} is also regarded as a relation stating that 
the entropy production is lower bounded by the \rhcomment{statistical distance $d(p(0),p(\tau))$}.
Such lower bounds of the entropy production are also actively investigated nowadays; {a closely related topic will be discussed briefly in Sec.~\ref{Sec:TUR}.}

For a (time-independent) random observable $A$ whose expectation value is given by $\braket{A}=\sum_ia_ip_i$,
its speed is bounded as $|d\braket{A}/dt|\leq \|A\|_\infty \sum_i|dp_i(t)/dt|$,
where $\|A\|_\infty =\max_i|a_i|$ \rhcomment{and we have used the H\"{o}lder inequality}.
Thus, we also have
\aln{\label{Eq:Sent}
\lrv{\fracd{\braket{A}}{t}}\leq \|A\|_\infty \sqrt{2{\mc{A}\dot{\Sigma}}}.
}

Another approach concerning the tradeoff relation for classical systems is the uncertainty relation between time and information.\cite{SBN20,SI20}
For any (possibly time-dependent) observable $\hat{A}$, its classical expectation value $\braket{A}(t)=\sum_ia_i(t)p_i(t)$ has a speed\cite{SBN20}
\aln{
{\fracd{\braket{A}}{t}}=-\mr{cov}\lrs{A,\fracd{I}{t}}+\braketL{\fracd{{A}}{t}},
}
where $\mr{cov}(A,B)=\braket{AB}-\braket{A}\braket{B}$ is the covariance, $\braket{dA/dt}=\sum_ip_i(da_i/dt)$, and \rhcomment{$I=\{I_i\}=\{-\ln p_i\}$,} which corresponds to the obtained information when we observe that the system is in the state $i$.
This identity is known as the Price equation.
Since $|\mr{cov}(A,B)|\leq \sigma_A\sigma_B$, we have\cite{SBN20}
\aln{\label{Eq:class_fis}
\lrv{{\fracd{\braket{A}}{t}}-\braketL{\fracd{{A}}{t}}}\leq \sigma_A\sqrt{\mc{F}_c},
}
where 
\aln{
\mc{F}_c=\sigma_{dI/dt}^2=\sum_ip_i\lrs{\fracd{\ln p_i}{t}}^2=\sum_i\frac{1}{p_i}\lrs{\fracd{p_i}{t}}^2
}
is the classical Fisher information.

Inequality \eqref{Eq:class_fis} is analogous to quantum speed limit based on the quantum Fisher information in \eqref{Eq:Fis_MT}; however, since we consider classical dynamics, the identification of the information as the energy fluctuation no longer applies.
Still, \eqref{Eq:class_fis} is regarded as the uncertainty relation between transition time and information.
Indeed, if we define a characteristic time $\tau_A^{-1}=\lrv{{\fracd{\braket{A}}{t}}-\braketL{\fracd{{A}}{t}}}/\sigma_A$, it satisfies $\tau_A\geq 1/\sqrt{\mc{F}_c}$, suggesting that the time and information cannot be simultaneously small.
\rhcomment{
Note that this inequality holds for a wide class of dynamics, indicating the universality of the time-information tradeoff relation. Indeed, while it was originally discussed for linear stochastic dynamics\cite{SBN20,SI20},
it can also be applied to nonlinear dynamics, such as population dynamics, which describe evolutionary and ecological processes of species.\cite{KA22,LPG22D}}

Similarly, we can also show the bound based on the total variation distance as
\aln{
\tau\geq \frac{2d(p(0),p(\tau))}{\av{\sqrt{\mc{F}_c}}}.
}
\rhcomment{This is proven by
\aln{
2d(p(0),p(\tau))=\sum_i\lrv{\int_0^\tau\dot{p}_i(t)dt}\leq \int_0^\tau dt\sum_i\lrv{\dot{p}_i(t)}
\leq\int_0^\tau dt\sqrt{\sum_i\frac{\lrv{\dot{p}_i(t)}^2}{p_i(t)}}=\tau\av{\sqrt{\mc{F}_c}}.}
}

\subsubsection{Bounds for open quantum dynamics}
Speed limits also exist for open quantum systems.
\rhcomment{Historically, this fact was demonstrated\cite{MMT13,AdC13,SD13} before the work on classical speed limits; for example, 
Ref. \refcite{MMT13} derived a geometrical bound based on quantum Fisher information, 
Ref. \refcite{AdC13} derived a bound defined from the relative purity.
}
We here detail the bound by Deffner and Lutz\cite{SD13}, who provided bounds based on the generator of an arbitrary dynamics $d\hat{\rho}/dt=\mathbb{L}[\hat{\rho}]$.
To see this, we assume that the initial state is pure $\hat{\rho}(0)=\ket{\psi_0}\bra{\psi_0}$,
which leads to $\mc{L}(\hat{\rho}(0),\hat{\rho}(\tau))=\arccos\sqrt{\braket{\psi_0|\hat{\rho}({\tau}
)|\psi_0}}$.
From this, we obtain
\aln{
2\cos(\mc{L})\sin(\mc{L})\fracd{\mc{L}}{t}\leq |\braket{\psi_0|\mbb{L}[\hat{\rho}(t)]|\psi_0}|\leq \|\mbb{L}[\hat{\rho}(t)]\|,
}
where $\|\hat{A}\|$ denotes the largest singular value of $\hat{A}$.
From this, we obtain the speed limit
\aln{
\tau\geq \frac{\sin^2(\mc{L}(\hat{\rho}(0),\hat{\rho}({\tau}
)))}{\av{\|\mbb{L}[\hat{\rho}]\|}}.
}
Note that this bound is related to the Mandelstam-Tamm and Margolus-Levitin bounds for specific cases {by applying different bounds on $\|\mathbb{L}[\hat\rho]\|$}.
For example, when we consider the unitary dynamics $\mbb{L}[\hat{\rho}]=-i[\hat{H},\hat{\rho}]/\hbar$,
we find that the energy fluctuation appears as
$\|\mbb{L}[\hat{\rho}]\|\leq \sqrt{\Tr[|\mbb{L}[\hat{\rho}]|^2]}=\sqrt{2}\Delta E/\hbar$ in a manner similar to the Mandelstam-Tamm bound.

Another approach is to generalize the time-information uncertainty relation in classical systems\cite{SBN20} to the quantum realm.
In fact, as a quantum extension of the results in the previous subsection, one can show\cite{LPG21}
\aln{
{\fracd{\braket{\hat{A}}}{t}}=\mr{cov}\lrs{\hat{\mathfrak{L}},\hat{A}}+\braketL{\fracd{\hat{A}}{t}},
}
where $\hat{\mathfrak{L}}=2\int_0^\infty ds e^{-\hat{\rho}s}\lrs{\fracd{\hat{\rho}}{t}} e^{-\hat{\rho}s}
$ satisfies $d\hat{\rho}/dt=\{\hat{\mathfrak{L}},\hat{\rho}\}/2$ ($\{\hat{A},\hat{B}\}=\hat{A}\hat{B}+\hat{B}\hat{A}$),
and $\mr{cov}(\hat{A},\hat{B})=\braket{\{\hat{A},\hat{B}\}}/2-\braket{\hat{A}}\braket{\hat{B}}$ is the symmetrized covariance.
From this, we have
\aln{\label{Eq:open_fis}
\lrv{{\fracd{\braket{\hat{A}}}{t}}-\braketL{\fracd{\hat{A}}{t}}}\leq \sigma_A\sqrt{\mc{F}_Q},
}
where 
\aln{
\mc{F}_Q=\braket{\hat{\mathfrak{L}}^2}=\sum_{ij}\frac{2}{q_i+q_j}|\braket{i|\mathbb{L}[\hat{\rho}]|j}|^2
}
is the {(symmetric logarithmic derivative)} quantum Fisher information.
Note that inequality \eqref{Eq:open_fis} is a generalization of inequality \eqref{Eq:URfis}.
\rhcomment{Furthermore, Ref.~\refcite{LPG21} decomposed the observable and the Fisher information into coherent (quantum) and incoherent (classical) parts and derived the speed limits similar to \eqref{Eq:open_fis} for each part, which can provide tighter bounds than \eqref{Eq:open_fis}.}

When we focus on a more specific setting, different types of speed limits\rhcomment{, i.e., bounds using the entropy production rate,} can be obtained.
Let us consider a quantum system coupled to a Markovian bath, whose dynamics is represented by the following Gorini-Kossakowski-Sudarshan-Lindblad (GKSL) equation:\cite{VG76,GL76}
\aln{
\fracd{\hat{\rho}}{t}=\mbb{L}[\hat{\rho}]=-\frac{i}{\hbar}[\hat{H},\hat{\rho}]+\mbb{D}[\hat{\rho}],
}
where the dissipator reads
\aln{
\mbb{D}[\hat{\rho}]=\frac{1}{2}\sum_{\mu,\omega}\gamma_\mu(\omega)\lrl{2\hat{L}_\mu(\omega)\hat{\rho}\hat{L}_\mu(\omega)^\dag-\lrm{\hat{L}_\mu(\omega)^\dag\hat{L}_\mu(\omega),\hat{\rho}}}.
}
Here, the jump operator $\hat{L}_\mu(\omega)$ describes  a quantum jump 
from one energy eigenstate of $\hat{H}$, $\ket{E_\alpha}$, to another eigenstate $\ket{E_\beta}$, where $E_\alpha-E_\beta=\omega$ is satisfied.
This means that $\hat{L}_\mu(\omega)^\dag=\hat{L}_\mu(-\omega)$ and $[\hat{L}_\mu(\omega),\hat{H}]=\omega\hat{L}_\mu(\omega)$.
We also assume that these dissipators satisfy the detailed-balance condition, meaning that 
$\gamma_\mu(-\omega)=\gamma_\mu(\omega)e^{-\beta\omega}$, where $\beta$ is the inverse temperature of the bath.
We can also diagonalize the state as $\hat{\rho}=\sum_iq_i\ket{i}\bra{i}$ and define $W_{ij}^{\mu,\omega}=\gamma_\mu(\omega)|\braket{i|\hat{L}_\mu(\omega)|j}|^2$.

Under this assumption, Ref.~\refcite{KF19} considered the constraint on the change of the {trace} 
distance,
$d(\hat{\rho}(0),\hat{\rho}(t))=\|\hat{\rho}(0)-\hat{\rho}(t)\|_1/2$, where $\|\hat{A}\|_1=\Tr[\sqrt{\hat{A}^\dag\hat{A}}]$ is the trace norm of $\hat{A}$.
Then, they found, as a generalization of inequality \eqref{Eq:speed_shiraishi},
\aln{
\tau\geq \frac{d(\hat{\rho}(0),\hat{\rho}({\tau}
))}{\frac{\av{\Delta E}}{\hbar}+\frac{\av{\Delta E_D}}{\hbar}+\sqrt{\frac{\av{\mc{A}}\av{\dot{\Sigma}}}{2}}},
}
where $\Delta E=\sigma_H$ is the energy fluctuation of the Hamiltonian, 
$\mc{A}=\sum'_{ij\mu\omega}(W_{ij}^{\mu,\omega}q_j+W_{ji}^{\mu,-\omega}q_i)/2$ and $\dot{\Sigma}=(1/2)\sum'_{ij\mu\omega}(W_{ij}^{\mu,\omega}q_j-W_{ji}^{\mu,-\omega}q_i)\ln(W_{ij}^{\mu,\omega}q_j/W_{ji}^{\mu,-\omega}q_i)$\footnote{Here, $\sum'_{ij\mu\omega}$ denotes the sum that excludes the indices that satisfy $(i=j)\land (\omega=0)$.}
correspond to the dynamical activity and the entropy production rate of this system, respectively.
Furthermore,
$\Delta E_D=\sigma_{H_D}$ is the additional term that comes from the off-diagonal term in $\mbb{D}$, where $\hat{H}_D=\sum_{q_i\neq q_j}\frac{i\hbar\braket{i|\mbb{D}[\hat{\rho}]]|j}}{q_i-q_j}\ket{i}\bra{j}$.
\rhcomment{Note that this final term is characterized by the speed of the unitary evolution due to the bath, which is unique to open quantum systems.}

\subsection{Macroscopic systems}\label{Sec:speed_macro}

\begin{figure}[bt]
\centerline{\psfig{file=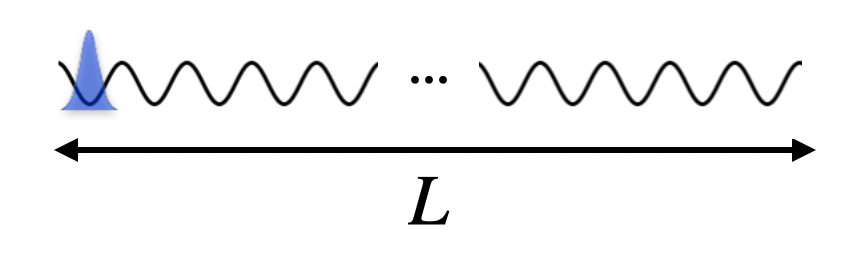,width=3in}}
\vspace*{8pt}
\caption{A single particle on a one-dimensional lattice with size $L$. 
If we initially place the wave packet at the left end, it will move to the right as time increases.}
\label{fig:1d}
\end{figure}

While speed limits discussed in the previous section have a wide variety of applications, 
they cannot provide a good bound for certain situations, such as processes involving macroscopic transitions.
As a simple example, let us consider a single particle on a one-dimensional lattice with size $L$ (Fig.~\ref{fig:1d}). 
If we initially place the particle at the left end, it will move to the right with the spreading of the wave packet.
In this case, how much time will it take for the particle to be transferred to a distant place, say, $x\sim L/2$?
Naively, it takes at least $t\gtrsim \mc{O}(L)$ for this process; this is indeed true for local Hamiltonians, 
which is also consistent with the Lieb-Robinson bound detailed in section~\ref{Sec:LRbound}.
On the other hand, many quantum speed limits, such as inequalities~\eqref{Eq:ML} or \eqref{Eq:general_MT}, cannot capture this macroscopic timescale that increases with $L$.
As a different viewpoint, let us consider the dynamics of, e.g., an observable $\hat{x}=\sum_{l=1}^Ll\ket{l}\bra{l}$.
While we expect that $d\braket{\hat{x}(t)}/dt$ is finite, inequalities~\eqref{Eq:UR} and \eqref{Eq:URfis} only provide diverging bounds with respect to $L$ or $t$.
For example, we typically have $\Delta x\sim t^\alpha\:(\alpha>0)$ during the spreading of the wave packet, which is divergent for $t=\mc{O}(L)$. 
Note that the situation becomes even worse for many-body systems, where $\Delta E$ also diverges with respect to the particle numbers.

There are at least two difficulties with the standard approaches.
First, many distance measures to define the quantum speed, such as the trace distance or the Kullback-Leibler divergence, cannot take account of the local structure of the lattice geometry (note that these measures are invariant under {arbitrary, typically non-local} unitary transformation).
For example, if we consider the case for  the one particle system, $\|\ket{1}\bra{1}-\ket{l}\bra{l}\|_1=\|\ket{1}\bra{1}-\ket{m}\bra{m}\|_1={2}$ {(here $\|\hat O\|_1\equiv\Tr[\sqrt{\hat O^\dag\hat O}]$)}
even when $|l-m|\gg 1$ as long as $l,m\neq 1$, which indicates that the trace distance is not appropriate for distinguishing  the transport from $1$ to $l$ and that from $1$ to $m$.
Second, even when the magnitude of the change of the distance is finite, we may obtain only divergent bounds for the speed of macroscopic observables from this.
For example, while the speed limits for $\|d\hat{\rho}/dt\|_1$\cite{KF19,SD17NJP} lead to the corresponding bound of an observable as $|d\braket{\hat{O}}/dt|\leq \|\hat{O}\|\cdot \|d\hat{\rho}/dt\|_1$ due to the H\"{o}lder inequality, the right-hand side can diverge when $\|\hat{O}\|$ diverges (e.g., $\|\hat{x}\|=L$ in the above example).

There are several different approaches to obtain more feasible timescales for such macroscopic systems.
One of such approaches is to invoke the Lieb-Robinson bound, which provides a state-independent general bound for information propagation (see section~\ref{Sec:LRbound}).
\rhcomment{Another approach is to 
evaluate a time for an initial state to be transferred to another state 
from a geodesic length imposed by the geometry of the eigenstate manifold set by the space of control parameters, as discussed theoretically\cite{MB19} and experimentally.\cite{MRL21}}

\subsubsection{Continuous systems}
Reference~\refcite{RH21} derived convergent bounds for dynamics with macroscopic transitions, employing the local conservation law of probability.
For intuitive understanding of the results, 
let us first derive speed limits for a mean-field type nonlinear Sch\"{o}dinger equation in a continuous space, which is given by
\aln{
i \hbar \frac{\partial \psi(\mbf{x},t)}{\partial t}=\left(-\frac{\hbar^{2}}{2 m} \nabla^{2}+V_\mr{e x t}+g|\psi(\mbf{x},t)|^{2}\right) \psi(\mbf{x},t).
}
This equation is also regarded as the Gross-Pitaevskii equation\cite{FD99} describing the dynamics of a Bose gas.
Here, we assume that $\psi$ is normalized such that $\int d\mbf{x}|\psi(\mbf{x},t)|^2=1$ without loss of generality; we will then regard $|\psi(\mbf{x},t)|^2$ as a probability density.
In this setting, the continuity equation holds as 
\aln{
\fracpd{\rho(\mbf{x},t)}{t}=-\nabla\cdot\mbf{j}(\mbf{x},t),
}
where $\rho=|\psi|^2$ is the probability density and $\mbf{j}=\rho\mbf{v}$ is the probability current.
The velocity $\mbf{v}$ is given by $\mbf{v}=\hbar\nabla\theta/m$, where $\nabla\theta$ is the gradient of the quantum phase {($\theta$ is determined from \rhcomment{the Madelung transformation\cite{EM27}} $\psi=\sqrt{\rho}e^{i\theta}$)}.
\footnote{\rhcomment{We here assume that  $\theta (\mathbf{x})$ can be expressed as a smooth single-valued function, which means that there are no quantum vortices.
When there are quantum vortices, $\theta (\mathbf{x})$ is a multi-valued function. 
Even under this situation, we can ensure that $\psi$  is a single-valued function by imposing a quantization condition $\oint_\mc{C}d\theta =2n\pi\:(n\in\mbb{Z})$ along any closed loop $\mathcal{C}$.\cite{TCW94}
In this case, it is often assumed that the gradient of the phase, $\nabla\theta(\mbf{x})$, can be as taken as a single-valued function.}}

We consider an observable $A$ defined on a space point $\mbf{x}$.
Its expectation value and the standard deviation are given by $\braket{A(t)}=\int d\mbf{x}a(\mbf{x})\rho(\mbf{x},t)$ and $\sigma_A(t)=\sqrt{\braket{A(t)^2}-\braket{A(t)}^2}$, respectively.
Let us derive speed limits for the expectation value.
If we use the Cram\'{e}r-Rao type bound discussed in the previous section, we have
\aln{\label{Eq:cont_fis}
\lrv{\fracd{\braket{A}}{t}}\leq \sigma_A\sqrt{\mc{F}},
}
where $\mc{F}=\int d\mbf{x}\rho(\mbf{x},t)(d\ln\rho(\mbf{x},t)/dt)^2$ (cf. inequality~\eqref{Eq:class_fis}).
Unfortunately, this bound becomes loose when $\sigma_A$ becomes large, which can occur for, e.g., {$a(\mbf{x})=x_1$, the position in the first direction}. 
Instead, Ref.~\refcite{RH21} first  used the continuity equation, performed the integral by parts, and applied the Cauchy-Schwarz inequality, obtaining
\aln{\label{Eq:cont_non}
\lrv{\fracd{\braket{A}}{t}}\leq \sqrt{\braket{(\nabla A)^2}}\sqrt{\int d\mbf{x}\frac{|\mbf{j}|^2}{\rho}}=\sqrt{\braket{(\nabla A)^2}}\sqrt{2E_\mr{kin}}=\frac{\hbar}{m}\sqrt{\braket{(\nabla A)^2}\braket{(\nabla \theta)^2}},
}
where $E_\mr{kin}=\frac{\hbar^2}{2m^2}\braket{(\nabla \theta)^2}$ is the kinetic energy of the gas.
Compared with \eqref{Eq:cont_fis}, inequality~\eqref{Eq:cont_non} has two distinguished points.
First, since $\sigma_A$ is replaced with $\sqrt{\braket{(\nabla A)^2}}$, the bound in 
\eqref{Eq:cont_non} can be much tighter than that in  \eqref{Eq:cont_fis}.
For example, if we consider a one-dimensional space and {$a(x)=x$}, 
$\sigma_x$ diverges with $t$ but $\braket{(\nabla x)^2}=1$ is always a constant.
This means that inequality  \eqref{Eq:cont_non} can be a useful bound for macroscopic transitions.
Second, inequality \eqref{Eq:cont_non} contains a term $E_\mr{kin}\propto \braket{(\nabla \theta)^2}$, which describes the averaged magnitude of the quantum phase gradient.
Thus, this inequality suggests that we cannot reduce the magnitude of the quantum phase gradient and the transition time simultaneously, which indicates a tradeoff relation different from the time-information uncertainty relation.
We note that a similar technique leads to the speed limit by the entropy production rate for classical stochastic dynamics described by the Fokker-Planck equation.\cite{AD18}
We also note that, while we do not introduce an explicit distance of two probability distributions here, inequality  \eqref{Eq:cont_non} may be understood by the change of the Wasserstein distance known in the optimal transport theory.\cite{CV09}

Furthermore, using the local conservation law of probability, we can even find a speed limit for the standard deviation as\cite{RH21}
\aln{\label{Eq:cont_nondev}
\lrv{\fracd{\sigma_A(t)}{t}}\leq \|\nabla A\|_\infty \sqrt{2E_\mr{kin}},
}
where $\|\nabla A\|_\infty=\max_\mbf{x}|a(\mbf{x})|$.
From the inequalities, we have, e.g.,
\aln{
\lrv{\braket{x(T)}-\braket{x(0)}}&\leq \av{\sqrt{2E_\mr{kin}}}T\nonumber\\
\lrv{\sigma_x(T)-\sigma_x(0)}&\leq \av{\sqrt{2E_\mr{kin}}}T
}
in one dimension.
From these inequalities, we find a concentration of probability. For example, when $\braket{x(0)}=0$, we have
\aln{\label{Eq:cont_conc1}
1-\int_{-\av{\sqrt{2E_\mr{kin}}}T-\Delta l}^{\av{\sqrt{2E_\mr{kin}}}T+\Delta l}\rho(x,T)dx\leq \frac{(\sigma_x(0)+{\av{\sqrt{2E_\mr{kin}}}}T)^2}{\Delta l^2}
}
for $l>0$.
Note that the following exponential-type concentration inequality also holds:
\aln{\label{Eq:cont_conc2}
1-\int_{-l}^l \rho(x,T)dx\leq 2\braket{\cosh \lambda x}(0)\times e^{\lambda(-l+\frac{\hbar}{m}\av{\|\nabla\theta\|_\infty} T)}
}
for $\lambda>0$, which indicates that the tail of the distribution $\rho(x,T)$ is exponentially suppressed for $l\gg \hbar\overline{\|\nabla \theta\|_\infty}T/m$.

\subsubsection{Discrete systems}
The discussion in the previous section can be extended to discrete systems, which include many-body systems.\cite{RH21}
We can map an arbitrary system onto a graph with vertices $\mc{V}$ and edges $\mc{E}$, where a vertex $i$ corresponds to a basis state $\ket{i}$ and  the edges between vertices $i$ and $j$ exist if and only if $H_{ij}\neq 0$ and $i\neq j$. 
We focus on observables given by $\hat{A}=\sum_ia_i\ket{i}\bra{i}$.
Assuming the Sch\"{o}dinger equation (with $\hbar=1$), we find a continuity equation of probability $dp_i/dt=-\sum_{{j:j\sim i}} 
J_{ji}$, where $p_i=\rho_{ii}$ and $J_{ij}=-i(H_{ij}\rho_{ji}-\rho_{ij}H_{ji})$\:(here, $X_{ij}=\braket{i|\hat{X}|j}$ and {$j\sim i$} 
means the set of $j$ such that $(j,i)\in\mc{E}$).
Using this relation, we find, as a counterpart of \eqref{Eq:cont_non},
\aln{\label{Eq:disc_non}
\lrv{\fracd{\braket{\hat{A}}}{t}}\leq \sqrt{\frac{\sum_{i\sim j}r^{ij}(a_i-a_j)^2}{\sum_{i\sim j}r^{ij}}}\sqrt{\lrs{\sum_{i\sim j}r^{ij}}^2-E_\mr{trans}^2}\leq \|\nabla A\|_\infty\sqrt{C_H^2-E_\mr{trans}^2},
}
where $r^{ij}=|H_{ij}|\sqrt{p_ip_j}$, $E_\mr{trans}=\sum_{i\sim j}H_{ij}\rho_{ji}=\braket{\hat{H}_\mr{trans}}$ is the expectation value of the off-diagonal (transition) part of the Hamiltonian $\hat{H}_\mr{trans}=\hat{H}-\sum_i H_{ii}\ket{i}\bra{i}$, $\|\nabla A\|_\infty=\max_{i\sim j}|a_i-a_j|$ is the maximum discrete gradient on the graph, and $C_H=\max_i\sum_{{j:j\sim i}} 
|H_{ij}|$ is the maximum strength of the transition.
Note that $r_{ij}$ only depends on the probability distribution (not off-diagonal coherent parts), $E_\mr{trans}$ is a standard expectation value of an observable, and $C_H$ is easily calculable if $\hat{H}$ is given.

There are several remarks about inequality \eqref{Eq:disc_non}.
First, The factor ${\sum_{i\sim j}r^{ij}(a_i-a_j)^2}$ (called the graph Laplacian in the graph theory) and $\|\nabla A\|_\infty$ only involve the gradient of the observable, which dramatically suppresses the bound when, e.g., $\|\nabla A\|_\infty \ll\Delta A, \|\hat{A}\|$.
For example, when we take a single particle and $\hat{x}=\sum_{l=1}^Ll\ket{l}\bra{l}$, $\|\nabla A\|_\infty=1$ and $\|\hat{A}\|=L$.
Second, for a simple case, the term $\sqrt{\lrs{\sum_{i\sim j}r^{ij}}^2-E_\mr{trans}^2}$ reduces to $\sqrt{2E_\mr{kin}}$ in the previous subsection in the continuous limit, indicating that inequality  \eqref{Eq:disc_non} is regarded as a discrete version of the tradeoff relation between time and the quantum phase difference.
Third, since  \eqref{Eq:disc_non} is a state-dependent inequality  through $r^{ij}$ or $E_\mr{trans}$, 
it can be tighter than the Lieb-Robinson velocity in section~\ref{Sec:LRbound}.
In particular, \eqref{Eq:disc_non} achieves the equality condition for certain situations.

An inequality similar to \eqref{Eq:cont_nondev} is also obtained.
We have
\aln{\label{Eq:disc_var}
\lrv{\fracd{\sigma_A(t)}{t}}\leq \|\nabla A\|_\infty \sqrt{C_H^2-E_\mr{trans}^2}.
}
When we choose $\hat{A}$ as additive observables, variance of $\hat{A}$ is known to relate with the measure of macroscopic quantum coherence.\cite{AS05,BY16}
Thus, in that sense, \eqref{Eq:disc_var} indicates the upper bound of how fast macroscopic quantum coherence can change via unitary time evolutions.

As a {simple} 
example, let us consider a single particle on a one-dimensional lattice whose Hamiltonian is given by $\hat{H}=-K\sum_{i}({\hat{a}_{i+1}^\dag}\hat{a}_i+\mr{h.c.})$.
In this case, $E_\mr{trans}=E$ becomes the total energy, and we have
\aln{
\lrv{\fracd{\braket{\hat{x}}}{t}}\leq \sqrt{\lrs{2K\sum_{i}\sqrt{p_ip_{i+1}}}^2-E^2}\leq \sqrt{4K^2-E^2}
}
and 
\aln{
\lrv{\fracd{\sigma_x(t)}{t}}\leq \sqrt{4K^2-E^2}.
}
Concentration inequalities similar to \eqref{Eq:cont_conc1} and \eqref{Eq:cont_conc2} {can also be} 
obtained.

Finally, we note that the above methods using the local conservation law of probability can be applied to speed limits for classical and quantum stochastic systems.
For example, one can show, for classical stochastic systems, that
\aln{
\lrv{\fracd{\braket{A}}{t}}\leq \|\nabla A\|_\infty \sqrt{\frac{\mc{A}\dot{\Sigma}}{2}},
}
where $\mc{A}$ is the dynamical activity and $\dot{\Sigma}$ is the entropy production rate.
Note that this becomes much tighter than inequality~\eqref{Eq:Sent} when $\|\nabla A\|_\infty\ll \|A\|_\infty$;
this is the case in, e.g., macroscopic transport  in the exclusion processes.\cite{BD98}

\section{Quantum thermalization} 
\label{Sec:QT}

In this section, we introduce the problem of quantum equilibration and thermalization {as well as the relevant bounds}. 
Here, the problem of quantum equilibration and thermalization asks the following:
does an initial nonequilibrium state relax to some stationary state by quantum dynamics? {If yes, how and when does this occur?}
What is the effective ensemble describing the stationary state?
These questions are of direct relevance with the foundation of quantum statistical mechanics.
For example, in isolated quantum macroscopic systems, 
statistical mechanics asserts that any nonequilibrium state relaxes to a stationary state described by the (micro)canonical ensemble.\cite{LDL80}
Is this principle derived only from the law of quantum mechanics, i.e., unitary dynamics?
We note that the problem has a long history dating back to von Neumann's seminal work in 1929.\cite{JVN29}
In the last decade, thermalization and equilibration have been tested in laboratories, thanks to the experimental development of artificial quantum systems.\cite{IMG14,IB08}

\subsection{Isolated systems}
Let us first consider a finite-dimensional isolated quantum many-body system described by a time-independent Hamiltonian $\hat{H}$ and discuss its equilibration and thermalization.
Since dynamics is described by the unitary evolution (we set $\hbar=1$ throughout this chapter), $\hat{\rho}(t)=\hat{U}\hat{\rho}(0)\hat{U}^\dag$ with the unitary operator $\hat{U}=e^{-i\hat{H}t}$,
the state cannot be equal to the thermal state $\hat{\rho}_\mr{therm}$, such as the microcanonical state $\hat{\rho}_{E,\delta E}$ (having mean energy $E$ and the energy window $\delta E$) or the canonical state $\hat{\rho}_\beta=e^{-\beta\hat{H}}/Z$ {with inverse temperature $\beta$} for any $t$.
Nevertheless, when we focus on few-body observables that are relevant for statistical mechanics (such as the local magnetization and spatial correlation function), $\hat{\rho}(t)$ and  $\hat{\rho}_\mr{therm}$ can lead to the same expectation values for sufficiently large $t$.
In other words, thermalization of a relevant observable $\hat{O}$ on a level of the expectation value
roughly means
\aln{\label{Eq:eqtherm}
\Tr[\hat{\rho}(t)\hat{O}]\simeq  \Tr[\hat{\rho}_\mr{therm}\hat{O}]
}
for sufficiently long $t$.
Here, $\simeq $ means that the difference becomes negligible in the thermodynamic limit, i.e.,
$|\Tr[\hat{\rho}(t)\hat{O}]- \Tr[\hat{\rho}_\mr{therm}\hat{O}]|/\|\hat{O}\|\ra 0$.
Because of the energy conservation, $E$ in $\hat{\rho}_{E,\delta E}$ or $\beta$ in $\hat{\rho}_\beta$ is determined from the condition
$\Tr[\hat{\rho}(0)\hat{H}]=\Tr[\hat{\rho}_{E,{\delta}
E}\hat{H}]=\Tr[\hat{\rho}_\beta\hat{H}]$.

\begin{figure}[bt]
\centerline{\psfig{file=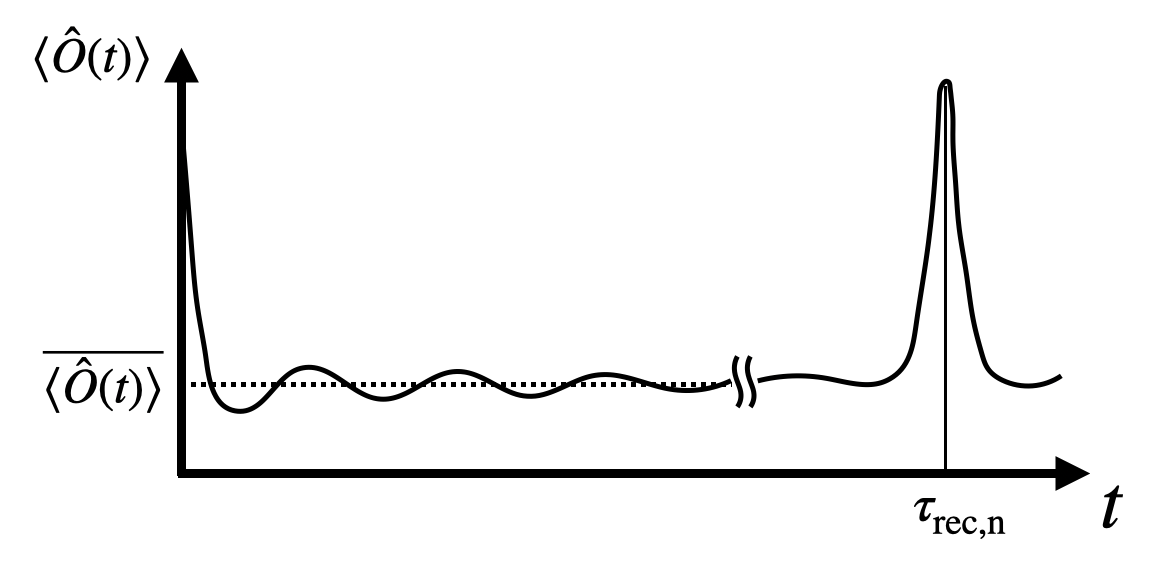,width=3in}}
\vspace*{8pt}
\caption{Schematic illustration of a unitary time evolution of an expectation value of $\hat{O}$. While  $\braket{\hat{O}(t)}\simeq \av{\braket{\hat{O}(t)}}$ for almost all $t$ in systems showing equilibration, this relation cannot hold true for all $t$ due to the recurrence phenomena at time $\tau_{\mr{rec},n}$\:($n\in\mbb{N}$).}
\label{fig:rec}
\end{figure}

\rhcomment{Now, let us clarify the meaning of ``sufficiently long $t$" mentioned above. Importantly,}
Eq.~\eqref{Eq:eqtherm} {cannot} be true for \textit{all} $t$ that satisfy $t\geq \tau$ with some timescale $\tau$ (see \ref{fig:rec}). 
This is due to the quantum recurrence phenomenon,\cite{PB57} which states that for any $\epsilon>0$ 
there exist recurrence times $\tau_{\mr{rec},n}\:(n=1,2,\cdots)$ such that $|\braket{\hat{O}(\tau_{\mr{rec},n})}-\braket{\hat{O}(0)}|<\epsilon$ in finite-dimensional isolated quantum systems, where $\braket{\hat{O}(t)}=\Tr[\hat{O}\hat{\rho}(t)]$.
However, such recurrence phenomena are rare in generic systems since $\tau_{\mr{rec},n}$ are very long.
\rhcomment{Thus, to discuss ``sufficiently long $t$," it is natural to 
focus on the behavior at a typical time in the long run, not at every time.
Namely, we usually say that thermalization occurs if Eq.~\eqref{Eq:eqtherm} holds true for \textit{almost all} $t$, instead of all $t$.
More precisely, we consider a uniform probability measure in the time domain $[0,T]$, and ask whether Eq.~\eqref{Eq:eqtherm} holds with probability one under this measure, when we take the long-time limit $T\ra\infty$.
}

To show the statement that Eq.~\eqref{Eq:eqtherm} holds for \textit{almost all} $t$,
it is sufficient to answer the following two questions:
\begin{enumerate}
\renewcommand{\labelenumi}{(\Roman{enumi})}
\item
Does the temporal fluctuation of $\hat{O}$ become sufficiently small around the stationary value? 
{The answer is affirmative if} 
\aln{\label{Eq:equilibration}
\Tr[\hat{\rho}(t)\hat{O}]\simeq  \Tr[\av{\hat{\rho}(t)}\hat{O}]
}
for almost all $t$.
Here,
\aln{
\av{\hat{\rho}(t)}=\lim_{T\ra\infty}\frac{1}{T}\int_0^Tdt\hat{\rho}(t)
}
is the long-time average of the time-evolving state,  regarded as a stationary state of the dynamics.
\rhcomment{
Note that, in this chapter, the overline $\av{\cdots}=\lim\frac{1}{T}\int_0^Tdt\cdots$ always denotes the infinite-time average, instead of the average over the finite duration in the previous section. }

\item
Is the stationary state well described by the thermal state?
{The answer is affirmative if} 
\aln{\label{Eq:thermalization}
\Tr[\av{\hat{\rho}(t)}\hat{O}]\simeq  \Tr[\hat{\rho}_\mr{therm}\hat{O}].
}

\end{enumerate}

When the condition Eq.~\eqref{Eq:equilibration} is satisfied, we often call that the equilibration (on average) occurs for the expectation value of $\hat{O}$.
When both of the two conditions Eqs.~\eqref{Eq:equilibration} and \eqref{Eq:thermalization} hold, we call that the thermalization occurs for the expectation value of $\hat{O}$.
Note that we can also consider equilibration and thermalization beyond the expectation value of a single observable. 
For example, we often consider an entire set of observables that nontrivially act on a subsystem $S$ and investigate their higher-order quantum fluctuations as well as the expectation value.
In this case, thermalization requires that $\|(\hat{\rho}(t))_S-(\hat{\rho}_\mr{therm})_S\|_1\ll 1$ for almost all $t$, where $(\hat{\rho})_S$ is the reduced density matrix of $\hat{\rho}$  to $S$.
Here, the usage of the trace norm is advantageous because $|\Tr[\hat{O}\hat{\rho}(t)-\hat{O}\hat{\rho}_\mr{therm}]|\leq \|\hat{O}\|\|(\hat{\rho}(t))_S-(\hat{\rho}_\mr{therm})_S\|_1$ holds for any $\hat{O}$ acting on the subsystem $S$ because of the H\"{o}lder inequality.
In the following, however, we focus on the  expectation value of a single observable for simplicity.

The above definition of equilibration on average and thermalization cannot take account of concrete relaxation timescales, since it relies on the statement for almost all $t$.
Thus, there are also works that try to {complement} 
this problem with asking
\begin{enumerate}
\renewcommand{\labelenumi}{(\Roman{enumi})}
\setcounter{enumi}{2}
\item
How can we define the timescale of equilibration?
How should we evaluate it?
\end{enumerate}

Plenty of works have addressed these three questions both numerically and analytically.
This review will not cover all of  the topics; we refer the interested reader to the previous reviews; Refs.~\refcite{AP11,JE15,MU20} provide the 
overview of this field;
Refs.~\refcite{CG16,LD16,TM18JP} are  comprehensive theoretical review papers.

\subsubsection{Equilibration}
We first discuss the question (I) above, i.e, the problem of equilibration on average.
First of all, the validity of Eq.~\eqref{Eq:equilibration} for almost all $t$ is equivalent to the smallness of the temporal fluctuation \rhcomment{of an expectation value $\braket{\hat{O}(t)}=\Tr[\hat{O}\hat{\rho}(t)]$ } defined by
\aln{\label{Eq:tfluc}
\delta O_t=\sqrt{\av{(\braket{\hat{O}(t)}-\av{\braket{\hat{O}(t)}})^2}}.
}
Indeed,  Chebyshev's inequality indicates
\aln{
\mr{Prob}_{t\in[0,\infty)}\lrl{|\braket{\hat{O}(t)}-\av{\braket{\hat{O}(t)}}|\geq\|\hat{O}\|\epsilon}\leq \frac{\delta O_t^2}{\epsilon^2\|\hat{O}\|^2}
} 
for any $\epsilon >0$, where $\mr{Prob}_{t\in[0,\infty)}[\cdots]$ is the probability with respect to the random time uniformly chosen from $[0,\infty)$.\footnote{To be more precise, this probability distribution is not normalizable. For a more rigorous treatment, we first consider the uniform measure in $[0,T]$ and take $T\ra\infty$.}
This means that vanishing the standard deviation $\delta O_t$ in the thermodynamic limit leads to the vanishing probability that $\braket{\hat{O}(t)}$ deviates from $\av{\braket{\hat{O}(t)}}$, i.e., the equilibration on average.

We next evaluate $\delta O_t$,
which was put forward by Refs.~\refcite{HT98,PR08,NL09,AJS11}. We assume  that the state is pure, i.e., $\hat{\rho}(t)=\ket{\psi(t)}\bra{\psi(t)}$, for simplicity.
We introduce the spectral decomposition of the Hamiltonian, $\hat{H}=\sum_{\alpha=1}^{\mr{dim}[\mc{H}]}E_\alpha\ket{E_\alpha}\bra{E_\alpha}$, where $\mc{H}$ is the Hilbert space, $E_\alpha$ is the $\alpha$th energy eigenvalue, and $\ket{E_\alpha}$ is the corresponding energy eigenstate.
Since
\aln{
\braket{\hat{O}(t)}=\sum_{\alpha,\beta}e^{i(E_\alpha-E_\beta)t}c_\alpha^*c_\beta \braket{E_\alpha|\hat{O}|E_\beta},
}
where $c_\alpha=\braket{E_\alpha|\psi(0)}$, 
we can explicitly evaluate $\delta O_t$ using the information of the spectrum.
Indeed, one can show that\cite{AJS12}
\aln{\label{Eq:infiniteshort}
\delta O_t\leq \sqrt{\frac{D_G\|\hat{O}\|^2}{d_\mr{eff}}},
}
where
\aln{
d_\mr{eff}=\frac{1}{\sum_\alpha|c_\alpha|^4}
}
is called the effective dimension 
and
\aln{
D_G=\max_E\lrl{\#\text{ of }(\alpha,\beta)\text{ satisfying }E_\alpha-E_\beta=E}
}
is the maximum degeneracy of the energy gap.
For generic systems without any symmetry, we expect that the gap is not degenerate and that $D_G=1$.

When we assume that $D_G$ is independent of the system size $N$, confirming $d_\mr{eff}^{-1}\ra 0$ is sufficient to show the equilibration.
Intuitively, $d_\mr{eff}\:(1\leq d_\mr{eff}\leq \mr{dim}[\mc{H}])$ measures how the initial state $\ket{\psi(0)}$ is spread over the energy eigenbasis.
For example, $d_\mr{eff}=k$ for $\ket{\psi(0)}=(\ket{E_1}+\ket{E_2}+\cdots+\ket{E_k})/\sqrt{k}$.
While $d_\mr{eff}$ is expected to grow exponentially with increasing $N$ for generic dynamics, it is not easy to show this generally.
On the other hand, when we assume some physically reasonable conditions, we can rigorously show that $d_\mr{eff}$ becomes vanishing for $N\ra\infty$.
Specifically, 
\rhcomment{let us consider a system on a $d$-dimensional hypercubic lattice with $N$ sites.}
We assume that the initial state $\hat{\rho}(0)$ (or, alternatively, its time average $\av{\hat{\rho}(t)}$)
satisfies the following exponential clustering property of correlations: there exist $N$-independent constants \rhcomment{$\xi>0$ and $K>0$} such that for any lattice regions $X$ and $Y$
\aln{\label{Eq:clustering}
\max_{\mr{supp}[\hat{A}]{\subseteq}
X,\mr{supp}[\hat{B}]{\subseteq}
Y}
\lrv{\frac{\braket{\hat{A}\hat{B}}-\braket{\hat{A}}\braket{\hat{B}}}{|X||Y|\cdot\|\hat{A}\|\|\hat{B}\|}}\leq Ke^{-\mr{dist}(X,Y)/\xi},
}
where {$\mr{supp}[\hat A]$ denotes the support of $\hat A$, $|X|$ is the cardinality (number of sites) of $X$, \rhcomment{${\rm dist}(X,Y)\equiv\min\{{\rm dist}(x,y)|x\in X,y\in Y\}$ denotes the distance between $X$ and $Y$, and}} $\braket{\cdots}$ denotes the average over $\hat{\rho}(0)$ (or $\av{\hat{\rho}(t)}$).
We also assume that the Hamiltonian $\hat{H}$ is $k$-local, i.e., $\hat{H}=\sum_{X:{|X|}
\leq k}\hat{h}_X$ with $\hat{h}_X$ having a support on $X$ ($k$ is independent of $N$), and that $\|\hat{h}_X\|$ is upper bounded {by} 
an $N$-independent constant.
Under these conditions of the exponential clustering property and the locality, we can show that\cite{TF17}
\aln{\label{Eq:Brandao}
\frac{1}{d_\mr{eff}}\leq \frac{CN\ln^{2d}N}{\Delta E^3},
}
with some $N$-independent constant $C$, where $\Delta E$ is the energy fluctuation for $\hat{\rho}(0)$ (or, equivalently, $\av{\hat{\rho}(t)}$).
The proof is made by quantum extension\cite{FGSL15} of the Berry-Esseen theorem, which evaluates the difference between the sum of random variables and the normal distribution.

We stress that $\Delta E$ is often easier to calculate than $|c_\alpha|$.
Moreover,  $\Delta E\propto \sqrt{N}$ for typical macroscopic systems, which indicates that $d_\mr{eff}$ indeed vanishes in the thermodynamic limit.
In conclusion, equilibration is justified just by evaluating $\Delta E$ for locally interacting Hamiltonians for the state with the exponential clustering property.

Although inequality \eqref{Eq:Brandao} applies to a broad class of systems without any assumption about energy eigenstates, its convergence  with respect to $N$ is very slow.
In fact, generic systems are expected to have exponentially small $\delta O_t$ with respect to $N$.
In Ref.~\refcite{HW19PRL}, this exponentially good convergence is shown to be ensured if we assume a certain volume law of entanglement for energy eigenstates.
Here, the volume law of entanglement intuitively means that $S_z(\hat{\rho}^\alpha_X)\sim |X|$ for sufficiently large subsystem $X$, where $\hat{\rho}^\alpha_X$ is the reduced density matrix of $\ket{E_\alpha}\bra{E_\alpha}$ and $S_z(\hat{\rho})=\ln\Tr[\hat{\rho}^z]/(1-z)$ is the R\'{e}nyi-$z$ entropy.
Roughly speaking, Ref.~\refcite{HW19PRL} showed that, when every eigenstate of a locally interacting Hamiltonian with finite energy density satisfies the volume law for R\'{e}nyi-2 entanglement entropy for some subsystem,
$1/d_\mr{eff}$ becomes exponentially small $\sim e^{-kN}$ with respect to $N$ for an initial product state with finite energy density  (see the reference for the rigorous statement).
Since generic many-body systems are expected to satisfy the volume law of entanglement entropy (see the next section), this theorem strongly supports  equilibration for such systems.

\subsubsection{Thermalization}
We next discuss  the question (II) above, i.e., the problem of thermalization.
In general, this problem is more challenging than the problem of equilibration in that fewer rigorous results have been  obtained.
One of the most important topics concerning thermalization is the eigenstate thermalization hypothesis (ETH), \rhcomment{which was essentially proposed in Ref.~\refcite{JVN29} for macroscopic observables, clearly stated in Refs.~\refcite{JMD91,MS94} on the basis of random-matrix argument and semiclassical analysis, respectively, and demonstrated in Ref.~\refcite{MR08} by a beautiful numerical simulation using hardcore bosons.}
The ETH provides a sufficient condition for the relaxation to thermal equilibrium from any initial state and is numerically verified in various chaotic systems.
Here, after briefly reviewing the ETH,
we refer to some {rigorous} 
results on related topics (see reviews~\refcite{LD16,TM18JP} for detailed and comprehensive discussions).

\paragraph{\rhcomment{The ETH and thermalization.}}
To simplify the discussion, let us assume that the Hamiltonian has no degeneracy.
In this case, Eq.~\eqref{Eq:thermalization} can be written as
\aln{\label{Eq:averaged_therm}
\sum_\alpha|c_\alpha|^2\braket{E_\alpha|\hat{O}|E_\alpha}\simeq \Tr[\hat{O}\hat{\rho}_{E,\delta E}]=\frac{1}{\mr{dim}[\mc{H}_{E,\delta E}]}\sum_{\alpha:\ket{E_\alpha}\in\mc{H}_{E,\delta E}} \braket{E_\alpha|\hat{O}|E_\alpha},
}
where we take the microcanonical state $\hat{\rho}_{E,\delta E}$ for the thermal state.
Here, $\mc{H}_{E,\delta E}$ is the microcanonical energy window, i.e., the Hilbert space spanned by energy eigenstates of $\hat{H}$ whose eigenvalues are within $[E,E+\delta E]$ (The choice of $\delta E$ is not important  as long as it is subextensive).
What is the condition for this equality to hold? 
If all $|c_\alpha|^2$ are approximated by $1/\mr{dim}[\mc{H}_{E,\delta E}]$, this equality holds;
however, it requires a specific choice of initial states, which is not satisfactory for understanding thermalization from a broader class of initial states.
Instead, it is now widely accepted that quantum thermalization for generic systems occurs because of the ETH, which states that
\aln{\label{Eq:ETH}
 \braket{E_\gamma|\hat{O}|E_\gamma}\simeq \Tr[\hat{O}\hat{\rho}_{E_\gamma,\delta E}]=\frac{1}{\mr{dim}[\mc{H}_{E_\gamma,\delta E}]}\sum_{\alpha:\ket{E_\alpha}\in\mc{H}_{E_\gamma,\delta E}} \braket{E_\alpha|\hat{O}|E_\alpha}
}
for \textit{every} $\gamma$ in the energy scale $E_\gamma\in[E_1,E_2]$ of interest.
This hypothesis intuitively indicates that every energy eigenstate $\ket{E_\gamma}$ serves as a thermal state by itself.
Note that the above definition that we adopt in this paper is again only for the expectation values of an observable $\hat{O}$. Instead, we can formulate the ETH for the subsystem $S$ by requiring that $\|(\ket{E}_\alpha\bra{E}_\alpha)_S-(\hat{\rho}_\mr{therm})_S\|_1$ is vanishingly small.

If we assume the ETH, thermalization is justified for all initial states with the corresponding energy scale $[E,E+\delta E]\subset [E_1,E_2]$ and the subextensive energy fluctuation $\Delta E$.
Indeed, since $\Delta E$ is small, all $\braket{E_\alpha|\hat{O}|E_\alpha}$ in the left-hand side {(lhs)} of Eq.~\eqref{Eq:averaged_therm} can be replaced with $\Tr[\hat{O}\hat{\rho}_{E_\alpha,\delta E}]\simeq \Tr[\hat{O}\hat{\rho}_{E,\delta E}]+c\Delta E+\:(\text{higher order terms})$ using Eq.~\eqref{Eq:ETH}, where $c$ is a constant depending on $\hat{O}$.
Since the term involving $\Delta E$ is negligible in the thermodynamic limit,
we have $\sum_\alpha|c_\alpha|^2\braket{E_\alpha|\hat{O}|E_\alpha}\simeq \sum_\alpha|c_\alpha|^2\Tr[\hat{O}\hat{\rho}_{E,\delta E}]=\Tr[\hat{O}\hat{\rho}_{E,\delta E}]$.
This discussion can be made more precise.\cite{GDP15}

\paragraph{\rhcomment{Off-diagonal matrix elements and random-matrix theory.}}
While the ETH predicts the behavior of the diagonal matrix elements of the energy eigenstates in the thermodynamic limit,
it can be generalized to off-diagonal matrix elements and finite system sizes.
Indeed, \rhcomment{it was conjectured (first for semiclassical systems\cite{MS99} and later general many-body systems\cite{EK13})} that matrix elements for chaotic systems take the following form:
\aln{\label{Eq:srednicki}
\braket{E_\alpha|\hat{O}|E_\beta}=\Tr[\hat{O}\hat{\rho}_{E_\alpha,\delta E}]\delta_{\alpha\beta}
+f_{\hat{O}}(\av{E}_{\alpha\beta},\omega_{\alpha\beta})e^{-S(\av{E}_{\alpha\beta})/2}R_{\alpha\beta},
}
where $\av{E}_{\alpha\beta}=(E_\alpha+E_\beta)/2$, $\omega_{\alpha\beta}=E_\alpha-E_\beta$,
$S(E)$ is the thermodynamic entropy at energy $E$, $f_{\hat{O}}(E,\omega)$ is a smooth function, and $R_{\alpha\beta}$ is a pseudo-random variable with zero mean and 
unit variance.
The distribution of $R_{\alpha\beta}$ is expected to be Gaussian for local observables, but can be non-Gaussian for some non-local many-body observables.\cite{RH1719,IMK19}
If we assume this conjecture, we immediately obtain the ETH by considering the diagonal elements and thermodynamic limit.
The conjecture \eqref{Eq:srednicki}  indicates that the pseudo-random fluctuations around 
$\Tr[\hat{O}\hat{\rho}_{E_\alpha,\delta E}]\delta_{\alpha\beta}$ are exponentially small with respect to the system size due to the factor $e^{-S(\av{E}_{\alpha\beta})/2}$, since $S(E)$ is extensive.

The hypotheses Eqs.~\eqref{Eq:ETH} and \eqref{Eq:srednicki} are deeply related to 
the analogy between quantum chaotic systems and random matrix theory.\cite{FH10}
In particular, it is widely believed that certain spectral statistics of quantum chaotic systems, such as \rhcomment{the statistics of nearest-eigenvalue spacings\cite{OB84}}, coincide with those of random matrices.
While the analogy is mainly verified by numerical simulations, it is verified analytically for the quantity called the spectral form factor for semiclassically chaotic quantum systems\cite{SM04,SM05} and Floquet systems described by dual unitary dynamics.\cite{BB18}

For energy eigenstates, assuming eigenstate distributions of random matrices\cite{MVB77} leads to analogous expressions to Eqs.~\eqref{Eq:ETH} and \eqref{Eq:srednicki}.
For example, if we pick an eigenstate of the Gaussian random matrix with dimension $D$,
we have $\braket{E_\alpha|\hat{O}|E_\beta}\simeq\frac{\Tr[\hat{O}]}{D}\delta_{\alpha\beta}+\sqrt{\frac{\Tr[\hat{O}^2]-\Tr[\hat{O}]^2}{D}}R_{\alpha\beta}$,\cite{LD16,RH1719}
which is similar to Eq.~\eqref{Eq:srednicki} but free from the energy dependence.
We note that the analogy with random matrix theory also results in the volume law of entanglement entropy for chaotic energy eigenstates.\cite{DN93}

While the ETH and the ansatz by Srednicki themselves remain conjectures, several analytical results related to these concepts have been investigated.
 First, let us discuss the behavior of $\braket{E_\alpha|\hat{O}|E_\beta}$ for large $|E_\alpha-E_\beta|$, which corresponds to the behavior of $f_{\hat{O}}(E,\omega)$ for large $\omega$ if we assume Eq.~\eqref{Eq:srednicki}.
 When we consider a local observable $\hat{A}$ and a few-body interacting Hamiltonian $\hat{H}$, the off-diagonal matrix elements are shown to decay exponentially with respect to the energy difference.
In particular, when $E_\alpha\in [E',\infty)$ and $E_\beta\in [0,E]$ with $E<E'$, it is sufficient to show that $\|\hat{\Pi}_{[E',\infty)}\hat{A}\hat{\Pi}_{[0,E]}\|$ is exponentially small, where $\hat{\Pi}_{[a,b]}$ is the projection operator onto the subspace with energy within the energy window $[a,b]$.
To be specific, we assume that the Hamiltonian is $k$-body interacting ($k$ is independent of $N$) with the form
$\hat{H}=\sum_{X:|X|\leq k}\hat{h}_X$, where $X$ denotes a set of lattice sites.
Let us also assume that local energy at each single site $i$ is bounded as $\sum_{X:i\in X}\|\hat{h}_X(t)\|\leq g$ with $N$-independent constant $g$.
We take a local observable $\hat{A}$ supported on finite lattice sites.
We define $\mc{A}$ as a subset of lattice sites composing the interaction term of $[\hat{H},\hat{A}]$ ($=\sum_{X\in \mc{A}}[\hat{h}_X,\hat{A}]$).
Then, defining $R=\sum_{X\in\mc{A}}\|\hat{h}_X\|$, one can show\cite{IA16}
\aln{
\|\hat{\Pi}_{[E',\infty)}\hat{A}\hat{\Pi}_{[0,E]}\|\leq \|\hat{A}\|e^{-\frac{1}{gk}\lrl{E'-E-R\lrs{1+\ln\frac{E'-E}{R}}}}\leq \|\hat{A}\|e^{-\frac{1}{2gk}(E'-E-2R)}.
}
Since $R$ is independent of $N$ in our setting, macroscopic energy difference $E'-E=\mc{O}(N)$ means the exponential decay of off-diagonal terms.

 What is the behavior of off-diagonal terms in the other limit, i.e., small $\omega$?
In the conjecture in Eq.~\eqref{Eq:srednicki}, it is often assumed from random matrix theory that $f_{\hat{O}}(E,\omega)$ is almost constant with respect to $\omega$ for $\omega \lesssim E_{\mr{Th}}$, 
which is regarded as a many-body Thouless energy.\cite{LD16}
Although this behavior is confirmed for several situations\cite{LD16,MS17}, we note that there exist arbitrary many local observables that do not satisfy this condition.
In fact, if we take a local observable $\hat{O}$ for which $f_{\hat{O}}(E,\omega)$ is a (non-zero) constant 
 for $\omega \lesssim E_{\mr{Th}}$, another local observable $i[\hat{H},\hat{O}]$ leads to $f_{i[\hat{H},\hat{O}]}(E,\omega)=i\omega f_{\hat{O}}(E,\omega)$, which is a non-constant with respect to $\omega$ even for $\omega \lesssim E_{\mr{Th}}$.\cite{RH1719,RH18PRL}
\rhcomment{These behaviors of $f(E,\omega)$ for large and small $\omega$ are not captured by the random matrix theory alone and reflect the specific structure (e.g., few-body property of interactions) of quantum many-body systems.}
 
\paragraph{\rhcomment{The weak version of the  ETH.}}
Another known result is about the number of energy eigenstates that are thermal.
While the standard ETH requires that Eq.~\eqref{Eq:ETH} holds for all $\gamma$ (at a given energy scale of interest), 
a weaker version of the ETH,\cite{GB10,TNI13,EI17} asserting that Eq.~\eqref{Eq:ETH} holds for almost all $\gamma$, can be rigorously proven under some conditions, \rhcomment{ such as locality of interactions, translation invariance, and the exponential decay of correlation functions.\cite{EI17}}
Reference~\refcite{TM16arxiv} used the large deviation technique\cite{HT09} to show that the number of energy eigenstates that are far from the microcanonical average decays exponentially.
To be precise, let us consider a locally interacting Hamiltonian with a finite-temperature state in one dimension or  a sufficiently high-temperature state in higher dimensions.
Then, one can show that, for any $\epsilon >0$, there exists $\Gamma >0 $ such that
\aln{\label{Ineq:weakETH}
\mr{Prob}_{\ket{E_\gamma}\in \mc{H}_{E,\delta E}}\lrl{\lrv{\braket{E_\gamma|\hat{O}|E_\gamma}-\Tr[\hat{O}\hat{\rho}_{E,\delta E}]}>\epsilon}\leq e^{-\Gamma N},
}
where $\mr{Prob}_{\ket{E_\gamma}\in \mc{H}_{E,\delta E}}$ means that we uniformly choose an eigenstate from the microcanonical energy shell.
The right-hand side vanishes in the thermodynamic limit, indicating that most energy eigenstates behave thermally.
Note that the exponential decay of the fraction of the exceptional states in \eqref{Ineq:weakETH} is numerically found for integrable systems; on the other hand, the fraction is found to decay double exponentially for non-integrable systems.\cite{TY18}
We also note that the weak version of the ETH is, in general, not sufficient to justify thermalization.

\paragraph{\rhcomment{The breakdown and the universality of the ETH.}}
The validity of the (standard) ETH deeply depends on the structure of the Hamiltonian.
Many nonintegrable systems are numerically found to satisfy the ETH;
\rhcomment{however, the ETH and thermalization are known to be violated for
several situations.
For example, 
integrable systems relax to the so-called generalized Gibbs ensemble\cite{MR07,MR09}, which takes the conserved quantities for integrable systems into account in addition to energy (review papers: Refs.~\refcite{FHL16,LV16}).
Another famous example is 
systems showing many-body localization\cite{VO07,MZ08M,AP10M}, where strong disorder prohibits the system from thermalizing  (review papers: Refs.~\refcite{RN15,DAA19}).
Recently,  new types of non-thermalizing systems attract recent attention, such as quantum many-body scars\cite{NS17PRL,CJT18,MS21} and the  Hilbert-space fragmentation.\cite{SP19,PS20,VK20}
}

Against this background, the natural question is whether the ETH and thermalization are universal or not.
One naive way to address this question is to assume that energy eigenstates $\lrm{\ket{E_\alpha}}$ of a generic Hamiltonian and the eigenstates $\lrm{\ket{o_s}}$ diagonalizing the {projected} observable (${\hat P_{E,\delta E}}\hat{O}{\hat P_{E,\delta E}}=\sum_{s=1}^{\dim[\mc{H}_{E,\delta E}]}o_s\ket{o_s}\bra{o_s}${, where $\hat P_{E,\delta E}=\sum_{E_\alpha\in[E,E+\delta E]}|E_\alpha\rangle\langle E_\alpha|$}) are uncorrelated to each other.
In fact, if we assume that the basis-transformation matrix between $\lrm{\ket{E_\alpha}}$ and  $\lrm{\ket{o_s}}$, denoted by $U$, is chosen uniformly from the unitary Haar measure, the ETH follows.\cite{JVN29,SG10PRE,SG10Eur,PR15}
However, it was shown that this argument cannot be justified unless the energy window $\delta E$ is exponentially small when we consider a set of few-body (or local) Hamiltonian and observables.
In such an experimentally relevant setup, a nontrivial correlation between $\lrm{\ket{E_\alpha}}$ and  $\lrm{\ket{o_s}}$ exists and  prevents us from identifying $U$ as a randomly sampled matrix from the Haar measure.\cite{RH18PRL}

Recently, universality of the ETH for local translation-invariant spin-1/2 Hamiltonians  was numerically addressed in Ref.~\refcite{SS21PRL}.
The authors introduced $\hat{H}=\sum_{i=1}^N\hat{T}^i\hat{h}_{(l)}\hat{T}^{-i}$, where {$\hat T$ is the lattice translation operator and} $\hat{h}_{(l)}$ acts nontrivially on $l$ consecutive spins, and took $\hat{h}_{(l)}$ as a $2^l\times 2^l$ random matrix.
They considered the indicator of the ETH as
\aln{
\Delta_\infty=\frac{\max_\gamma \lrv{\braket{E_\gamma|\hat{O}|E_\gamma}-\Tr[\hat{O}\hat{\rho}_{E_\gamma,\delta E}]}}{\max_so_s-\min_so_s},
}
where $\Delta_\infty \ra{0}
\:(N\ra\infty)$ indicates the ETH for a given $\hat{h}_{(l)}$.
They then focused on the following Markov inequality
\aln{
\mr{Prob}_{\hat{h}_{(l)}}[\Delta_\infty\geq\epsilon]\leq \frac{\mr{Ave}_{\hat{h}_{(l)}}[\Delta_\infty]}{\epsilon}{,}
}
{finding (numerically)} that the average $\mr{Ave}_{\hat{h}_{(l)}}[\Delta_\infty]$ with $l\geq 2$
decays as $\sim Ne^{-c_mN}\:(c_m>0)$ for large $N$.
By taking $\epsilon =e^{-cN}$ with $0<c<c_m$, we can show that the fraction of the local Hamiltonians for which $\Delta_\infty$ does not decay exponentially is exponentially suppressed with $N$.
This indicates the universality of the ETH for locally interacting systems.
If we give up locality and consider power-law decaying interactions ($\sim r^{-\alpha}$), $\Delta_\infty$ becomes larger as we decrease $\alpha$, which reflects the {permutation} symmetry at $\alpha=0$.\cite{SS21arxiv}
In particular, the signature of the universality of the ETH was not found for a system that consists of up to 20 spins with $\alpha\leq 0.5$, which is consistent with the experiment with trapped ions.\cite{BN17L}

\paragraph{\rhcomment{The ETH near the ground state.}}
Before ending this subsection, let us review a recent work that proves the ETH for a specific setting.
As mentioned above, the ETH in Eq.~\eqref{Eq:ETH} and the conjecture in Eq.~\eqref{Eq:srednicki} are often attributed to the analogy between quantum chaos and random matrix theory, which is expected only for the middle of the energy spectrum ($\min_\alpha E_\alpha \ll E_1<E_2\ll \max_\alpha E_\alpha$).
Counterintuitively, however, recent work~\refcite{TK20PRL} succeeded in a rigorous proof of the ETH in Eq.~\eqref{Eq:ETH} near the \textit{ground} states of a class of Hamiltonians and observables.
To be specific, they considered a general spin Hamiltonian $\hat{H}$ that consists of $N$  local terms as $\hat{H}=\sum_{i=1}^N\hat{h}_i$ on a $d$-dimensional hypercube and a global observable $\hat{O}$ that consists of $N$ local terms as 
$\hat{O}=\frac{1}{N}\sum_{i=1}^N\hat{o}_i$.
Here, $\hat{h}_i$ and $\hat{o}_i$ nontrivially act on {finite regions} 
whose diameters are less than $l_0$ and $l$, respectively.
We also set the energy of the ground state is zero.
Now, assume that the canonical ensemble with inverse temperature $\beta^*$
satisfies the clustering property (cf. Eq.~\eqref{Eq:clustering}) in the form $|\braket{\hat{A}_X\hat{B}_Y}_{\beta^*}-\braket{\hat{A}_X}_{\beta^*}\braket{\hat{B}_Y}_{\beta^*}|\leq \|\hat{A}_X\|\|\hat{B}_Y\|e^{-\mr{dist}(X,Y)/\xi}$ for $\mr{dist}(X,Y)>r$ with some constants $r$ and $\xi$, where $\hat{A}_X\:(\hat{B}_Y)$ are arbitrary operators nontrivially acting on regions $X\:(Y)$.
Then, Ref.\refcite{TK20PRL} showed the following bound
\aln{
\lrv{\braket{E_\alpha|\hat{O}|E_\alpha}-\braket{\hat{O}}_{\beta=\infty}}\leq \frac{1}{\sqrt{N}}\max\lrs{c_1\mc{A}_1,c_2\mc{A}_2}
}
for any energy eigenstate $\ket{E_\alpha}$, 
where $\braket{\hat{O}}_{\beta=\infty}$ is the expectation value over the ground state,  $\mc{A}_1=
{(\beta^* E_\alpha+\ln Z_{\beta^*})}^{(d+1)/2}$, $\mc{A}_2=[l^d(\beta^* E_{{\alpha}} 
+\ln Z_{\beta^*})]^{1/2}$ with $Z_\beta=\Tr[e^{-\beta\hat{H}}]$, and $c_1, c_2$ are constants depending on $r, d,\xi$, and $l_0$.
If we choose $\beta^*$ such that $\log Z_{\beta^*}=\mc{O}(1)$ and consider a low-energy regime $E_{{\alpha}} 
= \mc{O}({N}^c
)\:(c<\frac{1}{d+1})$, the right-hand side vanishes.
This means that the  low-lying energy eigenstates are indistinguishable from the ground state if we observe $\hat{O}$, meaning that the ETH holds near the ground state.
The above choice of $\beta^*$ is possible when the density of states in the low-temperature regime is sufficiently sparse, which typically holds true for gapped systems.
Note that the ETH near the ground state does not require nonintegrability of the Hamiltonian and is expected to occur by a different mechanism of the high-temperature ETH with the random-matrix-type matrix elements in Eq.~\eqref{Eq:srednicki}.

\subsubsection{Equilibration timescales}
\label{Sec:Eqtime}
In this subsection, we consider the question (III) raised at the beginning of the section, i.e., the problem of timescale of equilibration.
This problem is hard to tackle because of at least three difficulties.
First, the definition of timescales is less definitive than that of equilibration or thermalization{, i.e., there is no good consensus in the literature}.
Second, details of timescales seem to  depend on the details of the system highly. For example, certain systems exhibit prethermalization behavior with multiple timescales.
Third, a nonequilibrium state involves the complicated superposition of energy eigenstates, and
\rhcomment{we may need to deal with the process of many-body dephasing of such eigenstates.\cite{HW18Eq,TRO18}}
Thus, it is less easy to extract useful information for timescales than the case for thermalization, which is clearly explained by the ETH.

To demonstrate the complexity of considering timescales, we first introduce an analytical result 
showing that the timescale can be both extraordinar{ily} 
long and short depending on the choice of the  subspace to define the relaxation.
In Ref.~\refcite{SG13PRL}, the authors introduced a measure of equilibration to a subspace $\mc{H}_\mr{eq}$ as $p(T)=\frac{1}{T}\int_0^Tdt\braket{\psi(t)|\hat{P}_\mr{neq}|\psi(t)}$.
Here, $\hat{P}_\mr{neq}$ is the projection operator onto a nonequilibrium subspace $\mc{H}_\mr{neq}$, which is defined through $\mc{H}_{E,\delta E}=\mc{H}_\mr{eq}\oplus\mc{H}_\mr{neq}$.
Although the equilibrium subspace should be determined from physically relevant observables (such as local observables), we here take $\mc{H}_\mr{eq}$ and $\mc{H}_\mr{neq}$ at our will.
The condition  $p(T)\ll 1$ for $^\forall T>T_0$  indicates that the state reaches equilibrium after the timescale $T_0$.\footnote{Since the expression of $p(T)$ involves average over time, this quantity does not suffer from the problem of recurrence phenomena in evaluating the degree of equilibration.}
Under certain conditions, this is justified for infinitely large $T$.
On the other hand, reference~\refcite{SG13PRL} showed that the timescale can be very long if we take a specific $\mc{H}_\mr{neq}$.
They showed that, for any dimension $0<d\leq \mr{dim}[\mc{H}_{E,\delta E}]$, and any state $\ket{\eta}\in\mc{H}_{E,\delta E}$, there exists a subspace $\mc{H}_\mr{neq}\ni \ket{\eta}$ with $\mr{dim}[\mc{H}_\mr{neq}]=d$ such that, for any initial state  $\ket{\psi(0)}\in\mc{H}_\mr{neq}$, one has
\aln{
p(T)\geq\frac{3}{\pi}=0.95...
}
with any $T$ satisfying $T\leq \frac{\pi}{6\delta E}d$.
When we consider the case where $1\ll d= e^{\mathcal{O}(N)}\ll \mr{dim}[\mc{H}_{E,\delta E}]$, the inequality means that $T_0$ becomes exponentially large with respect to the system size for some artificially chosen subspace $\mc{H}_\mr{neq}$ (note that the subspaces with this property can be different from physically realistic nonequilibrium subspace).
In contrast, Ref.~\refcite{SG15NJP} showed that the extremely quick equilibration occurs for most of the subspaces with respect to the Haar measure.
Precisely speaking, we take an arbitrary basis state $\lrm{{\ket{\phi_s}}}_{s=1}^{d}$ and construct a random subspace $\hat{P}_\mr{neq}=\sum_{s=1}^{d}U\ket{\phi_s}\bra{\phi_s}U^\dag$ with a $\mr{dim}[\mc{H}_{E,\delta E}]\times \mr{dim}[\mc{H}_{E,\delta E}]$ random unitary matrix $U$ over the Haar measure.
Then, for almost all $\hat{P}_\mr{neq}$, one can show
\aln{
p(T)\lesssim \frac{\tau_B}{T}
}
for any initial state $\ket{\psi(0)}$ and any $T$ such that $T\leq \tau_B\min\lrm{(\mr{dim}[\mc{H}_{E,\delta E}]/d)^{1/4}, (\mr{dim}[\mc{H}_{E,\delta E}])^{1/6}}$.
Here, $\tau_B=h\beta$ is called the Boltzmann time, where the temperature $\beta^{-1}$ is determined from the energy derivative of the density of the states.
Since $\tau_B$ is typically small for high energies (e.g., $\tau_B\sim 10^{-13}$s at $\beta^{-1}=300$K),
this inequality indicates that the equilibration is achieved very quickly for short times satisfying $T\gg \tau_B$.
This fast timescale cannot explain macroscopic transport behavior or information propagation discussed in sections~\ref{Sec:speed_macro} and~\ref{Sec:LRbound}, since taking random $\hat{P}_\mr{neq}$ gives up the local or few-body structure of the system.
On the other hand, it is also argued\cite{PR16Nat} that the Boltzmann time may capture the rapid relaxation to the local thermal equilibrium, instead of the global thermal equilibrium. \textcolor{black}{We note that a very recent work Reference~\refcite{ZN22} also discussed a different but related timescale using the uncertainty principle for local observables.}

While the above discussion relies on a nonequilibrium subspace, we can discuss equilibrium timescale 
based on 
observable{s}.
In Ref.~\refcite{AJS12}, Short and Farrelly considered a measure of the finite-time equilibration of an observable as the temporal fluctuation over the range $[0,T]$,
\aln{
M(T)=\frac{1}{T\|\hat{O}\|^2}\int_0^Tdt\lrs{\braket{\hat{O}(t)}-\av{{\langle}\hat{O}(t){\rangle}}}^2, 
}
where $\av{\cdots}$ denotes the infinite-time average.
Note that $\lim_{T\ra\infty}M(T)=\delta O_t^2/\|\hat{O}\|^2$, where $\delta O_t$ is defined in Eq.~\eqref{Eq:tfluc}.
Reference~\refcite{AJS12} then showed the upper bound for $M(T)$ for a time independent Hamiltonian as
\aln{\label{Eq:finiteshort}
M(T)\leq \frac{N_G(\epsilon)}{d_\mr{eff}}\lrs{1+\frac{8\log_2d_E}{\epsilon T}}
}
for any $T>0$ and $\epsilon>0$.
Here, $d_E$ is the number of the distinct energies in the energy spectrum of $\hat{H}$,
and $N_G(\epsilon)$ denotes the maximum number of the energy gap at a density $\epsilon$, i.e., 
\aln{
N_G(\epsilon)=\max_E\lrl{\#\text{ of }(\alpha,\beta)\text{ satisfying }E_\alpha-E_\beta\in [E,E+\epsilon)}.
}
Since $\lim_{\epsilon\ra 0^+}N_G(\epsilon)=D_G$, the limit $\lim_{\epsilon\ra 0^+}\lim_{T\ra\infty}$ of inequality \eqref{Eq:finiteshort} reduces to inequality \eqref{Eq:infiniteshort}.

While inequality \eqref{Eq:finiteshort} addresses a general bound for finite-time fluctuations, the bound is in general loose for macroscopic systems.
Reference~\refcite{LPG17} derived timescales that are more plausible for realistic settings.
Here, we explain one of the simplest applications of their results, which corresponds to Theorem 1 in their paper.
Their approach is to expand $M(t)$ with energy eigenstates as
\aln{
M(T)=\sum_{a,b}\frac{v_av_b^*}{T}\int_0^Tdt e^{-i(G_a-G_b)t},
}
where $a,b,\dots$ denote the set of $(\alpha,\beta)$, $G_a=E_\alpha-E_\beta$ is the energy gap, and $v_a=\braket{E_\alpha|\hat{\rho}|E_\beta}\braket{E_\beta|\hat{A}|E_\alpha}/\|\hat{A}\|$.
They also define a probability distribution $p_a=|v_a|/Q$ with $Q=\sum_a|v_a|$.
Now, consider a situation where a small subsystem $S$ is attached to a large bath $B$, whose Hamiltonian can be written as $\hat{H}=\hat{H}_S+\hat{H}_I+\hat{H}_B$.
We assume that the initial state is given by a mixed product state $\hat{\rho}=\hat{\rho}_S\otimes \hat{\rho}_{B,\beta=0}$, where $\hat{\rho}_{B,\beta=0}$ is the infinite temperature state for the bath.
Then, for an observable of the subsystem $\hat{A}=\hat{A}_S\otimes\hat{\mbb{I}}_B$,
one can show that
\aln{
M(T)\leq \frac{T_\mr{eq}}{T}+4\pi\delta(\epsilon)Q^2
}
with
\aln{
T_\mr{eq}=\frac{4\pi a(\epsilon)\|\hat{A}_S\|^{1/2}Q^{5/2}}{\fracd{^2\braket{\hat{A}_S}}{t^2}|_{t=0}}
}
for any $\epsilon$, and that $Q$ is bounded as $Q\leq \sqrt{d_S\Tr_S[\hat{\rho}_S^2]}$ ($d_S$ is the Hilbert-space dimension of the subspace).
Here, $\delta(\epsilon)$ and $a(\epsilon)$ are quantities that are determined from the probability distribution of $\{p_\alpha\}$ (see Ref.~\refcite{LPG17} for explicit representations).

Let us assume that we can have $\delta(\epsilon)\ll 1$ and $a(\epsilon)\sim 1$ by an appropriate choice of $\epsilon$.
Since $Q$ is finite, we can see that $M(T)$ becomes sufficiently small for $T\gg T_\mr{eq}$, where $T_\mr{eq}$ is not too small or large.
As a caveat, the above assumption, e.g., non-large $a(\epsilon)$, can be violated for certain slow dynamics, such as a system that consists of a conserved part and a small conservation-breaking Hamiltonian.\cite{RH20PRX}

\subsection{Dissipative systems}
\subsubsection{GSKL equation and its spectral decomposition}
As discussed in the previous section, it is not easy to extract equilibration (or thermalization) timescales of physical observables from the system's spectrum alone in isolated systems.
On the other hand, the situation seems to be simplified for non-unitary systems, such as systems with external dissipation.
In this section, let us consider quantum dissipative systems described by the GKSL equation 
\aln{
\fracd{\hat{\rho}}{t}=\mbb{L}[\hat{\rho}]=-i[\hat{H},\hat{\rho}]+\sum_\mu\lrs{\hat{L}_\mu\hat{\rho}\hat{L}_\mu^\dag-\frac{1}{2}\lrm{\hat{L}_\mu^\dag\hat{L}_\mu,\hat{\rho}}}
}
and ask the timescale for the system to equilibrate to the stationary state.
Note that, in contrast with isolated systems, there is no quantum recurrence due to the Markovian environment.

\begin{figure}[bt]
\centerline{\psfig{file=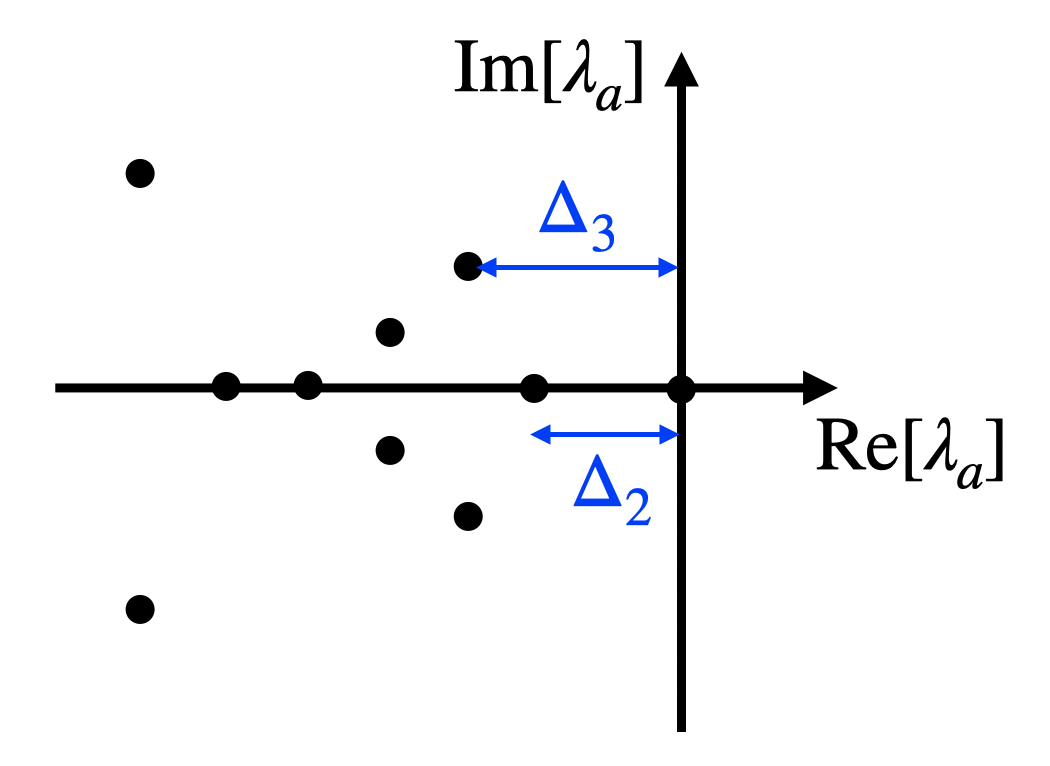,width=3in}}
\vspace*{8pt}
\caption{Typical example of the eigenvalues $\lambda_a$ of the GKSL super-operator. Assuming a unique stationary state, which corresponds to $\lambda_1=0$, all the other eigenvalues become negative. We can define the gap $\Delta_a$ as the (absolute value of) real part of $\lambda_a$.}
\label{fig:lind}
\end{figure}

Since the GKSL super-operator is a linear operator, we can consider its eigenstate equation.
\rhcomment{Assuming that the dimension of the Hilbert space is finite and that $\mbb{L}$ is diagonalizable, we have}
\aln{
\mbb{L}[\hat{\phi}_a]=\lambda_a\hat{\phi}_a,\quad\quad\mbb{L}^\dag[\hat{\chi}_a]=\lambda_a^*\hat{\chi}_a,
}
where 
\aln{
\mbb{L}^\dag[\hat{\rho}]=i[\hat{H},\hat{\rho}]+\sum_\mu\lrs{\hat{L}_\mu^\dag\hat{\rho}\hat{L}_\mu-\frac{1}{2}\lrm{\hat{L}_\mu^\dag\hat{L}_\mu,\hat{\rho}}}
}
is the conjugate of $\mbb{L}$, and
$\hat{\phi}_a$ and $\hat{\chi}_a$ are respectively the right and left eigenstate of $\mbb{L}$ with eigenvalue $\lambda_a\rhcomment{\in\mathbb{C}}$.
Note that $\rhcomment{\mathrm{Re}}[\lambda_a]\leq 0$ for every $a$\rhcomment{, since the mode with $\mathrm{Re}[\lambda_a]>0$ would grow indefinitely in time (see Eq.~\eqref{Eq:lindblad_expand}), which is incompatible with the relaxation to the stationary state (see e.g., Ref.~\refcite{AR11o} for a detailed discussion).}
\rhcomment{For simplicity, we assume that there is a single stationary state and that the eigenvalues are labeled by} $0=\lambda_1>\mr{Re}[\lambda_2]\geq\mr{Re}[\lambda_3]\geq \cdots\geq \mr{Re}[\lambda_{\dim[\mc{H}]^2}]$ (see Fig.~\ref{fig:lind}). 
Here, $\hat{\phi}_1=\hat{\rho}_\mr{ss}$ is regarded as the  stationary state of the GKSL equation\rhcomment{, which is obtained from the infinite-time limit of Eq.~\eqref{Eq:lindblad_expand}}.
We assume the normalization condition of eigenstates as $\|\hat{\phi}_a\|_1=\|\hat{\chi}_a\|_1=1$.

We can expand the time-evolved state as
\aln{\label{Eq:lindblad_expand}
\hat{\rho}(t)=e^{\mbb{L}t}[\hat{\rho}(0)]=\hat{\rho}_\mr{ss}+\sum_{a=2}^{\dim[\mc{H}]^2}c_ae^{\lambda_at}\hat{\phi}_a
}
with
\aln{\label{Eq:coefficient}
c_a=\frac{\Tr[\hat{\chi}_a^\dag\hat{\rho}]}{\Tr[\hat{\chi}_a^\dag\hat{\phi}_a]}.
}
From Eq.~\eqref{Eq:lindblad_expand}, one naively think that the longest timescale of equilibration to the stationary state is determined from 
\aln{
\Delta_a =|\mr{Re}[\lambda_a]|
}
with $a=2$, which is called the Liouvillian gap.\cite{MZ15}
Indeed, since the mode with $a=2$ exhibits the slowest decay rate $|\mr{Re}[\lambda_2]|$, the equilibration time toward the stationary state might be
\aln{\label{Eq:gap_time}
\tau_a\simeq \frac{1}{\Delta_a}
}
with $a=2$.
This is indeed true for many cases, including few-body and many-body ones.
Assuming \eqref{Eq:gap_time} (for general $a$) leads to various conclusions.
For example, the finite/vanishing Liouvillian gap in the thermodynamic limit often indicates finite/divergent equilibration times.
When $\lambda_a$ are vanishing for $a\leq a_c$ and gapped for $a>a_c$, there can appear a long-lived meta-stable state characterized by the modes with $a\leq a_c$ (the equilibration time for the meta-stable state will be $\tau_{a_c+1}$).\cite{KM16}

Similar arguments are often made for other dynamics and situations.
For example, if we consider classical Markovian systems, dynamics is described by $\partial_tp_i=\sum_jW_{ij}p_j$, where $W$ is the transition rate matrix and $\vec{p}
=(p_1,\cdots,p_{{|\mc{S}|}}
)^{\mathsf{T}}$ is the probability distribution over the state {set} 
$\mc{S}$.
\rhcomment{If we assume the diagonalizability of $W$, the eigenvalues of $W$, $\lambda_\alpha'\in\mbb{C}$, again satisfy $\rhcomment{\mathrm{Re}[\lambda_\alpha']}\leq 0$, which is understood from the eigenstate decomposition (similar to Eq.~\eqref{Eq:lindblad_expand})
\aln{
\vec{p}(t)=e^{Wt}[\vec{p}(0)]=\vec{p}_\mathrm{ss}+\sum_{\alpha=2}^{|\mathcal{S}|}c'_\alpha e^{\lambda_\alpha' t}\vec{\phi}_\alpha'
}
Here, we assume that the stationary state is unique and label $0=\lambda_1'>\mr{Re}[\lambda_2']\geq\mr{Re}[\lambda_3']\geq \cdots\geq \mr{Re}[\lambda'_{{|\mc{S}|}}
]$, $c_\alpha'$ is the expansion coefficient, and $\vec{\phi}_\alpha'$ is the right eigenstate of $W$. Note that $\vec{\phi}_\alpha'=\vec{p}_\mr{ss}$ is the stationary state, which is obtained by $t\ra\infty$ above.}
From this, the longest timescale towards relaxation is often regarded as $1/|\mr{Re}[\lambda_2']|$.

A similar discussion applies to discrete-time systems.
Let us consider quantum dynamics given by $\hat{\rho}_{n+1}=\mbb{K}[\hat{\rho}_n]$, where (time-independent) $\mbb{K}$ is a super-operator satisfying the completely positive and trace-preserving condition.
\rhcomment{Assuming the diagonalizability, we can perform the spectral decomposition of $\mathbb{K}$ for the time evolution:
\aln{
\hat{\rho}_n=\mathbb{K}^n[\hat{\rho}_0]=\hat{\rho}_\mathrm{ss}+\sum_{a=2}^{\dim[\mc{H}]^2}C_a\Lambda_a^n\hat{\Phi}_a,
}
where $\Lambda_a\in\mbb{C}$ and $\hat{\Phi}_a$ are the eigenvalue and the right eigenstate of $\mbb{K}$, respectively, and $C_a$ is the expansion coefficient.
Here, we assume that the stationary state $\hat{\rho}_\mr{ss}=\hat{\Phi}_a$ is unique.
In addition, since the eigenvalues of $\mbb{K}$ should satisfy $|\Lambda_a|\leq 1$ (otherwise a nonphysical growing mode would exist), we have labeled
$1=\Lambda_1>|\Lambda_2|\geq |\Lambda_3|\cdots$.\footnote{If $\mathbb{K}$ is given by $\mathbb{K}=e^{\mathbb{L}\tau}\:(\tau>0)$, we have $\Lambda_a=e^{\lambda_a \tau}$, and thus nonpositive $\lambda_a$ indicates $\Lambda_a\leq 1$.}
}
From the above expansion, the longest equilibration timescale is often considered as $-1/\ln|\Lambda_2|$ (a similar argument applies to classical discrete-time Markovian processes).

\subsubsection{Discrepancy between the inverse gap and the equilibration time}
Although the above estimate in Eq.~\eqref{Eq:gap_time} and the identification of the equilibration time as $\tau_2$ seem clear, 
recent analyses show that there is a caveat for this simple argument.
In fact, the coefficient $c_a$ in Eq.~\eqref{Eq:coefficient} can be so large that $c_ae^{\lambda_at}$ is not small even for $t\sim \tau_a$.
In this case, the equilibration time for an observable $\tau_{{\rm m}}$ (sometimes called the mixing time) can be different and even larger than the timescale $\tau_2$.
Here, $\tau_{{\rm m}}=\tau_{{\rm m}}(\epsilon)$ can be defined as
\aln{
\tau_{{\rm m}}(\epsilon) =\rhcomment{ \min t\text{\quad s.t.\quad} d(t)\leq \epsilon}
}
with some cutoff $\epsilon>0$,
where $d(t)=\max_{\hat{\rho}(0)}\|\hat{\rho}(t)-\hat{\rho}_\mr{ss}\|_1/2$.\footnote{In classical systems, $d(t)$ reduces to the maximized  total variation distance $d(t)=\max_{p(0)}\sum_i|p_i(t)-p_{\mr{ss},i}|/2$.}
The distinction between $\tau_2$ and $\tau_{{\rm m}}$ has been widely understood in classical Markovian systems\cite{DAL17} and recently realized in quantum Markovian systems.\cite{MJK12,MJK13c,EV20}

Then, what kind of physical conditions lead to such discrepancy between the inverse of asymptotic convergence rate $\tau_2$ and the mixing time $\tau_{{\rm m}}$?
This question is investigated quite recently\cite{TM20,TH21} motivated by the development of open quantum systems with engineered dissipation or measurement.
 
One of the situations was considered by Haga {\it et al.},\cite{TH21} who showed that ``skin effect" of Liouvillian eigenstates leads to the divergence of $\tau_{{\rm m}}$ despite finite $\tau_2$.
For simplicity, let us consider a dissipative single-particle system on   one-dimensional $L$ lattice sites with the open boundary condition.
The dynamics is assumed to be described by the GKSL equation, where the Hamiltonian and the jump operators are given by
$\hat{H}=-J\sum_l(\hat{b}_{l+1}^{{\dag}}\hat{b}_l+\hat{b}_{l}^{{\dag}}\hat{b}_{l+1})$, $\hat{L}_{R,l}=\sqrt{\gamma_R}\hat{b}_{l+1}^\dag\hat{b}_l$, and $\hat{L}_{L,l}=\sqrt{\gamma_L}\hat{b}_{l-1}^\dag\hat{b}_l$ ($\hat{b}_l$ is the annihilation operator of a particle at site $l$).
Note that the jump operators indicate the incoherent hopping of a particle, which becomes asymmetric for $\gamma_R\neq \gamma_L$.
The dynamics is exactly solvable for $L\ra\infty$ and $J=0$, and the Liouvillian gap is shown to be $\Delta_2=(\sqrt{\gamma_R}-\sqrt{\gamma_L})^2$, which is finite for the asymmetric hopping case, $\gamma_R\neq\gamma_L$.
On the other hand, $\tau_{{\rm m}}$ is proportional to {$L$}, 
which is much larger than $\tau_2=\mathcal{O}(L^0)$.
To see this, let us consider an initial state where the particle is at the left end ($l=1$) for the case with $\gamma_R>\gamma_L$.
Under the dissipative time evolution toward the stationary state, the particle is transferred to the right end ($l=L$) due to the asymmetric hopping.
If we take an observable as  $\hat{n}_L=\hat{b}_L^\dag\hat{b}_L$, the time for the observable to change from $\braket{\hat{n}_L(0)}=0$ to $\braket{\hat{n}_L(t)}\simeq 1$ takes $t=\mc{O}(L^\alpha)\:(\alpha\geq 1)$ (see Secs.~\ref{Sec:speed_macro} and~\ref{Sec:LRbound}), where $\alpha=1$ is expected for this model.
Thus, the discrepancy between $\tau_2$ and $\tau_{{\rm m}}$ appears; in this case, the  coefficient of $c_2$ should be large.
A similar situation was numerically found to appear even for $J\neq 0$.

Reference~\refcite{TH21} noticed that the left and right eigenstates of $\mbb{L}$ \rhcomment{(assumed to diagonalizable)} respectively localize at the left and right ends of the system, which makes the overlap between these states exponentially small with respect to $L$, i.e., $\Tr[\hat{\chi}_2^\dag\hat{\phi}_2]\sim e^{-\mc{O}(L/\xi)}$ with the localized length $\xi$.
Such a localization of eigenstates recently attract much attention for non-Hermitian systems and is called the skin effect.{\cite{SY18,FKK18,ZG18PRX,NO20}} 
Here, this is regarded as the Liouvillian counterpart of the non-Hermitian skin effect.\cite{TH21,FS19}
As a result of this Liouvillian skin effect, $c_2$ (and similarly other $c_a$) exponentially diverges as $e^{\mc{O}(L/\xi)},$ which leads to the mixing timescale for this system as
\aln{\label{Eq:mixing_skin}
\tau_{{\rm m}}(\epsilon)\geq \tau(\epsilon) \sim \frac{\ln(1/\epsilon)}{\Delta_2}+\frac{cL}{\Delta_2\xi}
}
with some constant $c$.
This formula is obtained by assuming that the relevant timescale is determined from the mode $\alpha=2$, $|c_2|e^{-\Delta_2\tau}\sim \epsilon$.
This timescale diverges as $L\ra\infty$, as expected.
We note that the above Liouvillian skin effect cannot occur for systems with microreversibility, which imposes a certain constraint on the right and left eigenstates.\cite{KK19s,RH20}

\begin{figure}[bt]
\centerline{\psfig{file=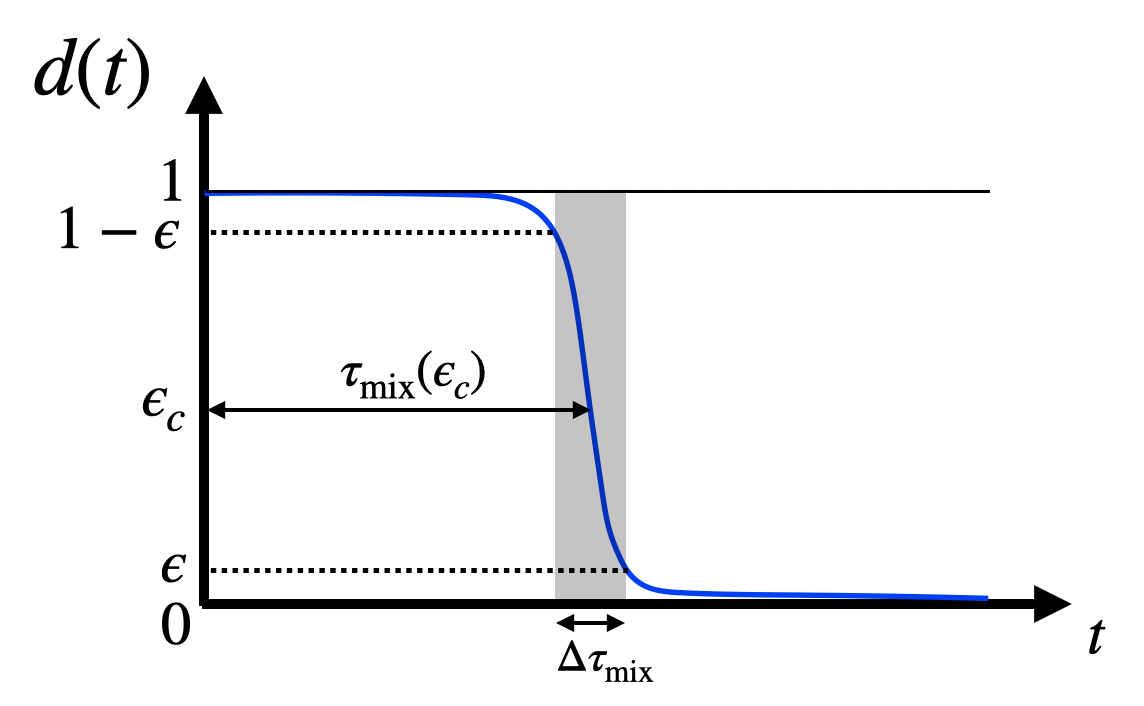,width=4in}}
\vspace*{8pt}
\caption{Schematic illustration of a cutoff phenomenon. At a mixing time $t\sim \tau_\mr{m}(\epsilon_c)$, the distance $d(t)$ rapidly changes from 1 to 0 for $L\ra\infty$, where $\Delta\tau_\mr{m}/\tau_\mr{m}(\epsilon_c)\ra 0$ is satisfied.}
\label{fig:cutoff}
\end{figure}

Interestingly, if we identify $\tau(\epsilon)$ as $\tau_{{\rm m}}(\epsilon)$ in Eq.~\eqref{Eq:mixing_skin}, we find that 
\aln{
\lim_{L\ra\infty}\frac{\tau_{{\rm m}}(\epsilon)}{\tau_{{\rm m}}(1-\epsilon)}=1
}
for all $0<\epsilon<1$ (see Fig.~\ref{fig:cutoff}).
This means that, the distance for the stationary state $d(t)$ can suddenly change from 1 to 0 when the time is measured in the unit of $\tau_{{\rm m}}(\epsilon_c)$ ($\epsilon_c$ is arbitrary, but let us take $\epsilon_c =0.5$ here).
Namely, the time interval $\Delta\tau_{{\rm m}}$ for the change of $d(t)$ from 1 to 0 vanishes compared with $\tau_{{\rm m}}(\epsilon_c)$, i.e., $\lim_{L\ra\infty}\Delta\tau_{{\rm m}}/\tau_{{\rm m}}(\epsilon_c)=0$, which is called the cutoff phenomenon\cite{PD96} originally known in classical systems.

While the Liouvillian skin effect is one mechanism to cause the discrepancy between $\tau_2$ and $\tau_{{\rm m}}$ as well as the cutoff phenomenon,
other mechanisms for these phenomena have also been found.
In Ref.~\refcite{TM20}, the authors considered quantum many-body chaotic systems coupled to boundary dissipation and numerically found that $c_a$ (including $a\gg 2$) can diverge as $\sim e^{\mc{O}(L^2)}$, which explains the diffusive timescale $\tau\sim \mc{O}(L^2)$ of the system even when $\tau_2\sim \mc{O}(L^2)$ is not satisfied.
Reference~\refcite{TM21} investigated classical many-body stochastic systems exhibiting metastability and demonstrated that the large coefficients $c_a$ are responsible for the metastability.
As a related observation, the authors in Ref.~\refcite{JB21} analyzed the entanglement dynamics of unitary random quantum circuits through some Markovian transfer matrix describing the averaged evolution of the entanglement.\cite{MZ08,WTK20}
It was found that the rate of the convergence of the entanglement toward the stationary value is different from the eigenvalue gap of the transfer matrix, which is attributed to the large expansion coefficients.

\section{Lieb-Robinson bounds}\label{Sec:LRbound}
Propagation of quantum information in locally interacting many-body systems is limited by an emergent ``soft" light cone, which is rigorously described by the Lieb-Robinson bound. Originally proved for strictly local systems nearly half a century ago, the Lieb-Robinson bound has undergone constant improvement and generalizations, especially to include algebraically decaying interactions. 
They have also been found to be useful for proving constraints on equilibrium properties (e.g., clustering condition), which we will also review briefly in this section. They will be extensively used in the following sections for proving constraints on dynamical properties arising from locality. \textcolor{black}{While our discussions here are mostly theoretical, we would like to mention a very recent review concerning the experimental tests of Lieb-Robinson bounds.\cite{MC22}}

\subsection{Short-range systems} 
We first review the Lieb-Robinson bound for systems with short-range interactions that decay no slower than exponentially. In this case, the correlations outside the ``soft" light cone are suppressed exponentially as well. For translation-invariant free systems, we will see that the Lieb-Robinson velocity is essentially the maximal group velocity determined by band dispersions.

\label{Sec:SRLR}
\subsubsection{Introductory example}
\label{Sec:IE}
Before introducing the general form of the Lieb-Robinson bound, it is instructive to first consider a simple situation --- a single particle in a 1D lattice with a finite hopping range. By assumption, denoting the Hamiltonian as $\hat H= \sum_{j,j'}h_{jj'}|j\rangle\langle j'|$, where $j\in\mathbb{Z}$ is the site label and $\|h\|$ (operator norm of matrix $h$) is assumed to be finite, we have  $h_{jj'}=0$ $\forall |j-j'|>\chi$ with $\chi$ being an $\mathcal{O}(1)$ constant.\footnote{\textcolor{black}{Hereafter, we use ``$\mathcal{O}(1)$ constant" to emphasize that the constant does not depend on the size or Hilbert-space dimension of the system or subsystem of interest. The exact meaning should be clear in the context where it appears.}} 
Suppose that the particle is initialized at site $0$, then the amplitude on site $r$ can be upper bounded by\cite{MFF15}
\begin{equation}
|\langle r| e^{-i\hat Ht} |0\rangle| \le \sum^\infty_{n=\left\lceil \frac{|r|}{\chi}\right\rceil} \frac{\|h\|^n}{n!}t^n< e^{e\|h\|t -\frac{|r|}{\chi}},
\label{rt0}
\end{equation}
where we have used the fact $\langle r|H^n|0\rangle=0$ $\forall n<\lceil |r|/\chi \rceil$ and the inequality $\sum^\infty_{n=m} x^n/n!< e^{ex-m}$ $\forall x\ge0.$\footnote{This result can be confirmed by comparing the coefficients in the Taylor series.} 

In the second quantization picture, $\hat H=\sum_{j,j'}h_{jj'}\hat a_j^\dag \hat a_{j'}$ and the lhs of Eq.~(\ref{rt0}) can be rewritten as $\|[\hat a^\dag_r(t),\hat a_0]\|$ ($\|\{\hat a^\dag_r(t),\hat a_0\}\|$) for bosons (fermions), where $\hat a_r^\dag(t)=e^{i\hat H t}\hat a_r^\dag e^{-i\hat H t}$. The rhs of Eq.~(\ref{rt0}) takes the form of $e^{-\kappa(|r|-vt)}$ with $\kappa=\chi^{-1}$ and $v=e\chi\|h\|$. One may interpret $\kappa$ as the inverse of correlation length, while $v$ is essentially determined by the hopping strength. 

\subsubsection{General form}
\label{Sec:GF}
Despite the situation discussed above is specific (1D and noninteracting), the observation turns out to be universal even for general interacting systems. It was originally proved by Lieb and Robinson that for a many-body quantum spin system $\hat H$ with finite interaction range, and for any two local operators $\hat O_X$ and $\hat O_Y$ supported on $X$ and $Y$, respectively, there exist constants $\kappa$, $v$ and $c$ such that\cite{EHL72}
\begin{equation}
\|[\hat O_X(t),\hat O_Y]\|\le c e^{-\kappa [{\rm dist}(X,Y)-vt]}.
\label{LRB}
\end{equation}
 Later this result was generalized to a broader class of short-range systems with exponential tails in the interactions.\cite{BN06} Let us give a precise statement of this generalized version, which will be extremely useful hereafter: 
\begin{theorem}
\emph{(Lieb-Robinson bound)} For a many-body quantum spin system living on a $d$-dimensional lattice $\Lambda$, suppose that the Hamiltonian takes the form of $\hat H= \sum_{A\subseteq\Lambda} \hat h_A$ and that for some $\kappa>0$
\begin{equation}
s\equiv \max_{x\in\Lambda}\sum_{A\ni x}|A| \|\hat h_A\| e^{\kappa {\rm diam}(A) }<\infty,
\label{sdef}
\end{equation}
where $\hat h_A$ is supported on $A$, $|A|$ denotes the cardinality (number of sites) of $A$ and ${\rm diam}(A)\equiv\max\{{\rm dist}(x,y):x,y\in A\}$ is the diameter of $A$. Then Eq.~(\ref{LRB}) is valid for
\begin{equation}
c=2\min\{|X|,|Y|\}\|\hat O_X\|\|\hat O_Y\|,\;\;\;\; 
v=2\kappa^{-1}s.
\label{cv}
\end{equation}
\label{Thm:LRB}
\end{theorem}
See Ref.~\refcite{MBH10} for a detailed proof. It is worth mentioning that the above result can be readily extended to time-dependent systems. Also, one can get rid of the factor $|A|$ in Eq.~(\ref{sdef}), at the price of making $v$ a bit more complicated:\cite{ZG20b}
\begin{theorem}
Consider the same setting as Thm.~\ref{Thm:LRB} except for additional time-dependence in $\hat h_A$ and that for some $\kappa>0$ and $\forall t'\in[0,t]$
 \begin{equation} 
 \mu(t')\equiv\max_{x\in\Lambda}\sum_{A\ni x} \|\hat h_A(t')\| e^{\kappa {\rm diam}(A)}<\infty.
 \label{mutp}
 \end{equation}
 Also, assume that the lattice geometry satisfies $|A|\le c_d(1+{\rm diam}(A))^d$ $\forall A\subseteq \Lambda$ for some constant $c_d$. Then Eq.~(\ref{LRB}) is valid for the same $c$ given in Eq.~(\ref{cv}) and
\begin{equation}
v=\frac{2c_d}{\kappa-\eta}\left(\frac{d}{e\eta}\right)^d \bar \mu(t),
\end{equation}
where $\eta\in(0,\kappa)$ can be chosen arbitrarily and $\bar\mu(t)\equiv t^{-1}\int^t_0 dt' \mu(t')$.
\label{Thm:tLRB}
\end{theorem} 

\begin{figure}[bt]
\centerline{\psfig{file=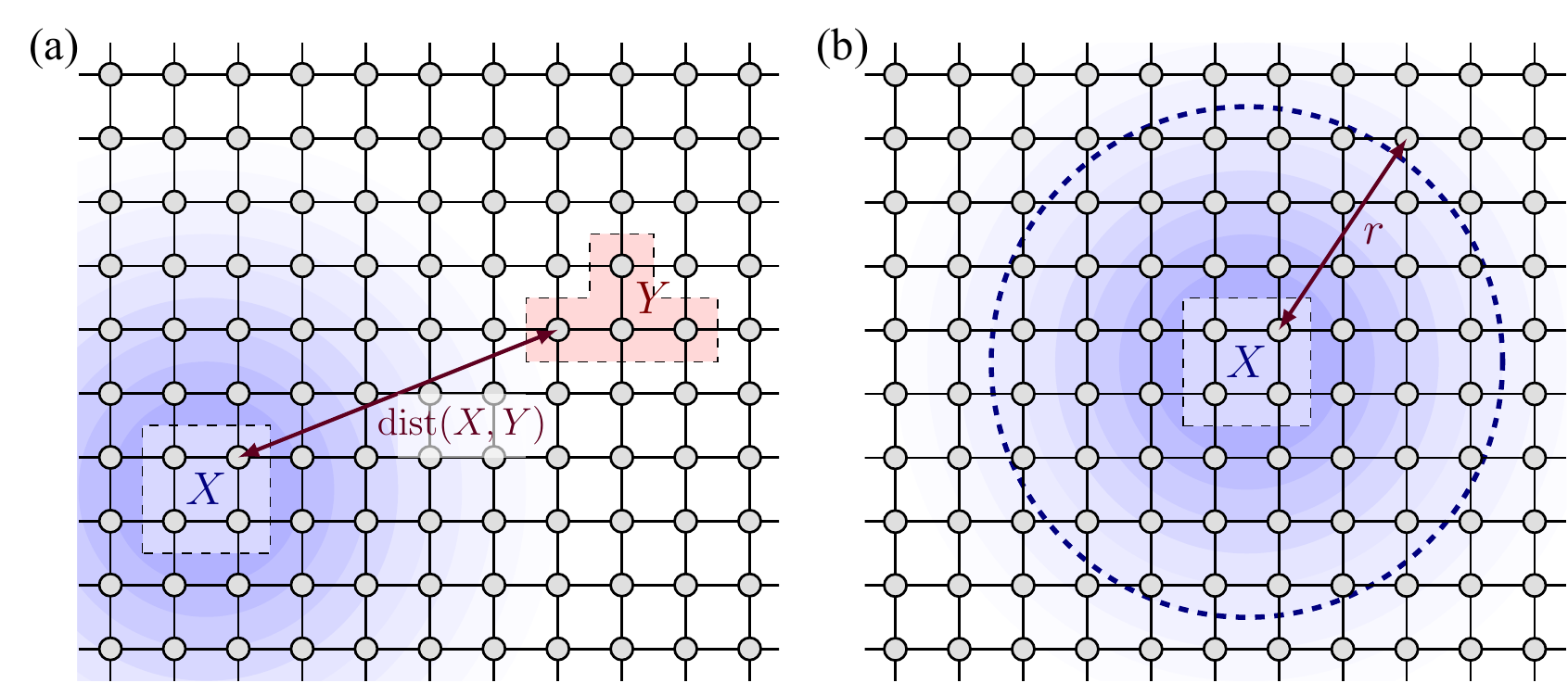,width=5in}}
\vspace*{8pt}
\caption{Two equivalent Lieb-Robinson bounds. (a) The commutator between a time-evolved local operator $\hat O_X(t)$ initially supported on $X$ and another local operator $\hat O_Y$ supported on $Y$ is exponentially suppressed by ${\rm dist}(X,Y)$ (cf. Eq.~(\ref{LRB})). (b) The error of approximating $O_X(t)$ by its truncation on $X^r$ (interior of the dashed circle) is exponentially suppressed by $r$ (cf. Eq.~(\ref{LRB2}))}.
\end{figure}

Note that the Lieb-Robinson bound (\ref{LRB}) only makes sense for $t<{\rm dist}(X,Y)/v$. Otherwise, due to the exponential dependence on $t$, it may be much looser than the trivial bound $\|[\hat O_X(t),\hat O_Y]\|\le2\|\hat O_X\|\|\hat O_Y\|$ at late time. It is worth mentioning that there is an equivalent statement that only involves $\hat O_X$:
\begin{equation}
\left\|\hat O_X(t) - \hat O^{[r]}_X(t)\right\|\le c e^{-\kappa(r-vt)},\;\;\;\;\hat O^{[r]}_X(t)\equiv\frac{\mathbbm{1}_{\overline{X^r}}}{{\rm Tr}\mathbbm{1}_{\overline{X^r}}}\otimes{\rm Tr}_{\overline{X^r}}[\hat O_X(t)],
\label{LRB2}
\end{equation}
where $X^r\equiv \{x:{\rm dist}(x,X)<r\}$ ($r\ge1$) and $\overline{X^r}\equiv\Lambda\backslash X^r$ is its complement. This result (\ref{LRB2}) can be derived from Eq.~(\ref{LRB}) by choosing $\hat O_Y$ to be a Haar random unitary supported on $\overline{X^r}$ and then taking the average.\cite{SB06} Conversely, one can reproduce Eq.~(\ref{LRB2}) from Eq.~(\ref{LRB2}) (up to a factor $2$ on the rhs) by using $\|[\hat O_X(t),\hat O_Y]\|=\|[\hat O_X(t)-\hat O^{[r]}_X(t),\hat O_Y]\|\le 2\|\hat O_X(t)-\hat O^{[r]}_X(t)\|\|\hat O_Y\|$ with $r={\rm dist}(X,Y)$.


While Thms.~\ref{Thm:LRB} and \ref{Thm:tLRB} assume spin systems, the generalization to short-range fermion systems is conceptually straightforward thanks to the Pauli exclusion principle (implying the finiteness of Hilbert space at each site), although a bit technically involved due to the anti-commuting nature of fermion operators.\cite{BN17} The results do not change at all if at least one of $\hat O_X$ and $\hat O_Y$ is bosonic. The commutator in Eq.~(\ref{LRB}) should be replaced by the anti-commutator if both $\hat O_X$ and $\hat O_Y$ are fermionic. Equivalently, one may simply determine which is the case depending on whether $[\hat O_X,\hat O_Y]=0$ or $\{\hat O_X,\hat O_Y\}=0$ for disjoint $X$ and $Y$. On the other side, the generalization to boson systems seems to be difficult since each local term $\hat h_A$ has a divergent operator norm. 
Nevertheless, by assuming the finiteness of the commutators between local terms, one can still obtain a bound like Eq.~(\ref{LRB}).\cite{IPS10} Very recently, variations of Eqs.~(\ref{LRB}) and (\ref{LRB2}) with respect to some low-density initial states have been proved for a class of Bose-Hubbard-like models.\cite{TK21b,JF22,CY22,TK22}

\subsubsection{Free systems}\label{Sec:free_LR}
While the Lieb-Robinson is rigorous and qualitatively tight, the velocity $v$ is typically overestimated, and thus the bound could be quantitatively very loose. This is especially the case for free-fermion (and also free-boson) systems with translation invariance, which can be described by band theory. It is widely believed, and indeed confirmed in many models\cite{PC11} and even some experiments,\cite{MC12,PJ14} that the optimal $v$ should be the maximal group velocity determined from the derivatives of band dispersions. Recently, this connection has finally been established in Ref.~\refcite{ZG19} for 1D systems with separable bands. The central idea lies in the analytic extension of the wave vector, a technique that had been used to prove the localization properties of Wannier functions.\cite{WK59,JDC64,CB07}  Here, we generalize the result to arbitrary dimensions. 

Without loss of generality, we consider a $d$-dimensional translation-invariant and number-conserving free-fermion (or boson) system living on a square lattice and with exponentially decaying hopping amplitudes. By assumption, the Bloch Hamiltonian $h(\boldsymbol{k})=\sum_{\boldsymbol{r}\in\mathbb{Z}^d} h_{\boldsymbol{r}} e^{-i\boldsymbol{k}\cdot\boldsymbol{r}}$ can be analytically extended to $h(\boldsymbol{k}+i\boldsymbol{\kappa})$ at least for $|\boldsymbol{\kappa}|<\xi^{-1}$, where $\xi$ is the localization length, since $\|h_{\boldsymbol{r}}\| \le Je^{-|\boldsymbol{r}|/\xi}$ for some constant $J$. Moreover, one can check $h(\boldsymbol{k}+i\boldsymbol{\kappa})^\dag = h(\boldsymbol{k}-i\boldsymbol{\kappa})$. Here $h(\boldsymbol{k})$ or $h_{\boldsymbol{r}}$ is a $|I|\times |I|$ matrix, where $I$ is the set of internal states at each site (unit cell). Denoting $\hat a_{\boldsymbol{r}s}$ ($\boldsymbol{r}\in\mathbb{Z}^d$, $s\in I$) as the creation operator of a particle with internal state $s$ at site $\boldsymbol{r}$, we know that $\|\{\hat a_{\boldsymbol{r}s}^\dag(t),\hat a_{\boldsymbol{0}s'}\}\|$ (or $\|[\hat a_{\boldsymbol{r}s}^\dag(t),\hat a_{\boldsymbol{0}s'}]\|$ for bosons) 
is upper bounded by
\begin{equation}
C(\boldsymbol{r},t) \equiv \left\| \int_{\rm B.Z.} \frac{d^{d}\boldsymbol{k}}{(2\pi)^d} e^{-i h(\boldsymbol{k})t + i\boldsymbol{k}\cdot\boldsymbol{r}} \right\|
\le e^{-\kappa |\boldsymbol{r}|}  \int_{\rm B.Z.} \frac{d^{d}\boldsymbol{k}}{(2\pi)^d} \|e^{-ih(\boldsymbol{k}+i\kappa\boldsymbol{e}_{\boldsymbol{r}})t}\|,
\end{equation}
where ${\rm B.Z.}=(2\pi\mathbb{R}/\mathbb{Z})^d$ is the $d$-dimensional Brillouin zone, $\boldsymbol{e}_{\boldsymbol{r}}\equiv \boldsymbol{r}/|\boldsymbol{r}|$ is the unit vector along $\boldsymbol{r}$ and $\kappa\in(0,\xi^{-1})$ can be chosen freely. This inequality can be understood by deforming the integral contour, which leaves the integral invariant due to the analyticity of the integrand. The analyticity simply follows short-range nature (i.e., the hopping amplitude decays exponentially) of the system.

Noting that $\|e^M\|\le e^{\|M+M^\dag\|/2}$ for an arbitrary matrix $M$,\footnote{To see this, first denoting $M_S\equiv (M+M^\dag)/2=M_S^\dag$ and $M_A\equiv M - M_S= -M_A^\dag$, we have $e^M=e^{M_A}\overleftarrow{\rm T}e^{\int^1_0 dt M_S(t)}$ with $M_S(t)= e^{-M_At}M_Se^{M_At}$. Using the unitary invariance of operator norm and $\|\overleftarrow{\prod}_j e^{M_j}\|\le \prod_j\|e^{M_j}\|\le e^{\sum_j\|M_j\|}$, we end up with $\|e^M\|\le e^{\int^t_0 dt\|M_S(t)\|}=e^{\|M_S\|}$.} we can further bound $C(\boldsymbol{r},t)$ by the Lieb-Robinson form $C(\boldsymbol{r},t) \le e^{-\kappa (|\boldsymbol{r}| -vt)}$, where 
\begin{equation}
v=\max_{\boldsymbol{k}\in{\rm B.Z.}}\|w(\boldsymbol{k},\kappa)\|,\;\;w(\boldsymbol{k},\kappa)= \frac{1}{2\kappa}[h(\boldsymbol{k}+i\kappa\boldsymbol{e}_{\boldsymbol{r}}) - h(\boldsymbol{k}-i\kappa\boldsymbol{e}_{\boldsymbol{r}})].
\end{equation}
We claim that $v$ is a non-decreasing function of $\kappa$. To see this, it is sufficient to show that for $\kappa<\kappa'$, $w(\boldsymbol{k},\kappa)$ is majorized by $w(\boldsymbol{k},\kappa')$, i.e., there exists a positive semi-definite kernel $K(\boldsymbol{k},\kappa;\boldsymbol{k}',\kappa')$ such that $\int_{\rm B.Z.} \frac{d^d\boldsymbol{k}}{(2\pi)^d}K(\boldsymbol{k},\kappa;\boldsymbol{k}',\kappa')=\int_{\rm B.Z.} \frac{d^d\boldsymbol{k}'}{(2\pi)^d}K(\boldsymbol{k},\kappa;\boldsymbol{k}',\kappa')=1$ and
\begin{equation}
w(\boldsymbol{k},\kappa)= \int_{\rm B.Z.} \frac{d^d\boldsymbol{k}'}{(2\pi)^d} K(\boldsymbol{k},\kappa;\boldsymbol{k}',\kappa') w(\boldsymbol{k}',\kappa').
\end{equation}
If this is true, then $\|w(\boldsymbol{k},\kappa)\|\le \int_{\rm B.Z.} \frac{d^d\boldsymbol{k}}{(2\pi)^d} K(\boldsymbol{k},\kappa;\boldsymbol{k}',\kappa')\max_{\boldsymbol{k}'\in{\rm B.Z.}}\|w(\boldsymbol{k}',\kappa')\|=\max_{\boldsymbol{k}'\in{\rm B.Z.}}\|w(\boldsymbol{k}',\kappa')\|$  $\forall\boldsymbol{k}\in{\rm B.Z.}$ and in particular $\max_{\boldsymbol{k}\in{\rm B.Z.}} \|w(\boldsymbol{k},\kappa)\|\le \max_{\boldsymbol{k}'\in{\rm B.Z.}}\|w(\boldsymbol{k}',\kappa')\|$. In fact, we can analytically figure out the expression of the kernel to be $K(\boldsymbol{k},\kappa;\boldsymbol{k}',\kappa')= \sum_{\boldsymbol{G}\in(2\pi\mathbb{Z})^d} Y(\boldsymbol{k} - \boldsymbol{k}' + \boldsymbol{G};\kappa,\kappa')$, where
\begin{equation}
Y(\boldsymbol{k};\kappa,\kappa')=\frac{\delta(\boldsymbol{k}- (\boldsymbol{k}\cdot\boldsymbol{e}_{\boldsymbol{r}})\boldsymbol{e}_{\boldsymbol{r}}) \sin(\frac{\kappa}{\kappa'}\pi)}{2\kappa[\cosh(\frac{\pi}{\kappa'}\boldsymbol{k}\cdot\boldsymbol{e}_{\boldsymbol{r}})+\cos(\frac{\kappa}{\kappa'}\pi)]}\ge0.
\end{equation}
This completes the confirmation of the monotonicity. Taking $\kappa\to0$, we obtain
\begin{equation}
v\ge \lim_{\kappa\to0} \max_{\boldsymbol{k}\in{\rm B.Z.}}\| w(\boldsymbol{k},\kappa) \| = \max_{\boldsymbol{k}\in{\rm B.Z.}} \| \boldsymbol{e}_{\boldsymbol{r}} \cdot\nabla h(\boldsymbol{k})\|.
\end{equation}

While the result appears not optimal except for the single-band case (so that $\boldsymbol{e}_{\boldsymbol{r}} \cdot\nabla h(\boldsymbol{k})$ is the group velocity), it does not require band separations and is thus more general than Ref.~\refcite{ZG19}. Moreover, by assuming band separation,\footnote{The spectrum of a Hermitian matrix is stable against arbitrary non-Hermitian perturbations. See, e.g., Ref.~\refcite{RB97}. Therefore, the complex-energy bands remain separated for a sufficiently small $\kappa$.} it is possible to make use of the approach in Ref.~\refcite{ZG19} to improve the result by spectral decomposing $e^{-ih(\boldsymbol{k}+i\kappa \boldsymbol{e}_{\boldsymbol{r}})t}=\sum^{|I|}_{\alpha=1}e^{-i\epsilon_\alpha(\boldsymbol{k}+i\kappa \boldsymbol{e}_{\boldsymbol{r}})t}{\bf u}^{\rm R}_\alpha(\boldsymbol{k}+i\kappa \boldsymbol{e}_{\boldsymbol{r}}) {\bf u}^{\rm L}_\alpha(\boldsymbol{k}+i\kappa \boldsymbol{e}_{\boldsymbol{r}})^\dag$ followed by applying a similar argument to each eigenenergy $\epsilon_\alpha(\boldsymbol{k}+i\kappa \boldsymbol{e}_{\boldsymbol{r}})$, which should be analytic in $\boldsymbol{k}$. In this case, $v$ reduces to $\max_{\boldsymbol{k},\alpha}\|\boldsymbol{e}_{\boldsymbol{r}} \cdot\nabla \epsilon_\alpha(\boldsymbol{k}) \|$ in the $\kappa\to0$ limit, which is the desired result. On the other hand, there will be an additional Petermann factor before $e^{-\kappa(|\boldsymbol{r}|-vt)}$ due to the discrepancy between left and right eigenvectors ${\bf u}^{\rm L/R}$ of a non-Hermitian matrix.\cite{YA20}  

Finally, we mention that the idea of deforming the contour integral can be applied to the single-particle Green's functions of translation-invariant interacting systems. Again, this approach can lead to much tighter bounds than Thm.~\ref{Thm:LRB}.\cite{ZW20} 



\subsection{Long-range systems}
Long-range interactions with power-law decay appear ubiquitously in nature. Well-known examples include Coulomb interactions ($\propto r^{-1}$), dipole-dipole interactions ($\propto r^{-3}$) and van der Waals interactions ($\propto r^{-6}$). They are also relevant to many quantum simulators, such as polar molecules\cite{BY13}, trapped ions\cite{PR14} and NV-centers.\cite{SC17t} Studying the Lieb-Robinson bound in long-range systems is thus not only fundamentally important for understanding the nature, but also has direct implications on implementing various quantum tasks (e.g., quantum state transfer\cite{ZE17} and quantum simulation\cite{MCT19}) on many atomic, molecular, and optical platforms.

\subsubsection{Logarithmic light cone}
Let us again start from the single-particle example in Sec.~\ref{Sec:IE}. The only difference is that now $\hat H$ involves long-range hoppings that decay algebraically in distance, i.e., $|h_{jj'}|\le J(1+|j-j'|)^{-\alpha}$ for some positive constants $J$ and $\alpha$. To ensure the finiteness of the single-particle energy in the thermodynamic limit, we assume $\alpha > 1$ so that $\|h\|\le\max_j (\sum_{j'}|h_{jj'}|)\le J(2\zeta(\alpha)-1)<\infty$ ($\zeta(s)\equiv\sum^\infty_{n=1} n^{-s}$ is the Riemann zeta function). Using the power-law decay bound on $|h_{jj'}|$, one can obtain
\begin{equation}
|\langle r| e^{-i\hat Ht} |0\rangle| \le \sum^\infty_{n=0} \frac{t^n}{n!}|\langle r|\hat H^n|0\rangle| \le \delta_{r0} + \frac{e^{pJ t}-1}{p(1+|r|)^\alpha},
\label{LRrt0}
\end{equation}  
where 
$p$ is a constant that validates $\sum_j (1+|j_1 - j|)^{-\alpha}(1+|j - j_2|)^{-\alpha} \le p (1+|j_1 - j_2|)^{-\alpha}$ $\forall j_{1,2}$.\footnote{For example, one can take $p = 2^{\alpha+1}(2\zeta(\alpha)-1)$. To see this, we only have to note that $\min\{(1+|j_1-j|)^{-\alpha},(1+|j-j_2|)^{-\alpha}\}\le (1+|j_1- j_2|/2)^{-\alpha}$, implying $\sum_j (1+|j_1 - j|)^{-\alpha} (1+|j - j_2|)^{-\alpha} < (1+\frac{|j_1 - j_2|}{2})^{-\alpha}\sum_j \left[(1+|j_1 - j|)^{-\alpha} + (1+|j - j_2|)^{-\alpha}\right] <  2^{\alpha+1}(2\zeta(\alpha)-1)(1+|j_1 - j_2|)^{-\alpha}$.} Clearly, this bound (\ref{LRrt0}) implies a logarithmic light cone, i.e., the rhs becomes small for $t\ll t_{\rm c}\sim (\alpha/pJ)\log|r|$. 

Just like the short-range case, by replacing the lhs with the norm of commutator (or anti-commutator for fermions), Eq.~(\ref{LRrt0}) applies to general many-body systems with algebraically decaying interactions. That is, for any two local operators $\hat O_X$ and $\hat O_Y$, there exist constants $c$ and $\lambda$ such that\cite{MBH06}
\begin{equation}
\|[\hat O_X(t) , \hat O_Y]\|\le c \frac{e^{\lambda t}-1}{(1+{\rm dist}(X,Y))^\alpha}.
\label{LRLR}
\end{equation}
Here, we used $(1+{\rm dist}(X,Y))^{-\alpha}$ instead of ${\rm dist}(X,Y)^{–\alpha}$ simply because the former is well-defined also for ${\rm dist}(X,Y))=0$. However, the latter is also commonly used in the literature (as we will soon encounter later), in which case $X$ and $Y$ are always assumed to be disjoint. More precisely, we have:\cite{MBH06} 
\begin{theorem}
\emph{(Hastings-Koma bound)} For a many-body quantum spin system living on a $d$-dimensional lattice $\Lambda$, suppose that the Hamiltonian takes the form of $\hat H= \sum_{A\subseteq\Lambda} \hat h_A$ with $\hat h_A$'s satisfying $\forall x,y\in\Lambda$
\begin{equation}
\sum_{A\ni x,y} \| \hat h_A\| \le \frac{J}{(1+{\rm dist}(x,y))^\alpha},
\label{xya}
\end{equation}
where $\alpha>d$ and $J$ is a constant. Further assuming the existence of constant $p$ such that $\forall x,y\in\Lambda$
\begin{equation}
\sum_{z\in\Lambda}\frac{1}{(1+{\rm dist}(x,z))^\alpha} \frac{1}{(1+{\rm dist}(z,y))^\alpha} \le \frac{p}{(1+{\rm dist}(x,y))^\alpha},
\label{xzy}
\end{equation}
then Eq.~(\ref{LRLR}) holds true for
\begin{equation}
c=2p^{-1}\|\hat{O}_X\| \|\hat{O}_Y\| |X| |Y|,\;\;\;\; \lambda = 2pJ. 
\end{equation}
\label{Thm:HKB}
\end{theorem}
For the hypercube lattice $\Lambda=\mathbb{Z}^d$, one can check that one valid $p$ could be $2^{\alpha+1} \sum_{\boldsymbol{r}\in\mathbb{Z}^d}(1+|\boldsymbol{r}|)^{-\alpha}$, where the sum is bounded  since it is roughly given by $\int_{|\boldsymbol{r}|\ge1} d^d\boldsymbol{r} |\boldsymbol{r}|^{-\alpha}\propto \int^\infty_1 dr r^{-(\alpha-d+1)}$, which converges for $\alpha>d$. 
In fact, the condition (\ref{xzy}) may be replaced by 
\begin{equation}
q\equiv\sum_{y\in\Lambda}(1+{\rm dist}(x,y))^{-\alpha}<\infty, 
\end{equation}
so that Eq.~(\ref{xzy}) is always validated by $p=2^{\alpha+1}q$.

As we will see in the following, Thm.~\ref{Thm:HKB} is qualitatively correct, in the sense of the logarithmic light cone, only for $\alpha \in (d,2d)$. This result is especially not satisfactory in the short-range limit $\alpha\to\infty$, in which case $p$ and thus $\lambda$ diverges. To fix this shortcoming, Gong {\it et al.} pointed out in Ref.~\refcite{ZXG14} that it is better to separately estimating those term in the lhs of Eq.~(\ref{xzy}) with $z$ close to $x$ or $y$ (precisely speaking, those with ${\rm dist}(z,x)\le 1$ or ${\rm dist}(z,y)\le 1$) from the remaining. At least for systems with two-body interactions, so that $\hat H=\sum_{x,y\in\Lambda} \hat h_{xy}$ and that Eq.~(\ref{xya}) is simplified into $\|\hat h_{xy}\|\le J(1+{\rm dist}(x,y))^{–\alpha}$, this leads to a ``hybrid-form" bound:
\begin{equation}
c_1(e^{\lambda_1 t}-1)e^{-\kappa r} + c_2\frac{e^{\lambda_2 t}-1}{[(1-\kappa)r]^\alpha},
\label{hbLR}
\end{equation}
where $r={\rm dist} (X,Y)\ge1$ (this notation simplification will be used hereafter), $\kappa\in(0,1)$ is \textcolor{black}{tunable} and $c_1,c_2,\lambda_1,\lambda_2$ are all constants that remain finite for $\alpha>d$. While the light cone predicted by Eq.~(\ref{hbLR}) is still logarithmic, this bound is obviously tighter than Eq.~(\ref{LRLR}) for large $\alpha$.


\subsubsection{Sublinear and linear light cone}
\label{sll}
While Eq.~(\ref{hbLR}) is after all a quantitative improvement compared to Eq.~(\ref{LRLR}), the idea of separating the short-range interacting part from the long-range one turns out to be very useful and can indeed lead to qualitative improvement for sufficiently large $\alpha$ (precisely speaking, $\alpha >2d$). That is, we can tighten the logarithmic light cone into a sublinear and even a linear one.

Let us briefly explain how to execute the idea mentioned above. For technical reasons, in addition to Eq.~(\ref{xya}), we will further assume $\forall x\in\Lambda$
\begin{equation}
\sum_{A:x\in A,{\rm diam}(A)\ge R} \|\hat h_A\| \le \frac{J_1}{(1+R)^{\alpha -d}},
\label{xad}
\end{equation}
where $J_1$ is some constant. Given a general many-body Hamiltonian $\hat H=\sum_{A\subseteq\Lambda} \hat h_A$, we can decompose it as $\hat H= \hat H_{\rm sr} + \hat H_{\rm lr}$, where
\begin{equation}
\hat H_{\rm sr} \equiv \sum_{A:{\rm diam}(A)< \chi}\hat h_A,\;\;\;\;
\hat H_{\rm lr} \equiv \sum_{A:{\rm diam}(A)\ge \chi}\hat h_A,
\end{equation}
and $\chi$ is a tunable parameter.  The time-evolution generated by $\hat H$ can thus be rewritten in the interaction picture:
\begin{equation}
e^{-i\hat H t} = e^{-i\hat H_{\rm sr} t} \overleftarrow{\rm T}e^{-i \int^t_0 dt' \hat H^{(\rm I)}_{\rm lr}(t') },\;\;\;\;
\hat H^{(\rm I)}_{\rm lr}(t')= e^{i\hat H_{\rm sr} t'}  \hat H_{\rm lr} e^{-i\hat H_{\rm sr} t'}. 
\end{equation}
For $\hat H_{\rm sr}$, one can apply the conventional short-range Lieb-Robinson bound discussed in Sec.~\ref{Sec:SRLR}. Choosing $\kappa = \chi^{-1}$ in Eq.~(\ref{sdef}), one will obtain a Lieb-Robinson velocity proportional to $\chi$, provided that $s$ is finite for the $\kappa=0$ version (which is true for $k$-body interactions with $k$ finite). Moreover, for sufficiently large $\chi$, each term in $\hat H_{\rm lr}$ will be of very small norm. In particular, the single-particle energy of $\hat H_{\rm lr}$ given by the $\kappa=0$ version of Eq.~(\ref{mutp}), which essentially determines $\lambda$ in the Hastings-Koma bound (\ref{LRLR}), becomes as small as $\chi^{-(\alpha - d)}$ according to Eq.~(\ref{xad}). After time-evolution by $\hat H_{\rm sr}$ in the interaction picture, the single-particle energy of $H_{\rm lr}^{(\rm I)}(t)$ roughly grows as $(\chi t)^d \chi^{-(\alpha -d)}= t ^d \chi^{-(\alpha - 2d)}$, which can be made arbitrarily small by choosing a sufficiently large $\chi$ for $\alpha >2d$.  

The first big progress along this line was made by Foss-Feig {\it et al.}, who proved that for any long-range systems with two-body interactions and $\alpha >2d$, the rhs of Eq.~(\ref{LRLR}) can be improved to\cite{MFF15}
\begin{equation}
c_1 e^{\lambda t - r t^{-\gamma}} + c_2 \frac{t^{\alpha(1+\gamma)}}{r^\alpha},\;\;\;\;\gamma = \frac{d+1}{\alpha -2d},
\label{sl} 
\end{equation}
by choosing $\chi = t^\gamma$. This result implies a sublinear light cone $t\propto r^{\frac{1}{\gamma +1}}$. Matsuta {\it et al.} then generalized this result to include arbitrary multiple-body interactions. The bound takes the form of\cite{TM17} 
\begin{equation}
\left(c_1 + c_2\frac{t}{r^{2\eta}}\right) e^{\lambda t - r^\eta} + c_3 \frac{t}{r^\eta} ,\;\;\;\; \eta=\frac{1}{1+\gamma}= \frac{\alpha -2d}{\alpha -d +1},
\label{mkn}
\end{equation}
which also predicts the same sublinear light cone $t\propto r^\eta$, but has a much slower spatial decaying tail compared to Eq.~(\ref{sl}) (i.e., $r^{-\eta}$ vs $r^{-\alpha}$). The above result was recently improved by Else {\it et al.} to\cite{DVE20}
\begin{equation}
c_1 e^{vt - r^{1-\sigma}} + \frac{c_2 t + c_3 t^{\frac{d+1}{1-\sigma}}}{r^{\sigma(\alpha -d)}},
\label{emny}
\end{equation}
where $\sigma\in (\eta \gamma,1)$ is tunable. The sublinear light cone is again consistent with the above results in the $\sigma\to\eta\gamma$ limit. Moreover, it improves the decaying exponent from $\eta <1$ to $\sigma(\alpha -d)>d$, which is still worse than Eq.~(\ref{sl}) but is good enough for confirming the existence of the so-called second-type light cone\footnote{This refers to the light cone outside which the change of the \emph{Hamiltonian} (almost) does not alter the operator dynamics. For short-range systems, it coincides with the usual first-type light cone, outside which the operators (almost) commute.} for arbitrary long-range systems with $\alpha >2d$. 

A second big progress was made by Chen and Lucas, who found that for 1D systems with two-body long-range interactions, there exists a linear light cone for $\alpha>3$.\cite{CFC19} Recently, this result was greatly generalized by Kuwahara and Saito to arbitrary long-range interactions and dimensions:\cite{TK20}  
\begin{theorem}
Consider the same setting as Thm.~\ref{Thm:HKB} except for an additional condition given in Eq.~(\ref{xad}) and $\alpha >2d+1$, then we have
\begin{equation}
\|[\hat O_X(t),\hat O_Y]\|\le c |X^{(vt)}| |Y^{(vt)}| \frac{t^{2d+1}\log^{2d}(r+1)}{(r -vt)^\alpha},\;\;\;\; t<\frac{r}{v},
\end{equation}
where $v$ and $c$ are constants that depend only on $\{d,\alpha,J,J_1\}$, $X^{(\xi)}$ is defined as the smallest subset of $X$ such that $X\subseteq\{x:{\rm dist}(x,X^{(\xi)})\le \xi \}$.
\end{theorem}
While the rigorous proof is very technical, one may intuitively understand why a linear light cone exists for $\alpha > 2d +1$ as follows:\cite{TK20} We recall the idea of separating short-range and long-range parts, but now make a cutoff for the long-range part at $2\chi$ and consider its influence on a \emph{fixed} short-range Hamiltonian. We would like to see whether the long-range contribution survives in the limit of $\chi\to\infty$. 
For a finite $\chi$, however, we may even apply the short-range Lieb-Robinson bound to this ``long-range" part. Following a similar argument explaining why the Hasting-Koma bound can be improved for $\alpha>2d$, we can evaluate the long-range contribution to the Lieb-Robinson velocity in the interaction picture to be $\chi\times t^d\chi^{-(\alpha-d)}\sim\chi^{-(\alpha - 2d -1)}$ for $t\sim\chi$, which vanishes for $\alpha > 2d+1$.



\subsubsection{Optimality}
There remains the problem whether the long-range Lieb-Robinson bound mentioned above is optimal in the qualitative sense, i.e., whether one can improve (enlarge) the exponent $\beta$ in the light cone $t\sim r^\beta$. The linear light cone for $\alpha > 2d +1$ is certainly optimal since even short-range Hamiltonians can generate linear correlation propagations. It turns out that the threshold $2d +1$ cannot be smaller, as it was found by Tran {\it et al.} that for a $d$-dimensional qubit lattice there exists a fast operator-spreading protocol generated by a time-dependent Hamiltonian with (no slower than) $r^{-\alpha}$ 
decay ($\forall \alpha>d$) such that $\|[\hat \sigma^x_{\boldsymbol{0}}(t), \hat \sigma^x_{\boldsymbol{r}}]\|\ge c t^{1+2d}|\boldsymbol{r}|^{-\alpha}$ for $t\in(3,c'|\boldsymbol{r}|^{\alpha/(1+2d)})$.\cite{MCT20} Here $\hat\sigma^x_{\boldsymbol{r}}$ is the Pauli-$x$ operator acting at site $\boldsymbol{r}$ and $c,c'$ are both constants.

For $\alpha\in(d,2d)$, the Hastings-Koma bound (\ref{LRLR}) is also optimal, if one does not distinguish $\log r$ from ${\rm poly}(\log r)$. This is because one can find an iterative protocol for preparing a Greenberger-Horne-Zeilinger state on a $d$-dimensional hypercube with length $r$ starting from a product state and with time cost $\propto \log^\kappa r$, where $\kappa$ is a constant depending only on $\alpha$ and $d$.\cite{MCT21} Applying the same construction to $\alpha\in (2d,2d+1)$, one obtains a protocol with time cost $\propto r^{\alpha - 2d}$. While the sublinear Lieb-Robinson bounds in Eqs.~(\ref{sl}), (\ref{mkn}) and (\ref{emny}) seem to suggest a much smaller time cost $t\propto r^{(\alpha - 2d)/(\alpha - d +1)}$, a very recent work by Tran {\it et al.} showed that the light cone can indeed be tightened as $t\propto r^{\alpha - 2d - \epsilon}$ with arbitrarily small $\epsilon>0$.\cite{MCT21b} The derivation of this optimal bound again involves an iteration technique: unlike simply decomposing the Hamiltonian into two parts (short-range and long-range) as done in Sec.~\ref{sll}, we may perform a more general decomposition $\hat H= \sum^n_{j=1} \hat H_j$ with $\hat H_j$ containing the interactions ranging from $\ell_{j-1}$ ($\ell_0\equiv0$) to $\ell_j$ ($\ell_n\equiv\infty$). By iteratively analyzing the effect of adding $\hat H_j$ to $\sum^{j-1}_{j'=1} \hat H_{j'}$, one will end up with the optimal light cone $t\propto r^{\alpha - 2d - \epsilon}$ for $\alpha\in (2d,2d+1)$, and also reproduce $t\propto r$ for $\alpha > 2d +1$.

Finally, it is worthwhile to mention that for noninteracting systems described by quadratic Hamiltonians, we have a linear light cone for $\alpha>d+1$ and a sublinear light cone $t\propto r^{\alpha - d}$ for $\alpha\in (d,d+1)$. These results are optimal because there exist single-particle transfer protocols with such types of time cost.\cite{MCT20} To understand the new threshold $d+1$, one may further assume the translation invariance such that the system can be described by a Bloch Hamiltonian with band dispersions $\epsilon_{\boldsymbol{k}}$'s. Just like the Hamiltonian, the Fourier components of $\epsilon_{\boldsymbol{k}}$ also decay as $|\boldsymbol{r}|^{-\alpha}$, so the group velocity $\nabla \epsilon_{\boldsymbol{k}}$ is ensured to converge only for $\alpha > d+1$ (cf. Sec.~\ref{Sec:free_LR}).


\subsection{Application to equilibrium physics}
While the Lieb-Robinson bound is a dynamical constraint, it has important implications on the equilibrium, typically ground-state properties of gapped quantum many-body systems. Here we pick up a few popular topics and explain the general ideas of gaining insights into equilibrium physics from the Lieb-Robinson bound. Readers with interest may refer to more expert reviews such as Refs.~\refcite{MBH10} and \refcite{MBH21}.

\subsubsection{Clustering condition}
A quantum state is said to obey the clustering condition if its correlation function of two observables  $\hat O_X$, $\hat O_Y$ always decays with increasing distance $r={\rm dist}(X,Y)$ between the observables. It was first noted by Hastings that the Lieb-Robinson bound can be used to prove the clustering condition for ground states of strictly local and gapped Hamiltonians.\cite{MBH04a} This result was later generalized to include exponential\cite{BN06} and even algebraic tails.\cite{MBH06} For simplicity, here we focus on quantum spin systems with unique ground states, although it is rather straightforward to generalize the results to fermion systems or/and degenerate ground states.\footnote{In the presence of degeneracy, we have to properly subtract some matrix elements between different ground states from the correlation function to obtain the clustering condition.} 

We consider a gapped many-body Hamiltonian $\hat H$ with either short-range or long-range (yet $\alpha >d$) interactions and, without loss of generality, assume its ground state $|\Psi_0\rangle$ to have zero energy. The central quantity is the correlation function:
\begin{equation}
{\rm Cor}(\hat O_X,\hat O_Y,|\Psi_0\rangle)\equiv \langle\Psi_0| \hat O_X\hat O_Y|\Psi_0\rangle - \langle\Psi_0| \hat O_X|\Psi_0\rangle\langle\Psi_0|\hat O_Y|\Psi_0\rangle.
\label{cor}
\end{equation}
In order to make a connection with the Lieb-Robinson bound, we observe that 
$\langle\Psi_0|[\hat O_X(t),\hat O_Y ]|\Psi_0\rangle= \langle\Psi_0| \hat O_X e^{-i\hat Ht}\hat O_Y|\Psi_0\rangle -  \langle\Psi_0| \hat O_Y e^{i\hat Ht}\hat O_X|\Psi_0\rangle$. Accordingly, 
suppose there is a filter function $f(t)$ whose Fourier transform $F(\omega)=\int^{\infty}_{-\infty} dt f(t) e^{-i\omega t}$ is exactly ${\rm sgn}(\omega)/2$, we have
${\rm Cor}(\hat O_X,\hat O_Y,|\Psi_0\rangle) =\int^\infty_{-\infty}dt \langle\Psi_0|[\hat O_X(t),\hat O_Y ]|\Psi_0\rangle f(t)$. However, even if $F(\omega)$ is not exactly ${\rm sgn}(\omega)/2$ but an odd function approximating it, we can bound the correlation function (\ref{cor}) by
\begin{equation}
{\rm Cor}(\hat O_X,\hat O_Y,|\Psi_0\rangle) \le 2\int^\infty_0 dt |f(t)| B(r,t) +  \|\hat O_X\|\|\hat O_Y\|\max_{\omega\in[\Delta,\infty)}|1-2F(\omega)|,
\label{corb}
\end{equation}
where $\Delta >0$ is the energy gap and $B(r,t)\ge \|[\hat O_X(t),\hat O_Y]\|$ is a Lieb-Robinson bound. In particular, we may choose $F(\omega)= {\rm erf}(\omega/\sigma)/2$ (${\rm erf}(x)\equiv 2\pi^{-1/2} \int^x_0 dy e^{-y^2}$ is the error function) and $f(t)=i e^{-\sigma^2t^2/4}/(2\pi t)$. For the short-range (long-range) Lieb-Robinson bound (\ref{LRB}) (Hastings-Koma bound (\ref{LRLR})) $B(r,t)\sim e^{\lambda t -\kappa r}$ ($B(r,t)\sim e^{\lambda t}/r^\alpha$),\footnote{Since the first term in Eq.~(\ref{corb}) is a time integral to infinity, it is better to use the trivial bound $B(r,t)=2\|\hat O_X\|\|\hat O_Y\|$ for late times.} we can set $\sigma\sim r^{-1/2}$ ($\sigma\sim\log r$) to obtain an exponential (algebraic) clustering property ${\rm Cor}\lesssim e^{-\kappa' r}$ (${\rm Cor}\lesssim r^{-\alpha'}$) with $\kappa'=\kappa/(1+2\lambda/\Delta)$ ($\alpha'=\alpha/(1+2\lambda/\Delta)$).\cite{MBH06} We mention that for long-range systems with only two-body interactions and $\alpha>2d$, we can use the improved long-range Lieb-Robinson bound in Eq.~(\ref{sl}) to show that the clustering property can be improved to ${\rm Cor}\lesssim r^{-\alpha + 0^+}$ (``$0^+$" means there could be ${\rm poly}(\log r)$ corrections).\cite{MCT20} On the other hand, it remains an open problem whether the correlation in the ground state essentially follows the same algebraic decay as the Hamiltonian also for $\alpha\in (d,2d]$, as is observed in the long-range Kitaev chain.\cite{DV16}

The Lieb-Robinson bound can also be used to prove the clustering condition for fermionic Gibbs states $\hat\rho_\beta = e^{-\beta \hat H}/{\rm Tr}[e^{-\beta \hat H}]$ and \emph{odd}-parity operators such as fermion creation and annhilation operators. This is based on the identity\cite{MBH04b,SHS17} 
\begin{equation}
\langle \hat O_X\hat O_Y \rangle_\beta =\frac{i}{\beta}\int^\infty_0 dt \frac{\langle\{\hat O_X(t) - \hat O_X(-t),\hat O_Y\}\rangle_\beta}{2\sinh (2\pi t/\beta)},
\label{XYb}
\end{equation} 
where $\langle\cdots\rangle_\beta\equiv {\rm Tr}[\cdots \hat\rho_\beta]$ and $X\cap Y=\emptyset$ so that $\{\hat O_X,\hat O_Y\}=0$. Also note that $\langle \hat O_X\rangle_\beta = \langle \hat O_Y\rangle_\beta=0$ due to the fermion superselection rule, implying that Eq.~(\ref{XYb}) is indeed the correlation function. Similar to Eq.~(\ref{corb}), Eq.~(\ref{XYb}) can be simply upper bounded by $\int^\infty_0 dt B(r,t)/(\beta \sinh(2\pi t/\beta) )$, where $B(r,t)$ is a Lieb-Robinson bound. It is worth mentioning that the clustering properties have been proved for general systems above critical temperatures\cite{MK14,TK20b} \textcolor{black}{(or at arbitrary temperatures, if one concerns entanglement\cite{TK22PRX})}, although they are more relevant to the ``imaginary-time" Lieb-Robinson bound rather than the real-time one.\cite{AMA22} 




\subsubsection{Boundary-condition insensitivity}
Intuitively, the bulk properties of locally interacting quantum many-body systems should be almost independent of the boundary condition, provided that the system size is large enough. This statement can be formalized as a perturbation problem, where the boundary terms are naturally identified as the perturbation part. Again, the Lieb-Robinson bound turns out to be very useful in this context, as we explain in the following.   

Consider a parametrized Hamiltonian $\hat H_h = \hat H_0 + h\hat V$ with $h\in[0,1]$. Suppose that the ground state $|0_h\rangle$ is non-degenerate, then for any observable $\hat O$ we have
\begin{equation}
\partial_h \langle \hat O\rangle_h = \langle (\hat O- \langle\hat O\rangle_h)(E_h - \hat H_h)^{-1} (\hat V -\langle \hat V\rangle_h) + {\rm H.c.}\rangle_h,
\label{ph}
\end{equation}
where $\langle\cdots\rangle_h\equiv \langle 0_h|\cdots |0_h \rangle$ and $E_h$ is the ground-state energy. Note that $(\hat O - \langle\hat O\rangle_h)|0_h\rangle$ is orthogonal to $|0_h\rangle$, so the action of $(E_h - \hat H_h)^{-1} $ is well-defined. Just like the correlation function (\ref{cor}), the rhs of Eq.~(\ref{ph}) can also be obtained from the nonequal-time commutator via $\int^\infty_{-\infty} dt\langle [\hat O(t),\hat V] \rangle_h f(t)$, although now the Fourier transform of the filter function $f(t)$ reads $F(\omega)=-1/\omega$. 
To suppress the long-time contribution, we can again use an approximate filter to obtain a bound like Eq.~(\ref{corb}), where the Lieb-Robinson bound appears.  For example, we can choose $F(\omega) =(e^{-\omega^2/\sigma^2} - 1)/\omega$ and set $\sigma\sim r^{-1/2}$ ($r={\rm dist}({\rm supp}\textcolor{black}{[\hat O]},{\rm supp}\textcolor{black}{[\hat V]})$) for short-range systems, leading to\cite{HW18,ZW21}
\begin{equation}
|\partial_h \langle \hat O\rangle_h|\le c \|\hat O\|\|\hat V\| \sqrt{r} e^{-\kappa r}
\label{ohb}
\end{equation}
where $c$ is a constant that does not depend on $r$ (but on $\Delta_h$, the gap of $\hat H_h$; roughly we have $c\propto\Delta_h^{-1}$), and $\kappa$ follows that in Eq.~(\ref{LRB}). An immediate implication of this result is that, for a system with finite size $L$, the deviation of a bulk local observable from the thermodynamic limit is exponentially suppressed by $L$, provided that $\Delta_h$ is no smaller than ${\rm poly}(L^{-1})$.\footnote{This includes the case of, e.g., 2D quantum Hall and topological insulators. Although their open boundary spectra are gapless in the thermodynamic limit due to the topological edge modes, there is generally a nonzero gap $\Delta\propto L^{-1}$ for finite $L$.} 

A more tricky situation appears for the so-called twisted boundary condition, which formally requires $\hat c_{\boldsymbol{r} + L_j \boldsymbol{e}_j } = e^{i\theta_j} \hat c_{\boldsymbol{r}}$ ($j=1,2,...,d$) on a hyperrectangle lattice, where $\boldsymbol{e}_j$ and $L_j$ are the unit vector and length along the $j$th direction. If the system has a global ${\rm U}(1)$ symmetry (such as total particle-number or spin-excitation-number conserving), an equivalent way of imposing the twisted boundary condition is to introduce a ${\rm U}(1)$ gauge field. Thanks to the gauge degree of freedom, this modification can be made local for finite-range interacting systems so that Eq.~(\ref{ohb}) applies. In particular, choosing $\hat O = \hat H_{\boldsymbol{\theta}}$ ($\boldsymbol{\theta}\equiv[\theta_j]^d_{j=1}$), we obtain the flatness of the ground state energy with respect to $\boldsymbol{\theta}$. This is because for each local term in $\hat H_{\boldsymbol{\theta}}$, we can choose a different gauge such that the perturbation is far away from its support. Further assuming the existence of a local operator $\hat O_0$ such that the $|\langle 1_h|\hat O_0 |0_h\rangle|$ is at least ${\rm poly}(L^{-1})$, where $|1_h\rangle$ is the first excited eigenstate of $\hat H_h$, one can show that the energy gap $\Delta_{\boldsymbol{\theta}}$ should also have very week dependence on $\boldsymbol{\theta}$ for sufficiently large $L$.\cite{HW18} This fact is crucial for proving the Lieb-Schultz-Mattis-Oshikawa-Hastings theorem,\cite{EHL61,MO00,MBH04} which rules out the possibility of a unique ground state for gapped ${\rm U}(1)$-symmetric short-range systems with fractional filling factors in arbitrary dimensions. It is worth mentioning that the insensitivity to the twisted boundary condition also applies to bulk response properties, such as the Hall conductance which is related to the Chern number.\cite{QN85} This fact implies that the many-body Chern can be obtained near a fixed twisted boundary condition without integration.\cite{KK19}    

\subsubsection{Entanglement area law}
\label{Sec:EAL}
It is widely believed that the ground states of a gapped Hamiltonian with sufficiently ``local" interactions obey the entanglement area law, i.e., the entanglement entropy of a subsystem $V$ scales like $|\partial V|$.\cite{JE10} A remarkable breakthrough on this topic was made by Hastings, who rigorously proved that for the ground states of finite-range interacting quantum spin Hamiltonians in 1D, the entanglement entropy is always bounded for any segment of the spin chain.\cite{MBH07}  
In the original proof, the Lieb-Robinson bound was explicitly used to show a useful lemma. 
Later, this entanglement upper bound, which has a polynomial dependence on the local Hilbert-space dimension, was improved by Arad {\it et al.} to have a poly-logarithmic dependence 
using a very different method called approximate ground state projection,\cite{IA13} which also applies to long-range systems.\cite{TK20c} It is also worth mentioning that the exponential clustering condition has been proved to imply the entanglement area law in 1D.\cite{FGSLB13,FGSLB15,JC18}

In higher dimensions, the entanglement area law has only been proved with rather strong further assumptions, such as noninteracting\cite{MBP05,MC06} or frustration-free.\cite{AA21} Physically, however, it is rather natural to expect the area law from the perspective of phases of quantum matter\cite{ZB19} --- any two many-body states in the same phase should be related to each other by a finite-depth quantum circuit of local unitaries (possibly with some rapidly decaying tails), which only generate $|\partial V|$ amount of entanglement for a subsystem $V$. Then provided that each phase has a representative obeying the area law, we know that the area law holds generally. This idea was partially made rigorous\footnote{This is not a final proof of the area law since one still has to rigorously show the second statement, i.e., there is always a representative obeying the area law in each phase. It may not be so difficult to demonstrate the area law for some specific states, but it is definitely difficult to show that those states form a complete set of representatives, in the sense that their parent Hamiltonians can be smoothly deformed to cover all the Hamiltonians of interest without closing the gap.} by Acoleyen {\it et al.} in Ref.~\refcite{KVA13} based on a tight bound on entanglement generation, as we will review in the next section, and \rhcomment{by} the quasi-adiabatic continuation technique,\cite{MBH05,TJO07} which relies heavily on the Lieb-Robinson bound and will be briefly reviewed in the following. 

Roughly speaking, the quasi-adiabatic continuation is a kind of shortcut-to-adiabaticity with locality constraint. For a smooth path of gapped Hamiltonians $\hat H(\lambda) =\sum_{A\subseteq\Lambda} \hat H_A(\lambda)$ ($\lambda\in[0,1]$), the non-adiabatic Hamiltonian $\hat H^{(\rm na)}(\lambda)$ that mimics the adiabatic evolution, i.e., satisfies $|\Psi_0(\lambda)\rangle=\overleftarrow{\Lambda }e^{-i\int^\lambda_0 d\lambda' \hat H^{(\rm na)}(\lambda')}|\Psi_0(0)\rangle$ with $|\Psi_0(\lambda)\rangle$ being the ground state of $\hat H(\lambda)$, can be constructed as\cite{SB10}
\begin{equation}
\hat H^{(\rm na)}(\lambda)= \sum_{A\subseteq\Lambda}\int^{\infty}_{-\infty} dt f(t) e^{i\hat H(\lambda)t}\partial_\lambda \hat H_A(\lambda)e^{-i\hat H(\lambda)t},
\label{Hna}
\end{equation}
where $f(t)$ is a real filter function satisfying $F(\omega)\equiv \int^{\infty}_{-\infty} dt f(t) e^{-i\omega t}=-F(-\omega)$ and $F(\omega)=i/\omega$ for $|\omega|\ge\Delta$, with $\Delta$ being the minimal gap along the path $\hat H(\lambda)$. Thanks to the degree of freedom of $F(\omega)$ for $|\omega|<\Delta$, one may make $f(t)$ decay rapidly for large $t$ so that each term in Eq.~(\ref{Hna}) roughly stays local, provided that $\hat H(\lambda)$ satisfies the Lieb-Robinson bound. The finite-time dynamics governed by $\hat H^{(\rm na)}(\lambda)$ should thus not be able to generate too much entanglement. Note that the above analysis applies not only to short-range systems, as originally considered by Acoleyen {\it et al.}, but also to long-range systems with sufficiently large $\alpha$.\cite{ZXG17} The later point will be discussed in further detail in Sec.~\ref{Sec:EGMB}.  

\section{Entanglement generation} 
\label{Sec:EG}
The Lieb-Robinson bound generally sets a limit on the propagation of correlation, which is not necessarily purely quantum. In this section, we focus on the dynamical behavior of entanglement, a genuinely quantum property. We will see that the locality of Hamiltonian has a very strong constraint on the entanglement growth --- that is, an area law for the growth rate. We will also discuss some specific situations where the entanglement growth has a lower bound, as well as a hydrodynamic description for coarse-grained entanglement dynamics, where a hierarchy of various entanglement velocities emerges. 

\subsection{Bipartite systems}
\begin{figure}[bt]
\centerline{\psfig{file=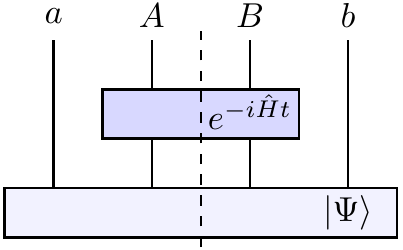,width=2in}}
\vspace*{8pt}
\caption{Entanglement generation in a general bipartite system $AB$ starting from $|\Psi\rangle$ and time evolved by $\hat H$. Here $a$ and $b$ are ancillas of $A$ and $B$, respectively, while $\hat H$ acts nontrivially only on $A$ and $B$. Our concern is the entanglement entropy of the reduced state on $aA$ (cf. Eq.~(\ref{GHP})).}
\label{Fig:SIEC}
\end{figure}

Let us start from the simplest, yet most general situation -- a bipartite system evolved by a general Hamiltonian. The minimal setup of entangling two qubits was studied by D\"ur {\it et al.} in Ref.~\refcite{WD01}. A remarkable observation is that the same Hamiltonian evolution may generate more entanglement, if ancillas are allowed. This motivates the following general formulation of the problem: Consider a bipartite Hamiltonian $\hat H$ acting on two subsystems $A$ and $B$, which are associated with auxiliary subsystems $a$ and $b$, respectively. For a quantum state $|\Psi\rangle$ on $aABb$, the entanglement rate is defined by\cite{SB07}
\begin{equation}
\Gamma (\hat H, |\Psi\rangle) \equiv \left.\frac{d S(\hat \rho_{aA}(t))}{dt}\right|_{t=0},\;\;\;\;\hat\rho_{aA}(t)={\rm Tr}_{Bb}[e^{-i\hat Ht}|\Psi\rangle\langle\Psi|e^{i\hat Ht}],
\label{GHP}
\end{equation}
where $S(\hat\rho)\equiv -{\rm Tr}[\hat\rho \log \hat\rho]$ is the von Neumann entropy. See Fig.~\ref{Fig:SIEC} for a schematic illustration of the setup. After straightforward calculations, one can obtain  the following explicit expression:
\begin{equation}
\Gamma(\hat H,|\Psi\rangle) =i{\rm Tr}_{aAB}[\hat H[\hat\rho_{aAB},\log\hat \rho_{aA}]],
\label{GHP2}
\end{equation}
where $\hat \rho_{aAB}={\rm Tr}_b |\Psi\rangle\langle\Psi|$ and $\hat\rho_{aA}={\rm Tr}_B\hat \rho_{aAB}$, $\hat H$ and $\log\hat \rho_{aA}$ should be understood as $\mathbbm{1}_a\otimes \hat H$ and $\log\hat \rho_{aA}\otimes \mathbbm{1}_B$, respectively. The entanglement capability of $\hat H$ can then be defined as $\Gamma (\hat H)\equiv\max_{|\Psi\rangle}\Gamma (\hat H, |\Psi\rangle)$, which is an intrinsic property of the bipartite Hamiltonian itself. Note that the maximization is carried out also to include arbitrarily large ancillas. Similarly, we can define the disentanglement capability as $\tilde \Gamma(\hat H)\equiv -\min_{|\Psi\rangle} \Gamma(\hat H,|\Psi\rangle)=\Gamma(-\hat H)$, which is generally not equal to $\Gamma(\hat H)$.\cite{NL09b}

It is not obvious whether $\Gamma(\hat H)$ is well-defined since it may diverge as the ancillas become larger and larger. This possibility was ruled out by Bennett {\it et al.}, who showed that $\Gamma (\hat H)\le c\|\hat H\|D^4$ for arbitrary ancillas.\cite{CHB03} Here $c$ is a universal $\mathcal{O}(1)$ constant and $D=\min\{\dim\mathcal{H}_A,\dim\mathcal{H}_B\}$ is the minimal Hilbert space of subsystems $A$ and $B$. The $D$-dependence of this upper bound was then constantly improved to $D^2$,\cite{SB07} $D$,\cite{EHL13} and finally to\cite{KVA13,KMRA14,MM16}
\begin{equation}
\Gamma (\hat H)\le c\|\hat H\|\log D,\;\;\;\; D=\min\{\dim\mathcal{H}_A,\dim\mathcal{H}_B\},
\label{SIE}
\end{equation}
which is called the small incremental entangling conjecture attributed to Kitaev.\cite{SB07} This bound is tight, in the sense that logarithmic dependence on $D$ cannot be further improved, unless additional constraint on $\hat H$ is assumed (for example, if $\hat H=\hat H_A\otimes \hat H_B$, then one can get rid of the $D$-dependence\cite{AC04}). When applied to quantum spins on a lattice, Eq.~(\ref{SIE}) implies that the entanglement growth rate will be at most proportional to the volume of the support. This fact turns out to be very useful in dealing with entanglement growth in many-body systems, as we will see later. 

Let us briefly explain how to derive Eq.~(\ref{SIE}) following Ref.~\refcite{KMRA14}. First, let us introduce the small incremental mixing conjecture attributed to Bravyi,\cite{SB07} which states that
\begin{equation}
\Lambda(\{p_j,\hat\rho_j\}^1_{j=0},\hat H)\equiv\left.\frac{dS(\hat \rho(t))}{dt}\right|_{t=0}\le c'\|\hat H\| h(\{p_j\}^1_{j=0}),
\label{SIM}
\end{equation}
where $\{p_j\}^1_{j=0}$ is a binary probability distribution, $\{\hat\rho_j\}^1_{j=0}$ are two normalized density operators on a finite Hilbert space, $\hat \rho (t) = p_0\hat \rho_0 + p_1 e^{-i\hat Ht}\hat \rho_1e^{i\hat Ht}$ and $h(\{p_j\}^1_{j=0})= -p_0\log p_0 - p_1\log p_1$  is the (classical) Shannon entropy. Although not obvious, Eq.~(\ref{SIM}) implies Eq.~(\ref{SIE}) with $c=4c'$ upon the substitution $p_0=D^{-2}=1-p_1$, $\hat\rho_0=\hat\rho_{AB}={\rm Tr}_{ab}|\Psi\rangle\langle\Psi|$ and $\hat\rho(0)={\rm Tr}_B[\hat\rho_{AB}]\otimes\mathbbm{1}_B/D $ (here $D={\rm dim}\mathcal{H}_B\le{\rm dim}\mathcal{H}_A$ is assumed without loss of generality).\cite{SB07} It thus suffices to derive Eq~(\ref{SIM}). After some straightforward calculations, one can show
\begin{equation}
S(\hat\rho(t)) - S(\hat \rho(0))= -\sum^1_{j=0}p_j\log p_j(S_{p_j}(\hat\rho_j||\hat\rho_{1-j}((-)^jt)) - S_{p_j}(\hat\rho_j||\hat\rho_{1-j})),
\label{StS0}
\end{equation}
where $S_\alpha(\hat\rho||\hat\sigma)\equiv S(\hat\rho||\alpha\hat\rho +(1-\alpha)\hat\sigma)/(-\log\alpha)$ ($\alpha\in(0,1)$) is the quantum skew divergence ($S(\hat\rho||\hat\sigma)\equiv{\rm Tr}[\hat\rho(\log\hat\rho -\log\hat\sigma)]$ is the usual quantum Kullback-Leibler divergence) and $\hat\rho_{0,1}(t)=e^{-i\hat H t}\hat\rho_{0,1}e^{i\hat Ht}$. Thanks to the following nice property satisfied by the quantum skew divergence:\cite{KMRA14}
\begin{equation}
S_\alpha (\hat\rho || e^{i\hat O}\hat\sigma e^{-i\hat O}) - S_\alpha (\hat\rho || \hat\sigma) \le 2\|\hat O\|,\;\;\;\;\forall \hat O^\dag =\hat O,
\end{equation}
we know that the rhs of Eq.~(\ref{StS0}) can be upper bounded by $2t \|\hat H\|h(\{p_j\}^1_{j=0})$. Taking the $t\to0$ limit, we obtain Eq.~(\ref{SIM}) with $c'=2$.



\subsection{Many-body systems}
\label{Sec:EGMB}
Having in mind the upper bound (\ref{SIE}) on entanglement generation for arbitrary bipartite systems, we are in a position to show some fundamental limit on the entangling power of many-body Hamiltonians with a certain kind of locality. We will also see some rare examples where the entanglement growth admits a lower bound.

\subsubsection{Area law for growth rate}
As is obvious from Eq.~(\ref{GHP2}), the entanglement rate satisfies the additivity $\Gamma(\hat H_1+\hat H_2,|\Psi\rangle)=\Gamma(\hat H_1,|\Psi\rangle) + \Gamma(\hat H_2,|\Psi\rangle)$. Since we can always make the decomposition $\hat H= \hat H_A +\hat H_B +\hat H_{AB}$, where only the third term $\hat H_{AB}$ acts simultaneous on $A$ and $B$, we have $\Gamma(\hat H,|\Psi\rangle)=\Gamma(\hat H_{AB},|\Psi\rangle)$ since $\Gamma(\hat H_A+\hat H_B,|\Psi\rangle)=0$. For a locally interacting many-body Hamiltonian, setting $A=V$ ($B=\bar V$) to be the subsystem of interest (its complement), we typically encounter the situation in which $\hat H_{V\bar V}=\sum^P_{j=1} J_j \hat v_j\otimes\hat u_j$,\cite{SB06} where $\hat v_j$'s ($\hat u_j$'s) act on $V$ ($\bar V$) with $\|v_j\|\le 1$ ($\|u_j\|\le1$) and $J_j$'s are uniformly bounded $\mathcal{O}(1)$ constants, and $P$ is of the order of $\mathcal{O}(|\partial V|)$. We can then use the $D$-independent version\cite{AC04} of Eq.~(\ref{SIE}) to obtain
\begin{equation}
\frac{dS}{dt}\le c\sum^P_{j=1} |J_j| \le c P\max_j|J_j|\sim\mathcal{O}(|\partial V|),
\label{dSdtdV}
\end{equation}
implying that the entanglement growth rate satisfies the area law. Accordingly, we know that the thermalization time (provided the validity of ETH) for local Hamiltonians should at least scale linearly with respect to the subsystem size.

The above argument by Bravyi {\it et al.} was made in 2006 before the proof of Kitaev's small incremental entangling conjecture. Now with Eq.~(\ref{SIE}) in hand, we can generalize the result to more general systems especially with long-range interactions.\cite{KVA13,ZXG17} For a general many-body Hamiltonian $\hat H=\sum_{A\subseteq\Lambda} \hat h_A$ on a lattice $\Lambda$ with local Hilbert-space dimension $d_s$ per site, the entanglement growth rate for subsystem $V$ should be bounded by
\begin{equation}
\frac{dS}{dt}\le c\log d_s\sum_{A:A\cap V\neq\emptyset,A\cap\bar V\neq\emptyset}|A|\|\hat h_A\|. 
\label{dsb}
\end{equation}
The sum in the rhs can be further bounded by
\begin{equation}
\sum_{A:A\cap V\neq\emptyset,A\cap\bar V\neq\emptyset}|A|\|\hat h_A\|\le \sum^\infty_{r=0} \sum_{x\in \Sigma_r}\sum_{A:x\in A,{\rm diam}(A)> r}|A|\|\hat h_A\|,
\end{equation}
where $\Sigma_0\equiv\partial V$ and $\Sigma_r\equiv\{x\in V: r\le {\rm dist}(x,\Sigma_0)<r+1\}$ for $r\in\mathbb{Z}^+$. By further assuming that the lattice geometry validates $|\Sigma_r|\le c'|\Sigma_0|$ for some $\mathcal{O}(1)$ constant $c'\ge 1$ and that 
\begin{equation}
\max_{x\in\Lambda} \sum_{A:x\in A,{\rm diam}(A)\ge R}|A|\|\hat h_A\| \le \frac{J}{(1+R)^{\alpha - d}},
\label{xaad}
\end{equation}
which is a locality constraint stronger than Eq.~(\ref{xad}), we obtain
\begin{equation}
\frac{dS}{dt}\le c c' J \log d_s|\Sigma_0| \sum^\infty_{r=0} \frac{1}{(1+r)^{\alpha - d}}\sim\mathcal{O}(|\partial V|),
\end{equation}
provided that $\alpha > d+1$. Note that for many-body systems involving up to $k$-body interactions, there is no essential difference between Eqs.~(\ref{xaad}) and (\ref{xad}).

Let us recall the entanglement-area-law problem for ground states, which can be related to the dynamical entanglement generation problem via the quasi-adiabatic-continuation technique (cf. Sec.~\ref{Sec:EAL}). It was pointed out by Gong {\it et al.} that, for any smooth path of a gapped two-body interacting Hamiltonian $\hat H(\lambda)=\sum_{x,y}\hat h_{xy}(\lambda)$ ($\lambda\in[0,1]$) with $\|\hat h_{xy}(\lambda)\|\le J(1+{\rm dist}(x,y))^{-\alpha}$ and $\|\partial_\lambda \hat h_{xy}(\lambda)\|\le J'(1+{\rm dist}(x,y))^{-\alpha}$, if $\alpha >2d+2$, then the ground state $|\Psi_0(\lambda)\rangle$ always satisfies the area law, provided that $|\Psi_0(\lambda=0)\rangle$ does.\cite{ZXG17} To see this, it suffices to show that the generator for quasi-adiabatic continuation (\ref{Hna}) is sufficiently local. Using the long-range Lieb-Robinson bound in Eq.~(\ref{sl}), we can decompose each quasi-local term in $\hat H^{(\rm na)}(\lambda)$ as
\begin{equation}
\hat h^{(\rm na)}_{xy}(\lambda) =\int^\infty_{-\infty} dt f(t) e^{i\hat H(\lambda) t} \hat h_{xy}(\lambda)e^{-i\hat H(\lambda) t} =\sum^\infty_{R=0} \hat k_{xy}(\lambda,R),
\end{equation}
where $\hat k_{xy}(\lambda,R)$ acts on $B^R_x\cup B^R_y$ ($B^R_x\equiv \{y\in\Lambda:{\rm dist}(y,x)\le R\}$) and satisfies $\|\hat k_{xy}(\lambda,R)\|\lesssim \mathcal{O} ((1+R)^{-(\alpha-d)})(1+{\rm dist}(x,y))^{-\alpha}$. Estimating the rhs of Eq.~(\ref{dsb}) for fixed $R\ge 1$ then yields $R|\partial V|\mathcal{O}(R^{-(\alpha -d)}) R^d$, where $R|\partial V|$ roughly gives the volume of $\{x\in V: B^R_x\cap\bar V\neq\emptyset\}$ and $R^d$ arises from $|B^R_x\cup B^R_y|$. After summing $R$ up, we end up with a finite result for $\alpha >2d+2$. As mentioned in Ref.~\refcite{ZXG17}, this result may not be optimal, in the sense that the entanglement area law may remain correct for some $\alpha \le 2d +2$.

\subsubsection{Specific lower bounds}
\label{Sec:LB}
In some specific situations, the entanglement growth rate may have a lower bound given by other physical quantities. Here we provide two such rare examples; both are short-range systems in 1D.

\begin{figure}[bt]
\centerline{\psfig{file=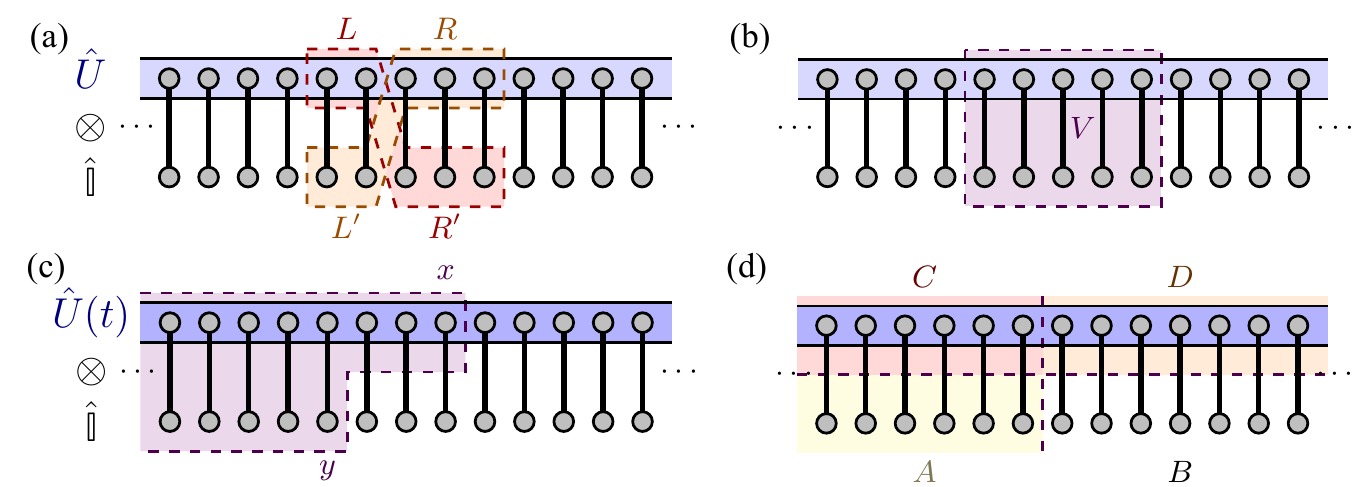,width=5in}}
\vspace*{8pt}
\caption{(a) Subsystems in the Choi-Jamio{\l}kowski state $|\hat U\rangle$ of a 1D QCA $\hat U$ for computing the index (\ref{ind}). Here $L$ and $R$ should be two adjacent intervals that are no smaller than the Lieb-Robinson length of QCA, which is the smallest integer $r$ such that $\hat U^\dag\hat O_j\hat U$ is supported on $[j-r,j+r]$ for any site $j$ and on-site operator $\hat O_j$. $L'$ and $R'$ are the corresponding regions in the replica lattice. (b) Subsystem $V$, which is the union of $LR'$ and $L'R$ in (a), for computing the operator entanglement of $\hat U$. (c) Subsystem in the Choi-Jamio{\l}kowski state $|\hat U(t)\rangle$ of a 1D time-evolution operator $\hat U(t)$ for computing the entanglement line tension $\mathcal{E}(v)$ with $v=(x-y)/t$ (\ref{SxtE}). (d) Subsystems for computing the tripartite information (\ref{IACD}).}
\label{Fig:Choi}
\end{figure}

The first example is quantum cellular automata (QCA),\cite{BS04,PA19,TF20} which are unitary transformations that map (strictly) local operators to local operators. In 1D, QCA are equivalent to matrix-product unitaries\cite{JIC17} and are fully classified by a topological invariant called (chiral) index,\cite{DG12} which is quantized as $\log\mathbb{Q}^+$ and is additive upon tensoring and composition. While there are several ways to define the index for a 1D QCA $\hat U$, one elegant definition is given by the following entropy formula:\cite{BRD18,DR20,ZG21}
\begin{equation}
{\rm ind}=\frac{1}{2}(S(\hat\rho_{LR'}) - S(\hat\rho_{L'R})), 
\label{ind}
\end{equation}
where 
$\hat \rho_{LR'}$ and $\hat \rho_{L'R}$ are two reduced density operators of the Choi-Jamio{\l}kowski state $|\hat U\rangle\equiv (\hat U\otimes\hat{\mathbb{I}})|\Phi_+\rangle$ ($|\Phi_+\rangle$: maximal entangled state on two copies of 1D lattices) shown in Fig.~\ref{Fig:Choi}(a). 
Using the triangle inequality, we immediately obtain a lower bound on the operator entanglement entropy (cf. Fig.~\ref{Fig:Choi}(b)):
\begin{equation}
S(\hat\rho_{LRL'R'})\ge|S(\hat\rho_{LR'}) - S(\hat\rho_{L'R})|=2|{\rm ind}|.
\label{Sind}
\end{equation} 
Recalling the additivity of the index, we know that, whenever $\rm ind\neq0$, the operator entanglement of $\hat U^t$ grows linearly in terms of the time step, ruling out the possibility of any sublinear behavior including many-body localization.\cite{HCP16,TZ17} We mention that Eq.~(\ref{Sind}) actually holds true for arbitrary R\'enyi entropies and is robust against exponential tails in $\hat U$.\cite{ZG21} \textcolor{black}{We also mention that QCA in higher dimensions can be efficiently represented by projected-entangled-pair unitaries and thus satisfy the area law for entanglement generation rate,\cite{LP20} just like finite-time evolution of local Hamiltonians as discussed previously.}

The second example is a ${\rm U}(1)$-symmetric (i.e., particle-number conserving) free-fermion chain with nearest-neighbor hopping and arbitrary on-site potential, whose Hamiltonian reads 
\begin{equation}
\hat H = -J\sum^M_{j=1} (\hat c^\dag_{j+1}\hat c_j + {\rm H.c.}) +\sum^M_{j=1} V_j\hat n_j,
\end{equation}
where $\hat n_j=\hat c_j^\dag \hat c_j$ is the fermion-number operator, $\hat c_j$ annihilates a fermion on site $j$ and $M$ is the total number of sites. Thanks to the free nature of $\hat H$, a fermionic Gaussian state remains Gaussian during the time evolution, allowing us to compute the entanglement dynamics from the covariance matrix\cite{IP09} and numerically access rather large system sizes. On top of the entanglement, another dynamical quantity of interest is the surface roughness,\cite{KF20,TJ20} which measures the particle-number fluctuation and is defined as the variance of the surface-height operator $\hat h_j = \sum^j_{j'=1} (\hat n_j - \nu)$, with $\nu=N/M$ ($N$: total particle number) being the filling factor. Under reasonable assumptions, one can show that this quantity is well approximated by the bipartite number fluctuation $\sqrt{\langle \hat h^2_{M/2}\rangle}\equiv\sqrt{\Tr[\hat h^2_{M/2}\hat \rho_{[1,M/2]}]}$ ($\hat \rho_{[1,M/2]}$: half-chain reduced state), which can be used to bound the half-chain entanglement entropy from both sides:\cite{KF21}
\begin{equation}
4\langle h^2_{M/2}\rangle\log 2 < S(\hat \rho_{[1,M/2]}) <\langle h^2_{M/2}\rangle \log M+1,\;\;\;\;\forall M>15.
\end{equation} 
This result implies that, if the surface roughness growth obeys a power-law scaling $\sim \sqrt{\langle \hat h^2_{M/2}\rangle} \sim t^\beta$, then the entanglement growth should follow the power law $\sim t^{2\beta}$.


\subsection{Coarse-grained dynamics}
On large spacetime scales, also referred to as the hydrodynamic level, it is possible to have a phenomenological description for the coarse-grained entanglement dynamics in generic locally interacting systems.\cite{CJ18} By generic, we mean the typical (chaotic) situation without any symmetry, integrability or other structures. Such a phenomenological theory is usually called entanglement membrane theory,\cite{TZ20} where the entanglement configuration is determined in a very similar way to minimizing the bending energy of an elastic membrane. For simplicity, we will focus on $(1+1)$D dynamics in homogeneous systems, although the theory also applies to higher dimensions\cite{AN18} and inhomogeneous systems.\cite{AN18b} We will then discuss some order relations between various entanglement velocities within the framework of the entanglement membrane theory.

\subsubsection{Entanglement-membrane theory}
\begin{figure}[bt]
\centerline{\psfig{file=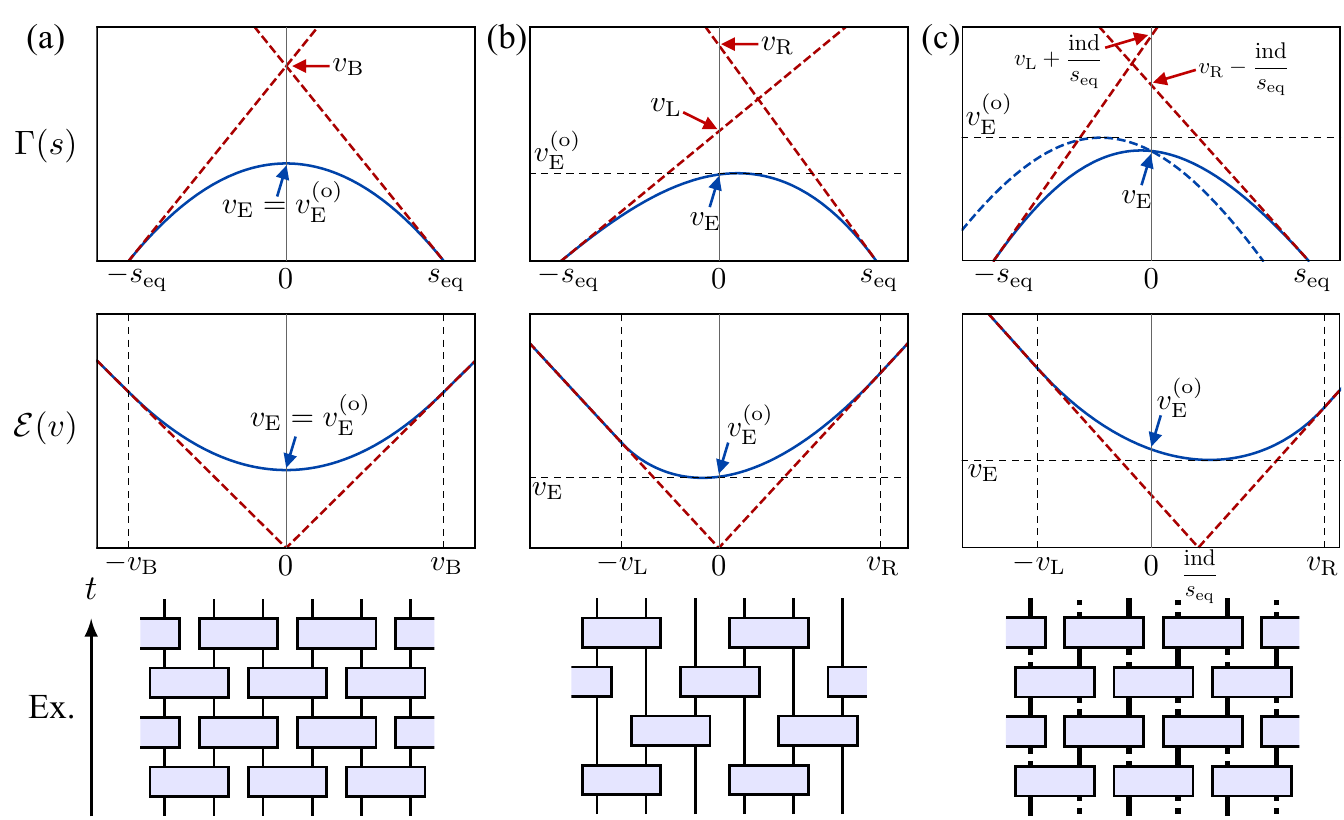,width=5in}}
\vspace*{8pt}
\caption{Typical entropy production function $\Gamma(s)$ (top), entanglement line-tension function $\mathcal{E}(v)$ (middle) and examples (bottom) for 1D normal dynamics (a) with or (b) without mirror symmetry, as well as (c) anomalous dynamics. In case (a), represented by random quantum circuits, both $\Gamma(s)$ and $\mathcal{E}(v)$ are even, leading to $v_{\rm R}=v_{\rm L}=v_{\rm B}\le v_{\rm E}=v^{(\rm o)}_{\rm E}$. In more general case (b) including, e.g., the staircase circuits, we have $v_{\rm L}\neq v_{\rm R}$ and $v_{\rm E}<v^{(\rm o)}_{\rm E}$ in general, yet the order relations in Eq.~(\ref{vEvB}) holds true. In the most general case (c), such as random QCA, the topological index (\ref{ind}) comes into play and modifies the order relations into Eq.~(\ref{vEvBind}). In all the figures, dashed vertical lines indicate the domain (i.e., $[-v_{\rm L},v_{\rm R}]$, outside which $\mathcal{E}(v)$ is plotted as $|v|$ or $|v-{\rm ind}/s_{\rm eq}|$) of $\mathcal{E}(v)$, dashed horizontal lines indicate $v^{(\rm o)}_{\rm E}$ (top) or $v_{\rm E}$ (middle), and red lines are tangents at the boundaries of $\Gamma(s)$ or $\mathcal{E}(v)$. In the top panel of (c), the dashed blue curve corresponds to $\Gamma(s)- {\rm ind} s/s_{\rm eq}^2$. In the bottom panel of (c), the dashed and solid legs indicate different Hilbert space dimensions.} 
\label{Fig:EMT}
\end{figure}

Consider a 1D quantum spin chain undergoing discrete (by quantum circuits) or continuous (by local Hamiltonians) time evolution. Denoting the entanglement entropy for the entanglement cut at position $x$ (i.e., the subsystem of interest consists of all the sites left to $x$) and time $t$ as $S(x,t)$, its equation of motion is suggested to be\cite{CJ18}
\begin{equation}
\partial_t S(x,t) = s_{\rm eq} \Gamma (\partial_x S(x,t)),
\label{Sxt}
\end{equation}
where $s_{\rm eq}$ is the equilibrium entropy density and $\Gamma(s)$ is the entropy production function that depends on the microscopic detail of the system. Here the time and space derivatives can be justified after a proper coarse graining. Formally, one can always write down the solution of Eq.~(\ref{Sxt}) as
\begin{equation}
S(x,t)= \min_y\left[ s_{\rm eq}t \mathcal{E}\left(\frac{x-y}{t}\right)+ S(y,0)\right],
\label{SxtE}
\end{equation}
where $\mathcal{E}(v)$ is the entanglement line tension, which is related to $\Gamma(s)$ via the Legendre transform
\begin{equation}
\mathcal{E}(v) = \max_s\left[\Gamma(s)+\frac{vs}{s_{\rm eq}}\right]\;\;\;\;\Leftrightarrow\;\;\;\;\Gamma(s) = \min_s\left[\mathcal{E}(v)-\frac{vs}{s_{\rm eq}}\right]. 
\label{EvGs}
\end{equation} 
The line tension function satisfies several properties: (i) it is defined on $[-v_{\rm L},v_{\rm R}]$, where $v_{\rm L}$ and $v_{\rm R}$ are the left and right butterfly velocities, respectively; (ii) it is non-negative ($\mathcal{E}(v)\ge0$) and convex ($\mathcal{E}''(v)\ge0$); (iii) $\mathcal{E}(-v_{\rm L})=v_{\rm L}$, $\mathcal{E}(v_{\rm R})=v_{\rm R}$ and $\mathcal{E}'(v_{\rm R})=-\mathcal{E}'(-v_{\rm L})=1$. 
Here the butterfly velocity can be roughly understood as the actual Lieb-Robinson velocity. Technically, it is determined by looking at the operator spreading dynamics,\cite{CWvK18} which may be characterized by the out-of-time-order correlator (cf. Sec.~\ref{Sec:OTOC}). Regarding the entropy production function, these properties are translated into: (i) it is defined on $[-s_{\rm eq},s_{\rm eq}]$; (ii) $\Gamma(s)\ge0$ and $\Gamma''(s)\le0$; (iii) $\Gamma(\pm s_{\rm eq})=0$, $v_{\rm L}=s_{\rm eq}\Gamma'(- s_{\rm eq})$ and $v_{\rm R}=-s_{\rm eq}\Gamma'(s_{\rm eq})$. See Fig.~\ref{Fig:EMT}(b) for an illustration of these relations.

Microscopically, the line tension can be obtained by computing the generalized operator entanglement of the time evolution operator $\hat U(t)$ with input and output cuts at different positions $y$ and $x$, as shown in Fig.~\ref{Fig:Choi}(c). More precisely, the line tension at velocity $v=(x-y)/t$ is equal to the generalized operator entanglement entropy divided by $s_{\rm eq}t$. For some specific models such as the Haar random quantum circuit consisting of nearest-neighbor gates (cf. Fig.~\ref{Fig:EMT}(a)),\cite{AN18,BB20} this can be done analytically for the R\'enyi-2 entropy.\footnote{Precisely speaking, the ensemble average is taken with respect to the purity, which is the exponentiated R\'enyi-2  entropy. The resulting difference is rather small for sufficiently large $q$.\cite{TZ19}} Suppose the local Hilbert space is $q$, we have $s_{\rm eq}=\log q$ due to the absence of any conserved quantity. The R\'enyi-2 line tension function is given by
\begin{equation}
\mathcal{E}_2(v)=\log_q\frac{q^2+1}{q}+\frac{1+v}{2}\log_q\frac{1+v}{2}+\frac{1-v}{2}\log_q\frac{1-v}{2},
\end{equation}
from which one can determined the corresponding entropy production function as
$\Gamma_2(s)= \log_q(\cosh s_{\rm eq}/\cosh s)$. Note that both of these functions are even, as a result of the underlying spatial reflection or mirror symmetry. By solving $\mathcal{E}'_2(v)=1$, one can obtain the butterfly velocity to be
\begin{equation}
v_{\rm B}=v_{\rm R}=v_{\rm L}=\frac{q^2-1}{q^2+1}.
\label{RCvB}
\end{equation}

Very recently, the entanglement-membrane theory has been generalized to anomalous dynamics which are still locality-preserving but are not Hamiltonian evolution or quantum circuits.\cite{ZG21b} Examples include discrete dynamics generated by nontrivial QCA, as discussed in Sec.~\ref{Sec:LB}, and edge dynamics of chiral many-body localized Floquet phases in 2D.\cite{HCP16} This is achieved by simply adding a velocity term into the equation of motion (\ref{Sxt}):
\begin{equation}
\partial_t S(x,t)+\frac{{\rm ind}}{s_{\rm eq}}\partial_x S(x,t) = s_{\rm eq}\Gamma(\partial_x S(x,t)),
\end{equation}
where ${\rm ind}$ is the index given in Eq.~(\ref{ind}) for a single time step. While we can mathematically recast $\frac{{\rm ind}}{s_{\rm eq}}\partial_x S$ into $\Gamma$, this is not physically appropriate since then we will not have $\Gamma(\pm s_{\rm eq})=0$ nor a clear picture of the background entropy current indicated by the stationary equation $\partial_t S +\frac{{\rm ind}}{s_{\rm eq}}\partial_x S=0$. The solution (\ref{SxtE}) to the entanglement dynamics remains formally the same, 
 although $\mathcal{E}(-v_{\rm L})=v_{\rm L}$, $\mathcal{E}(v_R)=v_R$ in property (iii) should be modified as $\mathcal{E}(-v_{\rm L})=v_{\rm L}+\frac{{\rm ind}}{s_{\rm eq}}$, $\mathcal{E}(v_{\rm R})=v_{\rm R}-\frac{{\rm ind}}{s_{\rm eq}}$. In addition, $v$ in the Legendre transform (\ref{EvGs}) should be replaced by $v-\frac{\rm ind}{s_{\rm eq}}$. See Fig.~\ref{Fig:EMT}(c) for an illustration of these modified relations.


\subsubsection{Order relations for entanglement velocities}
In the previous sections, we have mainly focused on the growth rate of state entanglement and also mentioned a little bit that of the operator entanglement. In the context of coarse-grained entanglement dynamics, it is more convenient to consider the so-called entanglement velocity, which is simply the entanglement growth rate divided by $s_{\rm eq}$. According to Eq.~(\ref{SxtE}), choosing a product initial state with $S(y,0)=0$, we know that the state entanglement velocity is given by
\begin{equation}
v_{\rm E}=\min_v\mathcal{E}(v)=\Gamma(0),
\label{vE}
\end{equation}
where the second equality arises from Eq.~(\ref{EvGs}). It is also obvious from the microscopic computation of $\mathcal{E}(v)$ that the operator entanglement velocity is given by
\begin{equation}
v^{(\rm o)}_{\rm E}=\mathcal{E}(0)=\max_s\Gamma(s).
\label{voE}
\end{equation}
See Fig.~\ref{Fig:EMT}(b) for an illustration of the above relations. Combining Eqs.~(\ref{vE}) and (\ref{voE}), we obtain the first order relation
\begin{equation}
v_{\rm E}\le v^{(\rm o)}_E.
\label{vEvoE}
\end{equation}
Note that the Haar random quantum circuit with mirror symmetry gives an example of saturation:\cite{AN18,CWvK18}
\begin{equation}
v_{\rm E}=v^{(\rm o)}_{\rm E}=\log_q\left(\frac{q^2+1}{2q}\right). 
\label{RCvE}
\end{equation}
Examples of strict inequality can be found in the systems without reflection symmetry, such as the staircase circuits and even some 2-local (nearest-neighbor interacting) Hamiltonians. Recalling properties (ii) and (iii) of the line-tension function, we obtain the following two order relations:
\begin{equation}
v_{\rm E}\le\min\{v_{\rm L},v_{\rm R}\},\;\;\;\; v^{(\rm o)}_{\rm E}\le \max\{v_{\rm L},v_{\rm R}\}.
\label{vEvB}
\end{equation}
For the random circuit, one finds an asymptotic saturation in the limit of $q\to\infty$ (cf. Eqs.~(\ref{RCvB}) and (\ref{RCvE})). We emphasize that all of these order relations are derived in a phenomenological manner. It would be interesting to see to what extent these relations can be justified microscopically, i.e., on the basis of quantum-information approaches alone as we did in the previous (sub)sections. 

Finally, we discuss how the above relations are changed by a nonzero index. First, Eq.~(\ref{vEvoE}) remains valid, although it seems unlikely to achieve the equality since ${\rm ind}\neq0$ enforces the mirror symmetry to be broken. In an extreme example of left or right translations, we have $0=v_{\rm E}<v^{(\rm o)}_{\rm E}=1$. On the other hand, Eq.~(\ref{vEvB}) should be modified into
\begin{equation}
v_{\rm E}\le\min\left\{v_{\rm L}+\frac{{\rm ind}}{s_{\rm eq}},v_{\rm R}-\frac{{\rm ind}}{s_{\rm eq}}\right\},\;\;\;\; v^{(\rm o)}_{\rm E}\le \max\{|v_{\rm L}|,|v_{\rm R}|\}.
\label{vEvBind}
\end{equation}
Again, right/left translations with $v_R=-v_L=\pm1$ are examples of saturation. It is worthwhile to mention another dynamical quantity called tripartite information, which is defined as a three-party quantum mutual information of the Choi-Jamio{\l}kowski state $|\hat U(t)\rangle\equiv (\hat U(t)\otimes\hat{\mathbb{I}})|\Phi_+\rangle$:\cite{PH16} 
\begin{equation}
I^{A:C:D}(t) \equiv I^{A:C}(t) + I^{A:D}(t) - I^{A:CD}(t), 
\label{IACD}
\end{equation}
where $I^{X:Y}(t)\equiv S(\hat\rho_X(t))+S(\hat\rho_Y(t))-S(\hat\rho_{XY}(t))$ ($\hat\rho_X(t)\equiv{\rm Tr}_{\bar X}|\hat U(t)\rangle\langle\hat U(t)|$) is the conventional bipartite quantum mutual information and the configurations of $A$, $C$, $D$ are shown in Fig.~\ref{Fig:Choi}(d). This quantity measures how much information from $A$ on the input side is scrambled into $C$ and $D$ on the output side, i.e., the information that cannot be retrieved by individually observing $C$ and $D$.  
At least for two different types of Haar random QCA in Ref.~\refcite{ZG21b}, the tripartite-information velocity $v_{\rm tri}\equiv -\lim_{t\to\infty}t^{-1}I^{A:C:D}(t)$ is found to be the difference between the operator-entanglement velocity and the scaled index:
\begin{equation}
v_{\rm tri} = v^{(\rm o)}_{\rm E} - \frac{|{\rm ind}|}{s_{\rm eq}}.
\label{vtri}
\end{equation}
Note that $v_{\rm tri}\ge0$ implies the entanglement lower bound in Eq.~(\ref{Sind}),\footnote{Here we do not have a factor $2$ simply because we are considering the half-infinite entanglement cut. For a segment embedded in a periodic ring or infinite chain, as is the case for Eq.~(\ref{Sind}), there are necessarily two boundaries.} and another obvious order relation is
\begin{equation} 
v_{\rm tri}\le v^{(\rm o)}_{\rm E}.
\end{equation}
While $v_{\rm tri}\le v_{\rm E}$ is also observed in the random QCA models,\cite{ZG21b} we expect that this is not always the case and can be violated by some normal (${\rm ind}=0$) dynamics with $v^{(\rm o)}_{\rm E}>v_{\rm E}$. It would be interesting to see to what extent Eq.~(\ref{vtri}) can be justified for anomalous dynamics.

\section{Error bounds}
\label{Sec:EB}
Exact descriptions of quantum dynamics, especially in many-body systems, are very difficult in general. Therefore, we usually carry out some approximations, whose precisions can be justified by estimating the error bounds. In this section, we pick up a few remarkable examples on this topic, including approximating time-continuous Hamiltonian evolutions by discrete quantum circuits, approximating periodically driven dynamics by truncating Floquet-Magnus expansions (which is closely related to Floquet prethermalization), and approximating energetically constrained dynamics by projection onto the relevant Hilbert subspace.

\subsection{Circuitization} 
One central task in digital quantum simulation is circuitization, which is, namely, to figure out a quantum circuit that approximates the unitary quantum dynamics one attempts to simulate.\cite{IMG14} To quantify how well the approximation could be for a specific problem, one usually needs to estimate an error bound that depends on the capability of implementing quantum-gate operations (e.g., gate locality and circuit depth) as well as the approach of circuitization. Concretely, focusing on the unitary evolution governed by a time-independent Hamiltonian $\hat H$, we would like to find an upper bound on
\begin{equation}
\epsilon\equiv\|e^{-i\hat Ht} - \hat U_{\rm c}\|,\;\;\;\;
\hat U_{\rm c}=\overleftarrow{\prod}^N_{n=1}\hat U_n,
\label{HtU}
\end{equation}
where $\{\hat U_n\}^N_{n=1}$ represents a sequence of unitary quantum gates for constructing the circuit approximation $\hat U_{\rm c}$. 

\subsubsection{Trotterization}
\label{Sec:Trot}
\begin{figure}[bt]
\centerline{\psfig{file=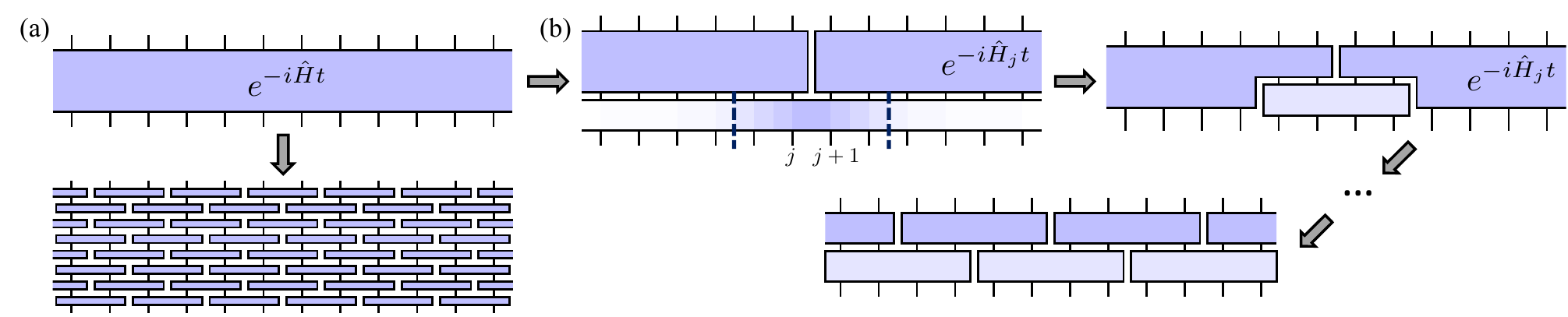,width=5in}}
\vspace*{8pt}
\caption{Schematic illustration of circuitizing the time evolution by nearest-neighbor interacting spin Hamiltonian $\hat H=\sum^L_{j=1}\hat h_{j,j+1}$ by (a) Trotterization and (b) cutting exponential tails. In the bottom of (a), each small block is a short-time evolution generated by $\hat h_{j,j+1}$. In the top of (b), the unitaries above $e^{-i\hat H_jt}$ before and after the cut are generated by $\hat h^{(\rm I)}_{j}(t)$ and $\hat h^{(\rm I)}_{j,r}(t)$ in Eq.~(\ref{TLh}), respectively.} 
\label{Fig:Circ}
\end{figure}

The arguably most popular method of circuitizaiton is Trotterization,\cite{HFT59,MS76,SL96} due partially to its simplicity in both concept and practice. The basic idea is based simply on the identity $e^{A+B}=\lim_{n\to \infty}(e^{\frac{1}{n}A}e^{\frac{1}{n}B})^n$ with $A$ and $B$ being arbitrary matrices. If we further impose anti-Hermiticity to $A$ and $B$, as is the case of simulating unitary dynamics, for any finite $n$, we have
\begin{equation}
\|e^{A+B} -  (e^{\frac{1}{n}A}e^{\frac{1}{n}B})^n \|\le n \|e^{\frac{1}{n}(A+B)} - e^{\frac{1}{n}A}e^{\frac{1}{n}B}\|\le \frac{1}{2n} \|[A,B]\|,
\end{equation}
where the middle result arises from the general relation 
\begin{equation}
\|V_1V_2...V_n - U_1U_2...U_n\|\le \sum^n_{m=1}\|V_m - U_m\| 
\label{VUn}
\end{equation}
for two sets of unitaries $V_m$'s and $U_m$'s, and the final bound can be obtained from the following identity (valid also for non-Hermitian $A$ and $B$):
\begin{equation}
e^{x(A+B)} - e^{xA}e^{xB} = \int^x_0 dx_1\int^{x_1}_0 dx_2 e^{(x-x_1)(A+B)}e^{x_2A}[B,A]e^{-x_2A} e^{x_1A}e^{x_1B}, 
\end{equation}
which implies $\|e^{x(A+B)} - e^{xA}e^{xB} \|\le x^2\|[A,B]\|/2$. 
This result can be applied to, e.g., nearest-neighbor interacting 1D spin Hamiltonian $\hat H=\sum^L_{j=1} \hat h_{j,j+1}$ by choosing $A$ ($B$) to be $\sum_{j:{\rm even}} \hat h_{j,j+1}$ ($\sum_{j:{\rm odd}} \hat h_{j,j+1}$), leading to an error bound $\mathcal{O}(Lt^2/l)$ for a circuit like the bottom of Fig.~\ref{Fig:EMT} with depth $l$ (so that $N=2l$, $\hat U_{2n-1}=\prod_{j:{\rm odd}}e^{-i\frac{t}{l}\hat h_{j,j+1}}$ and $\hat U_{2n}=\prod_{j:{\rm even}}e^{-i\frac{t}{l}\hat h_{j,j+1}}$ with $n=1,2,...,l$ in Eq.~(\ref{HtU}); cf. Fig.~\ref{Fig:Circ}(a)), provided that each local term $\hat h_j$ has $\mathcal{O}(1)$ norm.

In fact, what we mentioned above is just the first-order version of a large class of product formulas,\cite{AMC19} which are simply products of $e^{a_sA}$ and $e^{b_sB}$ with appropriately chosen coefficients $a_s$'s and $b_s$'s. For example, it is known that $S_2(x)= e^{\frac{x}{2}A} e^{xB} e^{\frac{x}{2}A}$ approximates $e^{x(A+B)}$ up to $\mathcal{O}(x^3)$, if one does not care about the order of the commutators between $A$ and $B$. This result has been further generalized recursively via\cite{MS91}
\begin{equation}
S_{2k}(x) = S_{2k-2}(s_k x)^2 S_{2k-2}((1-4s_k)x) S_{2k-2}(s_k x)^2,\;\;\;\;s_k=\frac{1}{4-4^{\frac{1}{2k-1}}}, 
\end{equation}
which, roughly speaking, suppresses the error to $\mathcal{O}(x^{2k+1})$. 
Rather surprisingly, the explicit dependence on commutators, which is simply proportional to $\|[A,B]\|$ in the first-order case, has not been rigorously established until recently by Childs {\it et al.}\cite{AMC21} The exact statement, which applies to the more general case of approximating $e^{x(\sum^\Gamma_{\gamma=1}A_\gamma)}$, is the following: 
\begin{theorem}
Let $A=\sum^\Gamma_{\gamma=1} A_\gamma$ be the sum of $\Gamma$ anti-Hermitian matrices. Denoting $\alpha_{\rm comm}=\sum^\Gamma_{\gamma_1,\gamma_2,...,\gamma_{p+1}=1}\|[A_{\gamma_{p+1}},\cdots[A_{\gamma_2},A_{\gamma_1}]\cdots]\|$, then $\forall x\ge0$ the Trotter error of a $p$th-order product formula $S_p(x)$ is bounded by
\begin{equation}
\|e^{xA} - S_p(x)\|\le c \alpha_{\rm comm} x^{p+1},
\end{equation} 
where constant $c$ relies only on $\Gamma$, $p$ but not on $\{A_\gamma\}^\Gamma_{\gamma=1}$.
\label{Thm:TE}
\end{theorem}
As a corollary, we know that the error in Eq.~(\ref{HtU}) scales like $\mathcal{O}(\alpha_{\rm comm} t^{p+1}/l^p)$, where $l$ is the depth of the circuit.\footnote{This is true if any function of $p$ or $\Gamma$ is considered as $\mathcal{O}(1)$. In practice, however, this result might not be so meaningful since the depth of each product-formula module can scale exponentially in $p$ and thus lightly exceed $l$.} 

It should be emphasized that this theorem cannot be obtained by truncating the Baker-Campbell-Hausdorff formula,\cite{AMC21} despite the seeming consistency in the appearance of higher-order commutators. This is because for a fixed $x$, the truncation error is not always dominated by the lowest-order commutators.\cite{DW14} Although there remains the problem of what is the optimal $c$ in Thm~\ref{Thm:TE}, this result is powerful enough for recovering or even outperforming the state-of-the-art error scaling behaviors in various quantum systems.\cite{RB18,GHL19,MCT19} In the latter (outperforming) case, the typical situation is that the previous error bounds can be reproduced by applying the triangle inequality to $\alpha_{\rm comm}$. This may lead to qualitatively loose results if there are a lot of vanishing commutators due to locality or other specific properties of the system. For example, if we apply the error bound $\mathcal{O}((\|\hat H\|t)^{p+1}/l^p)$ derived by Berry {\it et al.}\cite{DWB07} to nearest-neighbor spin chains, we will obtain $\mathcal{O}(L^2t^2/l)$ for the first-order approximation, which has an additional $L$ factor. 


\subsubsection{Cutting exponential tails}
Despite the wide applicability of Trotterization, it may not be so efficient since the error is only suppressed algebraically by the circuit depth. This is especially the case for short-range interacting Hamiltonians, for which one naturally expects from the Lieb-Robinson bound (\ref{LRB}) an exponentially small error by cutting the tails outside the light cone. This intuition was made rigorous by Osborne for 1D finite-range Hamiltonians.\cite{TJO06} Upon coarse graining, it suffices to  consider nearest-neighbor spin chains mentioned above. The basic idea of circuitization is to pick up a local term $\hat h_{j,j+1}$ and express the time evolution in the interaction picture (cf. Fig.~\ref{Fig:Circ}(b)):
\begin{equation}\label{Eq:exp_truncate}
e^{-i\hat Ht}= e^{-i\hat H_jt}\overleftarrow{\rm T}e^{-i\int^t_0dt'\hat h^{(\rm I)}_j(t')},
\end{equation}
where $\hat H_j=\hat H - \hat h_{j,j+1}$ and $\hat h^{(\rm I)}_j(t)= e^{i\hat H_j t} \hat h_{j,j+1} e^{-i\hat H_j t}$. Denoting $\hat H_{j,r}=\hat H_j - \hat h_{j-r,j-r+1} - \hat h_{j+r,j+r+1}$, we know that the $\hat h^{(\rm I)}_{j,r}(t)= e^{i\hat H_{j,r} t} \hat h_{j,j+1} e^{-i\hat H_{j,r} t}$ is supported on $[j-r+1,j+r]$ and so is the unitary it generates. Moreover, such a unitary turns out to be a good approximation for that generated by $\hat h^{(\rm I)}_j(t)$, with the error being bounded by
\begin{equation}
\|\overleftarrow{\rm T}e^{-i\int^t_0dt'\hat h^{(\rm I)}_j(t')} - \overleftarrow{\rm T}e^{-i\int^t_0dt'\hat h^{(\rm I)}_{j,r}(t')} \| 
\le\int^t_0dt_1\int^{t_1}_0dt_2\| [\hat L_{j,r}(t_2),\hat h_{j,j+1}]\|,
\label{TLh}
\end{equation}
where $\hat L_{j,r}(t)=e^{-i\hat H_jt}(\hat h_{j-r,j-r+1} + \hat h_{j+r,j+r+1})e^{i\hat H_jt}$ is the time-dependent generator of $e^{-i\hat H_jt}e^{i\hat H_{j,r}t}$. Since $\hat H_j$ only involves nearest-neighbor interactions, one can directly apply the Lieb-Robinson bound (\ref{LRB}) to the commutator in Eq.~(\ref{TLh}), obtaining a bound with order $\mathcal{O}(e^{-\kappa(r-vt)})$. Iterating the above procedure to $\hat H_j$ in Eq.~\eqref{Eq:exp_truncate} and so on, we will end up with a bilayer quantum circuit with each gates acting on $2r$ sites and the error in Eq.~(\ref{HtU}) is of order $\mathcal{O}(Le^{-\kappa(r-vt)}/r)$, where the factor $L/r$ arises from Eq.~(\ref{VUn}). This result has been exploited to give an alternative proof to the entanglement-growth-rate area law (\ref{dSdtdV}) in 1D\cite{JE06} and the robustness against Hamiltonian evolution of the entanglement lower bound (\ref{Sind}) for 1D QCA.\cite{ZG21}

Finally, we mention that an error quantity with more experimental relevance may depend on an observable:\cite{AMA21}
\begin{equation}
\epsilon_{\rm o}\equiv\|e^{i\hat Ht}\hat O e^{-i\hat Ht} - \hat U_{\rm c}^\dag\hat O\hat U_{\rm c}\|.
\label{HOUO}
\end{equation}
This kind of bound will be the main concern in Sec.~\ref{Sec:CD}. One can check that $\epsilon_{\rm o}\le 2\epsilon\|O\|$, so any bound on Eq.~(\ref{HtU}) applies straightforwardly. However, such 
bound could be very loose if  $\hat O$ enjoys some specific properties. For example, in the above circuitization based on the Lieb-Robinson bound, the error would be $\mathcal{O}(e^{-\kappa(r-vt)})$ for local observables. Conversely, if one only attempts to reduce the error in Eq.~(\ref{HOUO}), then the required resource could be much less. Taking the same example, we know that it suffices to circuitize near the support of $\hat O$.

\subsection{Floquet prethermalization} 


As discussed in {Sec.~\ref{Sec:Eqtime}}, no  general method is known for evaluating  timescales of thermalization for a given system.
On the other hand, for a periodically driven system, known as the Floquet system (see Refs.~\refcite{MB16,AE17,TO19,AH22} for reviews), 
notable lower bounds on timescales for thermalization exist.
The Hamiltonian for the Floquet system is given by $\hat{H}(t)$ that satisfies $\hat{H}(t)=\hat{H}(t+T)$ for the period $T$.
For simplicity, let us focus on times at the ends of the $s$ periods, i.e., $t_s=sT$.
Solving the Scr\"{o}dinger equation from an initial state $\ket{\psi(0)}$, we have
\aln{
\ket{\psi(t_s)}= {\hat U}_F^s\ket{\psi(0)},
}
where 
\aln{
\hat{U}_F=\overleftarrow{\rm T}\exp\lrl{-i\int_0^T\hat{H}(t)dt}
}
is called the Floquet operator.
Then, introducing the Floquet Hamiltonian $\hat{H}_F=(i/T)\ln\hat{U}_F$,
we can formally write $\ket{\psi(t_s)}=e^{-i\hat{H}_F\cdot sT}\ket{\psi(0)}$.

The Floquet Hamiltonian is, in general, not easy to calculate  from $\hat{H}(t)$.
For high-frequency driving (i.e., $T$ is small), the following Floquet-Magnus expansion may be useful to extract the effective dynamics of the Floquet system:
\aln{\label{Eq:FMexpansion}
\hat{H}_F\simeq \hat{H}_F^{(n)}=\sum_{m=0}^nT^m{\hat\Omega}_m,
}
where $\hat{H}_F^{(n)}$ denotes the truncated Hamiltonian for some truncation order $n$.
Low orders of $\hat{\Omega}_m$ are given by, e.g.,
\aln{
{\hat\Omega}_0 &=\frac{1}{T}\int_0^Tdt\hat{H}(t),\nonumber\\
{\hat\Omega}_1 &=\frac{1}{2iT^2}\int_0^Tdt\int_0^td\tau[\hat{H}(t),\hat{H}(\tau)];
}
see Ref.~\refcite{TK16} {for} the expression for general $n$.
While the Floquet-Magnus expansion dramatically simplifies the original problem, the approximate Hamiltonian $\hat{H}_F^{(n)}$ may not always describe the original dynamics.
From the mathematical perspective, the expansion Eq.~\eqref{Eq:FMexpansion} does not converge for large $n$ in general many-body systems in the thermodynamic limit even for small $T$.
From the physical perspective, the external driving will eventually heat the system and result in the infinite-temperature equilibrium in the long-time limit, which cannot be obtained by the energy-conserving dynamics generated by $\hat{H}_F^{(n)}$.
This is proven if we assume the Floquet eigenstate thermalization hypothesis,\cite{AL14,HK14,LDA14} which states that all of the energy eigenstates of the original Floquet Hamiltonian $\hat{H}_F$ {behave like} 
an infinite-temperature state for local observables.
Note that the 
Floquet eigenstate thermalization hypothesis is confirmed in many nonintegrable many-body systems.

From the above observation, the following questions arise:
{\begin{enumerate}
\renewcommand{\labelenumi}{(\roman{enumi})}
\item
What is the rate of the heating? 
Is it small enough to prohibit complete thermalization for a sufficiently long time?
\item
How should we take $n$? 
Can we well approximate the dynamics with $\hat{H}_F^{(n)}$ of local observables even for many-body systems?
\end{enumerate}}
These two questions were investigated in Refs.~\refcite{TK16,DAA15,TM16,DA17PRB,DA17,WWH18}.
In Ref.~\refcite{TM16}, the authors answered the question (i) by showing that $\hat{H}_F^{(n)}$ for $n\leq n_c\propto T^{-1}$ becomes an almost conserved quantity for an exponentially long time with respect to $T^{-1}$ in general many-body systems with few-body interactions.
More specifically, let us assume a $k$-body interacting Hamiltonian ($k$ is independent of $N$) in the form 
$\hat{H}(t)=\sum_{X:|X|\leq k}\hat{h}_X(t)$, where $X$ denotes a set of lattice sites.
We also assume that local energy at each single site $i$ is bounded from above as $\sum_{X:i\in X}\|\hat{h}_X(t)\|\leq g$, where $g$ is independent of $N$.
Then, for any $n$ satisfying $n\leq n_c=\lfloor 1/(8kgT)-1\rfloor$, we have
\aln{\label{Eq:prethermal_energy}
\frac{|\braket{\psi(t_s)|\hat{H}_F^{(n)}|\psi(t_s)}-\braket{\psi(0)|\hat{H}_F^{(n)}|\psi(0)}|}{N}\leq 16g^2k2^{-n_c}t_s+\mathcal{O}(T^{n+1})
}
for $T\leq 1/(8kg)$.
This means that the energy density obtained from $\hat{H}_F^{(n)}$ is almost conserved for exponentially long times with respect to $T^{-1}$, i.e., $t\lesssim \mathcal{O}(2^{n_c})= \mathcal{O}(e^{C/(kgT)})$ for some constant $C$.\cite{AR20}
In other words, even when the Floquet eigenstate thermalization holds, complete thermalization does not occur during exponentially long times with respect to $T^{-1}$.
We stress that this is obtained for general systems with few-body interactions, including long-range interacting systems with power-law decay whose exponent is $\alpha$.\footnote{For $\alpha\leq d$, where $d$ is the spatial dimension, the Hamiltonian should be appropriately normalized so that {the local energy is bounded, i.e.,} $g$ becomes $N$-independent.}

Let us next consider the second question.
For this purpose, we need to assume additionally that the system is composed of short-range interactions.
Then, for $ n'_c=\lfloor 1/(16kgT)-1\rfloor$
and any local observable $\hat{O}_X$ whose support is $X$, we have\cite{TK16}
{
\aln{\label{Eq:prethermal_observable}
\begin{split}
&|\braket{\psi(t_s)|\hat{O}_X|\psi(t_s)}-\braket{\psi^{(n'_c)}(t_s)|\hat{O}_X|\psi^{(n'_c)}(t_s)}| \\
\leq& \lrs{12g\cdot 2^{-n_c'/2}+\frac{2c}{T}e^{-\kappa(l_0-vt)}}\|\hat{O}_X\|\cdot|X|t
\end{split}
}}
for $T\leq 1/(16kg)$.
Here, 
\aln{
\ket{\psi^{(n'_c)}(t)}=e^{-i\hat{H}_F^{(n'_c)}t}\ket{\psi(0)}
}
denotes the quantum state obtained from the truncated Hamiltonian,
$l_0\propto 2^{n_c'/(2d)}\sim e^{-1/T}$,
and $c,\kappa$, and $v$ are constants appearing in the Lieb-Robinson bound in {Eq.~\eqref{LRB}}.
This inequality means that 
the approximate Hamiltonian 
$\hat{H}_F^{(n'_c)}$ can almost faithfully describe the exact dynamics for 
exponentially long times with respect to $T^{-1}$, i.e., $t\lesssim \mathcal{O}(e^{C'/(kgT)})$ for some constant $C'$.
Since this inequality is obtained using the Lieb-Robinson bound, it cannot be {applied to}
, e.g., long-range interacting systems with power-law decay whose exponent satisfies $\alpha\leq 2d$ (the inequality still holds for $\alpha >2d$).\cite{FM20}

The above results of exponentially long-time stability of the truncated Hamiltonian have direct relevance with the engineering of quantum matter by periodic drivings.
For example, the rigorous inequalities ensure that the effective Hamiltonians obtained from  periodic driving are useful for realizing systems that are difficult to obtain otherwise as long as the driving frequency is sufficiently large.
Furthermore, during the prethermalization {regime} 
in the periodic driving, the system may exhibit a novel dynamical phase such as discrete time crystals.\cite{DVE17}

Interestingly, the above conclusions are modified in periodically driven \textit{classical} systems.
Reference~\refcite{TM18} considered locally interacting spin-$S$ systems and took $S\ra\infty$, which corresponds to the classical limit of the spin systems.
By decomposing each spin-$S$ spin with $2S$ numbers of spin-$1/2$ spins, the author showed that the exponentially long-time prethermalization holds even for those classical systems from an inequality similar to \eqref{Eq:prethermal_energy}.
On the other hand, Ref.~\refcite{TM18} also pointed out that inequality similar to \eqref{Eq:prethermal_observable} no longer applies.
This is formally due to the emergent long-range interactions accompanied by the decomposition of the original spins into spin-$1/2$ ones.
Intuitively, the exponential sensitivity of dynamical trajectories in classical chaos leads to the breakdown of the approximate dynamics in \eqref{Eq:prethermal_observable}.
We note, however, that the truncated Hamiltonian may approximate original dynamics as an ensemble of trajectories, instead of each trajectory;
for example, we can take an initial state as a distribution that spreads on the  classical  phase space, not a single point.
This possibility was investigated in Refs.~\refcite{AP21,AP21PRB,BY21}, where novel classical prethermal phases of matter were discovered.

\rhcomment{
We have discussed prethermalization caused by the high-frequency driving above.
On the other hand, it is recently discovered that driving with high amplitude (not necessarily high-frequency) can also lead to non-thermal behavior of the system.\cite{AH18,AH21}
While the generality of this phenomenon was discussed in terms of the emergent conservation law, rigorous bounds on non-thermal behavior as in ~\eqref{Eq:prethermal_energy} and~\eqref{Eq:prethermal_observable}  are not known.
It would be interesting to uncover such constraints for high-amplitude driving since this type of driving may open another route to stabilize non-equilibrium phase of matter.
}

\subsection{Constrained dynamics} 
\label{Sec:CD}
It is widely believed on the basis of Fermi's golden rule that energy mismatch in isolated quantum systems prohibits coherent transitions. A well-known example of this statement is the Rydberg blockade, which forbids multiple Rydberg excitations due to the large energy detuning arising from strong dipole-dipole interactions. 
The prototypical model of quantum many-body scars --- the PXP model,\textcolor{black}{\cite{CJT18}} motivated by a recent Rydberg-atom-array experiment,\textcolor{black}{\cite{HB17}}  explicitly exploits  such an approximation. As indicated by the name, the Hamiltonian simply takes the form of $\hat H=\hat P\hat X\hat P$, where $\hat X$ is the (global) Rabi coupling between the ground states and Rydberg states, and $\hat P$ is the projector onto the constrained many-body Hilbert space without adjacent Rydberg excitations. In the following, we generally formalize the problem based on such kinds of projection approximations and review the related error bounds.

\subsubsection{Setup and main result}
Let us consider a Hamiltonian $\hat H=\hat H_0 + \hat V$, where $\hat H_0$ admits an energy band $\mathcal{H}_0$ with corresponding spectrum $\ell_0$ separated from the remaining by gap $\Delta_0$, and $\hat V$ is more like a perturbation with small $\|\hat V\|$ (or $\mu$ in Eq.~(\ref{mutp}) in a many-body setting) compared to $\Delta_0$. Denoting the projector onto $\mathcal{H}_0$ as $\hat P_0$, one may naturally expect that the constrained dynamics $e^{-i\hat P_0\hat H\hat P_0t}$ within $\mathcal{H}_0$ well approximates the actual dynamics $e^{-i\hat Ht}$, provided that the initial state lies in $\mathcal{H}_0$ and $\Delta_0$ is large enough. To quantity the precision of such a constrained-dynamics approximation, Gong {\it et al.} studied the following observable-based error similar to Eq.~(\ref{HOUO}):\cite{ZG20,ZG20b} 
\begin{equation}
\epsilon=\|\hat P_0(e^{i\hat Ht} \hat Oe^{-i\hat Ht} - e^{i\hat P_0\hat H\hat P_0t} \hat Oe^{-i\hat P_0\hat H\hat P_0t})\hat P_0\|. 
\label{POtP}
\end{equation}
Here the additional projectors outside the operator difference arise from the further assumption that the initial state lies within $\mathcal{H}_0$. As expected, there is indeed a rigorous bound which is suppressed by $\Delta_0$ and vanishes in the infinite gap limit $\Delta_0\to\infty$:\cite{ZG20b}
\begin{theorem}
\emph{(Universal error bound for constrained dynamics)} For the setup stated above and with $\Delta_0>2\|\hat V\|$, the error in Eq.~(\ref{POtP}) for any normalized observable with $\|\hat O\|=1$ is always upper bounded by
\begin{equation}
\epsilon\le\frac{4\|\hat V\|}{\Delta_0-2\|\hat V\|} + 2f\left(\frac{2\|\hat V\|}{\Delta_0-2\|\hat V\|}\right)\|\hat V\|t,\;\;\;\;
f(x)=\frac{(x-1)e^x+1}{x}.
\label{ebcd}
\end{equation}
\label{Thm:EB}
\end{theorem}
Noting that $f(x)\simeq x/2$ for small $x$, we have the following simple asymptotic bound for large gaps:
\begin{equation}
\epsilon\lesssim \frac{4\|\hat V\|}{\Delta_0} +  \frac{2\|\hat V\|^2}{\Delta_0} t.
\label{aseb}
\end{equation}
The intercept $4\|\hat V\|/\Delta_0$ and slope $2\|\hat V\|^2/\Delta_0$ in the above turn out to be tight and can be saturated (separately) in very simple models.\cite{ZG20} It is also worth mentioning that even without assuming $\Delta_0> 2\|\hat V\|$, we still have a time-linear bound like Eq.~(\ref{ebcd}) with the slope replaced by $2(e^{2\|\hat V\|/\Delta_0}-1)\|\hat V\|$.\footnote{This bound leads to a similar asymptotic bound like Eq.~(\ref{aseb}) but with the slope doubled. On the other hand, this bound is tighter than Eq.~(\ref{ebcd}) if $\|V\|>y^*\Delta_0$, where $y^*=0.1887...$ is the solution to $e^{2y}=f(2y/(1-2y))$.} 

Without going into the details of the derivation, which will be briefly mentioned in the next subsection, let us give some intuitions into the form of the asymptotic bound (\ref{aseb}). The order of the intercept can be understood from the standard perturbation theory.\cite{JJS11} Note that any initial state $|\psi_0\rangle\in\mathcal{H}_0$ can be expanded in the energy basis $\{|\varphi_n\rangle\}_n$ (of $\hat H$) by a dominant component within the perturbed energy band and those outside. In the simplest case with $\dim\mathcal{H}_0=1$, the coefficients of these components take the form $\langle\varphi_n |\hat V|\psi_0\rangle/(E_n - E_0)$ ($\hat H|\varphi_n\rangle=E_n|\varphi_n\rangle$ and $|\varphi_0\rangle$ is the eigenvector adiabatically connected to $|\psi_0\rangle$) and are thus of order $\mathcal{O}(\|\hat V\|/\Delta_0)$. These coefficients will rapidly oscillate with frequency of 
$\mathcal{O}(\Delta_0)$, leading to an initial rapid growth of error that constitutes the intercept. As for the slope, the order can be understood from the effective Hamiltonian determined by Green’s function.\cite{EB07} In the frequency domain, the discrepancy between the actual and the constrained dynamics arises clearly from the self-energy $\hat\Sigma(\omega)=\hat P_0\hat V\hat Q_0 (\omega - \hat Q_0\hat H\hat Q_0)^{-1}\hat Q_0\hat V\hat P_0$ ($\hat Q_0=1-\hat P_0$ is the complement projector), which accounts for the transient leave from and reentrance into the energy band and is of order $\mathcal{O}(\|\hat V\|^2/\Delta_0)$ for $\omega\in\ell_0$. On the other hand, we would like to emphasize that the explicit factors (i.e., $4$ and $2$) in Eq.~(\ref{aseb}) are far from trivial --- even their finiteness cannot be deduced from the above argument.

\subsubsection{Derivation and generalization}
Despite the universality of Thm.~\ref{Thm:EB}, the bound becomes meaningless for many-body systems, where $\hat V$ is typically a sum of local terms and thus $\|\hat V\|$ diverges in the thermodynamic limit. This result is somehow reasonable since for highly local observables it is likely that the error immediately blows up. On the other hand, for short-range Hamiltonians and local observables, the numbers of relevant terms in $\hat V$ should grow polynomially according to the light-cone picture. It is thus natural to expect that, even in the thermodynamic limit, the error (\ref{POtP}) is still suppressed by $\Delta_0$ and grows no faster than polynomially in time.

To make the above argument explicit, we have to dive a bit into the derivation of Thm.~\ref{Thm:EB} (precisely speaking, the asymptotically looser version without requiring $\Delta_0>2\|\hat V\|$), which relies heavily on the Schrieffer-Wolff transformation (SWT).\cite{JRS66,SB11} The latter refers to a series of unitary transformations that gradually block-diagonalize $\hat H=\hat H_0 + \hat V$ with respect to $\mathcal{H}_0$ and the complement. In its lowest-order version,  the transformation is given by $\hat S=e^{\hat T}$ with the anti-Hermitian generator $\hat T$ determined by the  $[\hat H_0,\hat T]=\hat P_0\hat V\hat Q_0 + \hat Q_0\hat V\hat P_0$ and $\hat T= \hat P_0\hat T\hat Q_0 + \hat Q_0\hat T\hat P_0$. Each off-block-diagonal component in $\hat T$ thus follows a Sylvester equation (i.e., a matrix equation of $X$ in the form $AX+XB=C$), which in turn implies $\|\hat T\|\le \|\hat V\|/\Delta_0$.\cite{RB97} Moreover, the transformed Hamiltonian reads $\hat H'_1\equiv\hat S\hat H\hat S^\dag = \hat H_1 + \hat V'$ ($\hat H_1=\hat P_0\hat H\hat P_0 + \hat Q_0\hat H\hat Q_0$), where the norm of $\hat V'$ can be asymptotically upper bounded by $2\|\hat V\|^2/\Delta_0$. Using the SWT, we can rewrite the error (\ref{POtP}) into 
\begin{equation}
\epsilon=\|\hat P_0[\hat S_{H_1}(t)^\dag \hat L(t)\hat S \hat O\hat S^\dag\hat L(t)^\dag \hat S_{H_1}(t) -\hat O]\hat P_0\|,
\label{ecd2}
\end{equation}
where $\hat S_{H_1}(t) = e^{\hat T_{H_1}(t)}= e^{e^{-i\hat H_1t}\hat Te^{i\hat H_1t}}$ is the SWT in the rotating frame with respect to $\hat H_1$ and $\hat L(t) =e^{-i\hat H_1 t}e^{i\hat H'_1t}$ is a Loschmidt-echo operator.\cite{TG06} Applying the general relation $\|[\hat U_1\hat U_2...\hat U_n,\hat O]\|\le\sum^n_{m=1}\|[\hat U_m,\hat O]\|$ to Eq.~(\ref{ecd2}), we obtain
\begin{equation}
\begin{split}
\epsilon&\le \|[\hat S,\hat O]\| + \|[\hat L(t),\hat O]\| + \|[\hat S_{H_1}(t),\hat O]\| \\
&\le \|[\hat T,\hat O]\| + \int^t_0 dt'\|[\hat V'_{H_1}(t'),\hat O]\| + \|[\hat T_{H_1}(t),\hat O]\|.
\end{split}
\label{eb3}
\end{equation}
This decomposition implies heuristically that the error production consists \rhcomment{of} three steps --- initial SWT, middle Loschmidt echo, and final SWT in the rotating frame. Further using the simple inequality $\|[A,B]\|\le2\|A\|\|B\|$, we can upper bound the first and third terms by $2\|\hat{T}\|\|\hat O\|\le 2\|\hat V\|/\Delta_0$ (recalling that $\|\hat O\|=1$), and the second by $2\|\hat V_1\|\|\hat O\|t\lesssim 4\|\hat V\|^2t/\Delta_0$, leading to the desired result.


\begin{figure}[bt]
\centerline{\psfig{file=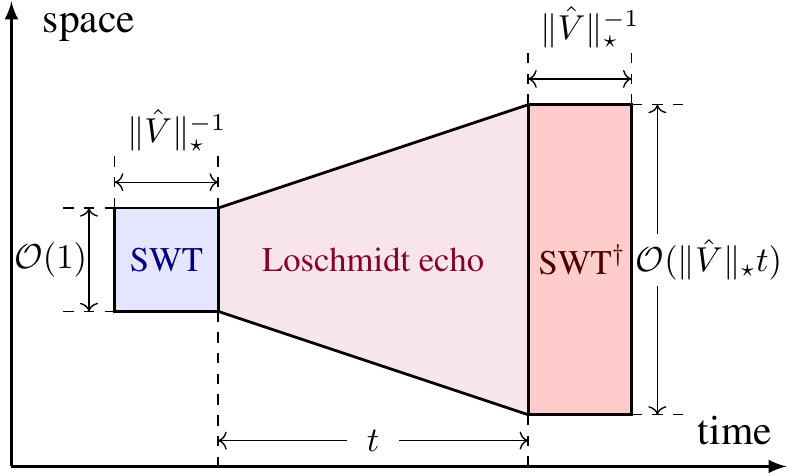,width=3in}}
\vspace*{8pt}
\caption{Schematic illustration of the error production in constrained many-body systems from three steps --- initial SWT, Loschmidt echo, and final SWT. The total error can be estimated to be the total spacetime volume in color multiplied by $\|\hat V\|^2_\star/\Delta_0$. Note that the space arrow only shows one of $d$ directions, so the total spacetime volume is of  $\mathcal{O}(\|\hat V\|^d_\star t^{d+1})$.} 
\label{Fig:MBerror}
\end{figure}

When it comes to many-body systems and local observables, we can still upper bound the error by Eq.~(\ref{eb3}), but should proceed more carefully by making full use of the locality. At least for a frustration-free, commuting and finite-range $\hat H_0$ with (local) gap $\Delta_0$ and finite-range $\hat V$, we can perform a many-body SWT such that $\hat T$ is a sum of local terms (i.e., strictly finite-ranged) and $\hat V'$ is short-ranged (i.e., with exponentially decaying tails),\cite{SB11,ND96} allowing a well-defined local interaction strength as given in Eq.~(\ref{mutp}). Moreover, denoting the $\kappa=0$ version as $\|\hat H\|_\star\equiv \max_{x\in\Lambda}\sum_{A\ni x}\|\hat h_A\|$, we can upper bound $\|\hat T\|_\star$ and $\|\hat V'\|_\star$ by some $\mathcal{O}(\|\hat V\|_\star/\Delta_0)$ and  $\mathcal{O}(\|\hat V\|^2_\star/\Delta_0)$ quantities, respectively. These results are well-analogous to the previous ones and the only difference is that all the interaction strengths are defined locally. For a normalized local observable $\hat O_X$ supported on $X$ with $|X|\sim\mathcal{O}(1)$, we can bound the first term in Eq.~(\ref{eb3}) by
\begin{equation}
\|[\hat T,\hat O_X]\|\le \sum_{x\in X}\sum_{A\ni x}\|[\hat T_A,\hat O_X]\|\le 2|X|\|\hat T\|_\star\lesssim \mathcal{O}\left(\frac{\|\hat V\|_\star}{\Delta_0}\right).
\end{equation}
Regarding the second term in Eq.~(\ref{eb3}), we can recast the time-dependence into $\hat O_X$ and exploit the fact that the volume of its effective support grows no faster than $t^d$ in $d$ dimensions as limited by the Lieb-Robinson bound, obtaining
\begin{equation}
 \int^t_0 dt'\|[\hat V'_{H_1}(t'),\hat O_X]\|= \int^t_0 dt'\|[\hat V',\hat O^{H_1}_X(t')]\|\lesssim\mathcal{O}\left(\frac{\|\hat V\|_\star}{\Delta_0}(\|\hat V\|_\star t)^{d+1}\right).
\end{equation}
Here $\hat O^{H_1}_X(t)= e^{i\hat H_1 t}\hat O_Xe^{-i\hat H_1t}$ and the Lieb-Robinson velocity of $\hat H_1$ has been estimated as $\mathcal{O}(\|\hat V\|_\star)$ (cf. Thm.~\ref{Thm:tLRB} in Sec.~\ref{Sec:GF}) since $\hat H_0$ is commuting and thus almost does not contribute to operator spreading. Similarly, we can upper bound the third term in Eq.~(\ref{eb3}) by
\begin{equation}
\|[\hat T_{H_1}(t),\hat O_X]\|=\|[\hat T,\hat O^{H_1}_X(t)]\|\lesssim\mathcal{O}\left(\frac{\|\hat V\|_\star}{\Delta_0}(\|\hat V\|_\star t)^d\right).
\end{equation}
Overall, we know that there exists a polynomial $p(x)$ with $\mathcal{O}(1)$ coefficients and degree $d+1$ such that
\begin{equation}
\epsilon \lesssim \frac{\|\hat V\|_\star}{\Delta_0} p(\|\hat V\|_\star t).
\label{mbeb}
\end{equation}
The above error-estimation procedure is schematically illustrated in Fig.~\ref{Fig:MBerror}. Note that Eq.~(\ref{aseb}) is included in the above result (\ref{mbeb}) as a special case in $d=0$D. 

It is worthwhile to mention that the $t^{d+1}$ scaling, which is closely related to the light-cone picture, may emerge in other situations concerning short-range interacting many-body systems such as the adiabatic theorem\cite{SB17} and Floquet prethermalization.\cite{DA17} In fact, our setup shares some similarity with the latter if we go into the rotating frame with respect to $\hat H_0$. Indeed, as mentioned in Ref.~\refcite{DA17} if $e^{-i\hat H_0 T}=1$ for some $T\in\mathbb{R}^+$, then the Floquet analysis applies to such time-independent systems. On the other hand, our result does not rely on this condition. Finally, we mention that both error bounds (ours and that in Ref.~\refcite{DA17}) have found applications to analyze the reliability of the quantum simulations of lattice gauge theory,\cite{JCH20,JCH21,JCH21b,JCH22} where some gauge-protection terms with large energy gaps are used to suppress gauge-symmetry-breaking errors.


\section{Miscellaneous topics} 
\label{Sec:MT}
No need to say, there are still many interesting bounds concerning nonequilibrium quantum dynamics that cannot be well categorized into the previous sections. Here we pick up a few examples that are of particular recent interests, including the no-go theorem for quantum time crystals, bound on chaos, and bounds related to quantum (as well as classical stochastic) thermodynamics. 

\subsection{Absence of quantum time crystals} 
Rigorous bounds on time-evolving systems also provide a no-go theorem concerning the existence of a dynamical phase of matter.
One of the most important examples is the absence of continuous-time crystals for an equilibrium state.
A time crystal was  originally proposed by Wilczek\cite{FW12} as a dynamical many-body phase that spontaneously breaks continuous time-translation symmetry, which is analogous to the standard crystal with spontaneously broken continuous space-translation symmetry.
Although the idea of spontaneous breaking of the continuous time-translation symmetry seems appealing, Watanabe and Oshikawa\cite{HW15} rigorously showed that this cannot happen for the ground state in an isolated system.
Their strategy is to characterize the time crystal with the presence of the long-range order in temporal direction.
The long-range order for the ordinary crystal is defined by $\lim_{V\ra\infty}\braket{\hat{\phi}(x)\hat{\phi}(x')}_0={f}
(x-x')$ for sufficiently large {$|x-x'|$}, where $V$ is the size of the system, $\hat{\phi}(x)$ denotes the local order parameter at {position} $x$, $\braket{\cdots}_0{\equiv}
\braket{0|\cdots|0}$ denotes the expectation value with respect to the ground state, and {$f$} 
is some nontrivial periodic function.
Likewise, the long-range order for the time crystal can be defined by 
$\lim_{V\ra\infty}\braket{\hat{\phi}(x,t)\hat{\phi}(x',0)}_0={g}
(t)$ for sufficiently large $t$ and with another periodic function ${g}
(t)$.\footnote{Two remarks are in order. First, we can consider the periodicity of {$g$} 
along the space direction, which defines the so-called space-time crystal. The absence of the space-time crystal can similarly be shown. Second, when we consider an equal-space correlation function $\braket{\hat{\phi}(x,t)\hat{\phi}(x,0)}_0$, it can exhibit time-periodic oscillation in, e.g., a set of noninteracting oscillators. However, this example is trivial and should be ruled out from the definition of time-crystalline order that would coexist with many-body interactions.}
 Defining a global order parameter $\Phi(t)=\int d^dx\hat{\phi}(x,t)$, we can rewrite this condition  as
 \aln{\label{Eq:WatanabeOshikawa}
 \lim_{V\ra\infty}\frac{\braket{\hat{\Phi}(t)\hat{\Phi}(0)}_0}{V^2}=g(t).
 }
 Now, following Ref.~\refcite{HW15},  we show that the right-hand side in Eq.~\eqref{Eq:WatanabeOshikawa} should be time-independent for short-ranged interacting systems, indicating the absence of time crystals.
 To see this, we first note that
{\aln{
 \begin{split}
 &|\braket{\hat{\Phi}(t)\hat{\Phi}(0)}_0-\braket{\hat{\Phi}(0)\hat{\Phi}(0)}_0| \\ 
 =& |\braket{\hat{\Phi}(0)e^{-i(\hat{H}-E_0)t}\hat{\Phi}(0)}_0-\braket{\hat{\Phi}(0)\hat{\Phi}(0)}_0|\\
 \leq& \int_0^tds|\braket{\hat{\Phi}(0)(\hat{H}-E_0)e^{-i(\hat{H}-E_0)t}\hat{\Phi}(0)}_0|\\
  =& \int_0^tds|\braket{\hat{\Phi}(0)(\hat{H}-E_0)^{1/2}e^{-i(\hat{H}-E_0)t}(\hat{H}-E_0)^{1/2}\hat{\Phi}(0)}_0|\\
  \leq& t |\braket{\hat{\Phi}(0)(\hat{H}-E_0)\hat{\Phi}(0)}_0|=\frac{t}{2}|\braket{[\hat{\Phi}(0),[\hat{H},\hat{\Phi}(0)]]}_0|\\
  \leq&\frac{t}{2}||[\hat{\Phi}(0),[\hat{H},\hat{\Phi}(0)]]||.
  \end{split}
}}
For short-ranged interacting systems,  
 $||[\hat{\Phi}(0),[\hat{H},\hat{\Phi}(0)]]||$
 is at most $\mathcal{O}(V^1)$ and thus
$ \lim_{V\ra\infty}\frac{\braket{\hat{\Phi}(t)\hat{\Phi}(0)}_0}{V^2}= \lim_{V\ra\infty}\frac{\braket{\hat{\Phi}(0)\hat{\Phi}(0)}_0}{V^2}$ for  $t=\mathcal{O}(V^0)$, which is independent of time.
Note that the discussion is easily generalized to systems with power-law decaying interactions.

Absence of time crystals has been shown also for finite-temperature Gibbs states ($\braket{\cdots}_0$ is replaced with $\braket{\cdots}_\beta=\Tr[e^{-\beta\hat{H}}\cdots]/Z$).
While the original derivation\cite{HW15} contains some technical problems,\cite{VK19} the complete derivation was carried out by Ref.~\refcite{HW20} using the Lieb-Robinson bound.
In addition, Ref.~\refcite{YH19TC} showed that the time-crystalline long-range order is absent for general states with correlation decay.

Although the no-go theorem discussed above prohibits the spontaneous symmetry breaking of continuous-time-translation symmetry for closed equilibrium, 
several ways are proposed to realize time crystals by evading this theorem (see, e.g., Refs.~\refcite{VK19,KS17,DVE20} for comprehensive reviews).
First, time crystals can exist out of equilibrium:
even when we drive the system with a period $T$, the system can respond with a different period  $nT\:(n=2,3,\cdots)$, which can be regarded as a spontaneous symmetry breaking of \textit{discrete}-time-translation symmetry.
Such ``discrete (Floquet) time crystals" have attracted great attention, both theoretically\cite{DVE16,NYY17} and 
experimentally.\cite{JZ17,SC17t,JR18,HK21,AK21}
Note that discrete time crystals can exist for infinitely long times only {(in the thermodynamic limit and)} in the presence of certain mechanisms to prohibit heating from external driving, such as many-body localization\cite{DVE16,NYY17} or dissipation\cite{ZG18,BZ19}; otherwise, the time crystals appear at {the} prethermal regime.\cite{DVE17}
Second, even when we consider time-independent systems without external driving, time-crystalline behavior can appear under dissipation.{\cite{FI18,FMG19,BB19}}
Third, if we allow non-local many-body interactions for the Hamiltonian, the no-go theorem no longer applies, and the long-range order can emerge even for isolated equilibrium.\cite{VKK19}

\subsection{Bound on chaos} 
\subsubsection{Out-of-time-order correlator}
\label{Sec:OTOC}
As discussed in the previous section, given two operators $\hat{A}$ and $\hat{B}$, their different-time commutator $i[\hat{A}(t),\hat{B}]$ can characterize how information of perturbation $\hat{A}$ propagates and affects $\hat{B}$ through the time evolution.
For a given state $\hat{\rho}$, such information propagation is measured by the following correlator\cite{AIL69,AK14,AK15,SHS15}:
\aln{
C(t)=\braket{|[\hat{A}(t),\hat{B}]|^2}=D(t)+I(t)-2\mr{Re}[F(t)],
\label{otoc}
}
where  $\braket{\cdots}=\Tr[\hat{\rho}\cdots]$, $D(t)=\braket{\hat{B}^\dag\hat{A}(t)^\dag\hat{A}(t)\hat{B}}$, $I(t)=\braket{\hat{A}(t)^\dag\hat{B}^\dag\hat{B}\hat{A}(t)}$, and $F(t)=\braket{\hat{A}(t)^\dag\hat{B}^\dag\hat{A}(t)\hat{B}}$.
These commutators can be distinct from the conventional ones in that we need a time-reversal protocol to measure them.
For example, $F(t)$ is evaluated at time 0, $t$, 0, and again $t$.
Thus, this correlator (as well as $C(t)$) is called the out-of-time-ordered correlator (OTOC).
We note that, while $D(t)$ is the standard time-ordered correlator, $I(t)$ is, in general, the OTOC, but becomes time-ordered when $\hat{\rho}$ and $\hat{H}$ commute.
There are universal relations between $C, D, I$, and $F$, such as
$|\sqrt{C}-\sqrt{D}|\leq \sqrt{I}$, $|\sqrt{D}-\sqrt{I}|\leq \sqrt{C}$, $|\sqrt{I}-\sqrt{C}|\leq \sqrt{D}$, and
$|F|\leq \sqrt{ID}$.\cite{RH18}

The OTOC is related to not only information propagation but also quantum chaos.
To see this, let us consider a quantum system that has a well-defined semiclassical limit on the phase space parametrized by two canonical variables, i.e., position ${q}$ and momentum ${p}$.
By taking $\hat{A}=\hat{q}$ and $\hat{B}=\hat{p}$, which satisfy $[\hat{q},\hat{p}]=i\hbar$ (where we explicitly write $\hbar$), we can evaluate $C(t)$ in the semiclassical limit as (using the truncated Wigner approximation, which is reviewed in Ref.~\refcite{AP10})
\aln{
C(t)\ra \hbar^2\braketL{\lrm{q_t,p}{^2}_\mr{Poisson}}=\hbar^2\braketL{\lrs{\fracpd{q_t}{q}}^2},
}
where $\lrm{A,B}_\mr{Poisson}=\fracpd{A}{q}\fracpd{B}{p}-\fracpd{A}{p}\fracpd{B}{q}$ is the Poisson braket and $\braket{\cdots}$ denotes the average over the Wigner function for $\hat{\rho}$.
Here, $(q_t, p_t)$ is a phase-space point at time $t$, which is obtained by solving  Hamilton’s classical equations from the initial point $(q, p)$. 
For the case where Hamilton’s equation generates a chaotic dynamics, we expect that the initial deviation of the phase-space point grows exponentially.
This leads to $\fracpd{q_t}{q}\sim e^{\lambda_Lt}$ with a positive Lyapunov exponent $\lambda_L>0$ after certain times.\cite{AIL69}
We thus conclude that $C(t)$ exponentially grows with the chaotic Lyapunov exponent  and some positive constant $a$ as
\aln{\label{Eq:semiclassical_OTOC}
C(t)\sim a\hbar^2 e^{2\lambda_Lt}
}
until the so-called Ehrenfest time $t_*$,  at which the semiclassical approximation breaks down.
The Ehrenfest time is estimated from the timescale where the Wigner function starts to interfere with itself, $t_*\sim -\lambda_L^{-1}\log(S_0/\hbar)$\cite{AIL69} ($S_0$ is the characteristic action of the system). The exponential increase of $C(t)$ has been found in many quantum systems with semiclassical limits.\cite{RH18,EBR17,JK18,JSC18,RAJ18,TS19}
Note, however, that the exponential increase may not always be attributed to classical chaos.\cite{EBR17,TX20,KH20,WK21}
We also note that the exponential growth of $C(t)$ comes from the exponential behavior of either $F(t)$ or $I(t)$, which depends on the initial state.\cite{RH18}

\subsubsection{Maldacena-Shenker-Stanford bound}
\label{Sec:MSSB}
In the last decade, 
it turns out that the OTOC can exhibit exponential growth even for many-body systems without a well-defined semiclassical limit, i.e., 
\aln{\label{Eq:quantum_OTOC}
C(t)\sim a\epsilon e^{\lambda t}
}
with $\lambda >0$
 for certain timescales $t\lesssim t_*$, where $t_*$ is called the scrambling time.\cite{JM16,SHS14}
Because of the analogy with Eq.~\eqref{Eq:semiclassical_OTOC}, systems satisfying Eq.~\eqref{Eq:quantum_OTOC} are called quantum many-body chaotic systems.\footnote{Note that there are many different definitions for quantum many-body chaos. For example, in Ref.~\refcite{PH16}, a system is said to show quantum chaos when $F(t)$ decays to zero without any assumption on exponential behavior. In this case, the behavior in Eq.~\eqref{Eq:quantum_OTOC} is called \textit{strong} chaos.}
Moreover, Maldacena, Shanker, and Stanford\cite{JM16} conjectured that there exists a universal upper bound on the exponent $\lambda$ for a thermal state with temperature $\beta^{-1}$.
Technically, they introduced the following regularized OTOC (assuming that $\hat{A}$ and $\hat{B}$ are Hermitian operators)
\aln{\label{Eq:reg_OTOC}
 C_\mr{reg}(t)=-\Tr[\sqrt{\rho_\beta}[\hat{A}(t),\hat{B}]\sqrt{\rho_\beta}[\hat{A}(t),\hat{B}]],
}
with $\hat{\rho}_\beta=e^{-\beta\hat{H}}/Z$, which is assumed to grow exponentially as in Eq.~\eqref{Eq:semiclassical_OTOC}.
Then, using techniques from complex analysis, they showed
\aln{\label{Eq:MSS}
\lambda \leq  \frac{2\pi}{\beta\hbar}
}
under some physical assumptions (e.g., rapid factorizaton of $D(t)$ into $\braket{\hat{A}^2}_\beta\braket{\hat{B}^2}_\beta$).
Note that a different derivation of \eqref{Eq:MSS} is performed by Ref.~\refcite{NT18}, which introduced a one-parameter family of the regularized OTOC (which includes Eq.~\eqref{Eq:reg_OTOC} as a special case) and showed that, if all of those regularized OTOCs exhibit exponential growth, the rate of the growth is independent of the regularization and bounded by $ \frac{2\pi}{\beta\hbar}$.

The inequality~\eqref{Eq:MSS}, called a bound on chaos {or the Maldacena-Shenker-Stanford bound}, {sets} 
a fundamental {limit} 
on the rate for information scrambling in quantum systems.
Notably, the upper bound of \eqref{Eq:MSS} is achieved for, e.g., the strong-coupling and low-temperature limit of the Sachdev-Ye-Kitaev model,\cite{AK15,SS93,JM16b,JP16,JSC17,YG17}
which describes $N$ Majorana fermions with random four-body interactions.
Note that Sachdev-Ye-Kitaev model has recently  been studied extensively as a solvable model for strange metals,\cite{SS15} where no quasi-particles exist and the resistivity is proportional to the temperature unlike ordinary Fermi liquids {(which follow a square law)}.


\subsubsection{Absence of fast scrambling due to locality}
The dynamical behaviors of OTOC have also been actively studied in condensed matter systems, which typically exhibit certain kind of locality. For short-range interacting systems, to which the conventional Lieb-Robinson bound (\ref{LRB}) applies, one can readily upper bound the OTOC (\ref{otoc}) for two local operators by
\begin{equation}
C(t)\le c^2 \min\{e^{-2\kappa (r-vt)},1\},
\label{LROTOC}
\end{equation}
where $c=2\|\hat A\|\|\hat B\|\min\{|{\rm supp}\textcolor{black}{[\hat A]}|,|{\rm supp}\textcolor{black}{[\hat B]}|\}$ (${\rm supp}\textcolor{black}{[\hat A]}$: support of $\hat A$), $r={\rm dist}({\rm supp}\textcolor{black}{[\hat A]}, {\rm supp}\textcolor{black}{[\hat B]})$ and the H\"older inequality $\Tr[\hat\rho\hat O]\le \|\hat\rho\|_1\|\hat O\|$ has been used. This result indicates a linearly long (in terms of system size) time for quantum information to spread over the whole system, thus ruling out the possibility of fast scrambling with a \emph{logarithmically} long scrambling time, just like the Sachdev-Ye-Kitaev model. It is worth mentioning that even if $\hat A$ and $\hat B$ are sums of local operators such that the density of OTOC can grow unboundedly in the thermodynamic limit, the growth is no faster than polynomially in time, and the degree is at most $3d$ in $d$ dimensions.\cite{IK17} This result can be shown by assuming the clustering condition for $\hat \rho$ and using the Lieb-Robinson bound (\ref{LRB}).

A natural question then arises: would fast scrambling be possible for long-range interacting systems? Having in mind the optimality of logarithmic light cones for $\alpha\in (d,2d)$, one may conjecture that fast scrambling occurs for $\alpha <2d$. Surprisingly, this turns out to be \emph{not} the case --- it is very recently shown by Kuwahara and Saito that fast scrambling is forbidden for any thermodynamically stable systems with $\alpha > d$.\cite{TK21} Here the crucial point is that the one may directly bound the OTOC based on the Frobenius norm instead of applying the H\"older inequality to relate it to the operator norm. Such a Frobenius-version of Lieb-Robinson bound reads
\begin{equation}
\|\hat A(t) - \hat A^{[r]}(t)\|_F \le c \frac{t^{\alpha - \frac{d-1}{2}}(r+r_0)^{\frac{d-1}{2}}}{r^{\alpha -\frac{d+1}{2}}},
\end{equation}
where $r_0={\rm diam}({\rm supp}\textcolor{black}{[\hat A]})$, $\|\hat O\|_F\equiv \sqrt{\Tr [\hat O^\dag\hat O]/\Tr \hat{\mathbb{I}}}$ with $\hat{\mathbb{I}}$ being the global identity and the definition of $\hat A^{[r]}(t)$ follows that in Eq.~(\ref{LRB2}) (i.e., obtained from $\hat A(t)$ by truncating the components away from ${\rm supp}\textcolor{black}{[\hat A]}$ by distance $\ge r$). Accordingly, for the infinite temperature OTOC we have the bound
\begin{equation}
C(t)\le 4\|\hat B\|^2 \|\hat A(t) - \hat A^{[r]}(t)\|^2_F\le c' \left(\frac{t}{r^{\frac{2(\alpha -d)}{2\alpha - d +1}}}\right)^{2\alpha - d+1},
\end{equation}
where $c'$ is a constant depending on $r_0$, $r={\rm dist}({\rm supp}\textcolor{black}{[\hat A]}, {\rm supp}\textcolor{black}{[\hat B]})$ and the inequality $\|\hat O_1\hat O_2\|_F\le \|\hat O_1\|\|\hat O_2\|_F$\footnote{This can be derived by choosing $\hat O=\hat O_1^\dag\hat O_1$ and $\hat\rho = \hat O_2\hat O_2^\dag$ in the H\"older inequality below Eq.~(\ref{LROTOC}) followed by taking the root.} has been used. Therefore, whenever $\alpha > d$, the scrambling time would be sublinear in terms of the system size.


\subsection{Bounds in quantum thermodynamics} 
Quantum thermodynamics is an emergent field that aims at extending the conventional thermodynamics to the quantum regime, where unique quantum effects like coherence and entanglement play crucial roles. Here we just pick up a few notable bounds in this comprehensive field. Readers with special interest may refer to the topical review Ref.~\refcite{JG16} and the references therein.

\subsubsection{Unitary work extraction}
The arguably simplest setting in quantum thermodynamics is work extraction by unitary operations.\cite{AEA04} Consider a closed quantum system, usually referred to as quantum battery in this context,\cite{RA13} described by Hamiltonian $\hat H$ and initialized in state $\hat\rho$. For simplicity, we assume a finite Hilbert-space dimension $n$ so that the Hamiltonian can be diagonalized as $\hat H=\sum^{n-1}_{m=0} E_m|m\rangle\langle m|$ with $0=E_0\le E_1\le ...\le E_{n-1}$. After applying a unitary transformation $\hat U$, the extracted work reads
\begin{equation}
W=\Tr[\hat H(\textcolor{black}{\hat\rho - \hat U\hat\rho \hat U^\dag)}].
\label{WHU}
\end{equation}
Given $\hat\rho$ and $\hat H$, there exists a maximal extractable work
$W_{\max}\equiv\max_{\hat U}\Tr[\hat H(\textcolor{black}{\hat\rho - \hat U\hat\rho \hat U^\dag})]$, 
which is sometimes called ergotropy. We call $\hat\rho$ passive if $W_{\max}=0$, 
and the sufficient and necessary condition for the passivity turns out to be $\hat\rho=\sum^{n-1}_{m=0}p_m|m\rangle\langle m|$ (i.e., diagonal under the energy basis) with $p_1\ge p_2\ge...\ge p_n$. Interestingly, if $\hat\rho$ is completely passive, in the sense that $\hat\rho^{\otimes N}$ is passive with respect to $\hat H^{\oplus_{\rm K} N}$ $\forall N\in\mathbb{Z}^+$, where ``$\oplus_{\rm K}$" is the Kronecker sum defined as $\hat O_X\oplus_{\rm K}\hat O_Y\equiv \hat O_X\otimes\hat{\mathbbm{1}}_Y + \hat{\mathbbm{1}}_X\otimes \hat O_Y$, then the only possibility is thermal states, i.e., $\hat\tau_\beta=e^{-\beta\hat H}/Z$ with $Z=\Tr[e^{-\beta\hat H}]$ for some $\beta\ge0$.\cite{WP78,AL78} Moreover, for an ensemble of passive but not completely passive states, it is always possible to perform an operation that achieves maximal work extraction but generates no entanglement.\cite{KVH13} While we will not go into further detail, It is worth mentioning that the notion of passivity has been generalized to partially accessible systems and quantum channels.\cite{MF14,AMA19} \textcolor{black}{Another generalization achieved recently is to unitary operations restricted by continuous symmetries.\cite{YM22} In this case, the completely passive states turn out to be the generalized Gibbs ensembles involving the corresponding conserved charges, which may not commute with each other.\cite{NYH16}}

Things become more nontrivial if we consider the work extraction from correlated quantum states rather than a direct product of $N$ copies. To make correlation the only resource for work extraction, Perarnau-Llobet {\it et al.} studied the so-called locally thermal states.\cite{MPL15} As the name indicates, denoting a locally thermal states as $\hat\rho_{\rm LT}$, we have $\Tr_{\bar\alpha} \hat\rho_{\rm LT} = \hat\tau_\beta$ (defined below Eq.~(\ref{WHU})) for any $\alpha=1,2,...,N$ and some $\beta\ge0$. Note that a locally thermal state could even be pure, such as $\hat \rho_{\rm LT}=|\Psi_{\rm LT}\rangle\langle\Psi_{\rm LT}|$ with $|\Psi_{\rm LT}\rangle = Z^{-1/2}\sum^{n-1}_{m=0}e^{-\beta E_m/2}|m\rangle^{\otimes N}$. In this case, one can achieve the maximal work extraction $W_N= NE_\beta$ with $E_\beta\equiv \Tr[\hat\tau_\beta\hat H]$. On the other hand, if we fix the entropy of $\hat \rho_{\rm LT}$ to be $Ns$, then the extractable work can be upper bounded by
\begin{equation}
W_N \le N (E_\beta - E_{\beta'}),
\label{WNE}
\end{equation}
where $\beta'$ is determined from $s=S(\hat\tau_{\beta'})$ with $S(\hat\rho)=-\Tr[\hat\rho\log\hat\rho]$ being the von Neumann entropy. This bound simply arises from the fact that the thermal state minimizes the energy for a fixed entropy. What is nontrivial is that this bound is tight, in the sense that there always exits some $\hat\rho_{\rm LT}$ that can be unitarily transformed into $\hat\tau_{\beta'}^{\otimes N}$. Another remarkable observation in Ref.~\refcite{MPL15} is that the trivial bound $W_N/N\le E_\beta$ is asymptotically (in the limit of $N\to\infty$) saturated by a separable locally thermal state $\hat\rho_{\rm LT}=Z^{-1}\sum^{n-1}_{m=0}e^{-\beta E_m}(|m\rangle\langle m|)^{\otimes N}$, implying that, at least for the current setup, quantum entanglement does not seem to have a significant advantage in extracting more work.


\subsubsection{Second law and Maxwell's demon}
In a general situation, the system of interest is not closed but coupled to one or several heat baths, with which the energy exchange should be identified as heat. To be compatible with the first law, we only have to define work as the difference between system energy change and heat. In the weak coupling regime, this also implies that the total energy change of the system and heat baths should be identified as work.\cite{MC11} What remains less clear is the second law, the undoubtedly most important bound in thermodynamics. While there are several different formalisms to derive the second law microscopically,\cite{SV16} such as the resource theory,\cite{MH13,FB15,EC19} we would like to introduce the arguably most comprehensible one by focusing on the case of a single heat bath. 

Consider a system $S$ described by a parameterized Hamiltonian $\hat H_S(\lambda)$ coupled to a heat bath $B$ described by $\hat H_B$ via a weak, time-dependent interaction $\hat H_{SB}(t)$. Here $\lambda$ is an external work parameter that can be controlled to vary in time. For a protocol $\lambda_t$ ($t\in[0,\tau]$), the entire system evolves under $\hat H(t)=\hat H_S(\lambda_t) + \hat H_B + \hat H_{SB}(t)$, where the system-bath interaction is assumed to be turned off initially and finally, i.e., $\hat H_{SB}(0)=\hat H_{SB}(\tau)=0$. Suppose that the initial state is  at thermal equilibrium with inverse temperature $\beta$: 
\begin{equation}
\hat\rho_0 = \hat \tau_\beta(\lambda_0),\;\;\;\;
\hat\tau_\beta(\lambda)\equiv \frac{e^{-\beta\hat H_S(\lambda)}}{Z_S(\lambda)}\otimes \frac{e^{-\beta \hat H_B}}{Z_B},
\end{equation}
where $Z_S(\lambda)=\Tr[e^{-\beta \hat H_S(\lambda)}]$ and $Z_B=\Tr [e^{-\beta\hat H_B}]$. 
Denoting the final entire state as $\hat\rho_\tau = \hat U_\tau \hat\rho_0\hat U_\tau^\dag$ with $\hat U_\tau = \overleftarrow{\rm T} e^{-i\int^\tau_0 dt\hat H(t)}$, we have
\begin{equation}
S(\hat \rho_0) = S(\hat \rho_\tau) \le -\Tr [\hat \rho_\tau \log \hat \tau_\beta (\lambda_\tau)],
\label{rhotau}
\end{equation}
where the rightmost bound arises simply from the non-negativity of the (quantum) Kullback-Leibler divergence. Since the total energy change of the system and bath should be identified as work, the above relation (\ref{rhotau}) implies
\begin{equation}
W\le -\Delta F_S = \beta^{-1} \log \frac{Z_S(\lambda_\tau)}{Z_S(\lambda_0)}.
\label{WDF}
\end{equation}
This result is nothing but the second law for isothermal processes -- the extracted work cannot exceed the free-energy reduction. It is worth mentioning that Eq.~(\ref{WDF}) can be derived from the quantum Jarzynski equality\cite{CJ97,HT00,JK00}
\begin{equation}
\langle e^{-\beta w} \rangle = e^{-\beta \Delta F_S}
\end{equation}
by using the Jensen's inequality. Here $w$ is the random variable of quantum work, which is determined from the joint two-point energy measurements on both the system and the heat bath.\cite{MC11}

Since the birth of the second law, it was challenged by Maxwell's demon, an imaginary intelligence that can reverse the heat flow by sophisticated microscopic manipulations.\cite{KM09} Later, the Gedankenexperiment was simplified by Szilard into the Szilard engine, which is compatible with the isothermal work-extraction setup discussed above.\cite{LS29} From a modern point of view, Maxwell's demon can be formalized as measurement and feedback control,\cite{JMRP15} as first pointed out by Sagawa and Ueda.\cite{TS08} One can retrieve the second law by noting that measurement and erasing the results in the memory require additional work cost.\cite{TS09}  

More concretely, let us consider the same setup as mentioned above. However, there are two crucial differences: (i) at time $t_1$, a quantum measurement ${\rm M}$ specified by a set of Kraus operators $\{\hat M_\alpha\}_\alpha$, is performed on the system; (ii) according to the measurement outcome $\alpha$, one may carry out different work protocols $\lambda_\alpha(t)$ during $t\in[t_1,\tau]$. Despite the large freedom of performing various feedback control, the extracted work turns out to be always upper bounded by\cite{TS08} 
\begin{equation}
W\le -\Delta F_S + \beta^{-1} I(\hat \rho_{S1}:{\rm M}),
\label{SLFB}
\end{equation}
where $\hat\rho_{S1}$ is the system state at $t_1$ right before the measurement, and  $I(\hat\rho:{\rm M})$ is the so-called QC-mutual information defined as
\begin{equation}
I(\hat \rho:{\rm M}) \equiv S(\hat\rho) - \sum_\alpha p_\alpha S(\hat\rho_\alpha),\;\;\;\; p_\alpha=\Tr[\hat M_\alpha^\dag \hat M_\alpha\hat\rho],\;\;\hat\rho_\alpha = \hat M_\alpha\hat\rho\hat M_\alpha^\dag/p_\alpha,
\label{IQC}
\end{equation}
which differs from the Holevo information $\chi=S(\sum_\alpha p_\alpha \hat\rho_\alpha) - \sum_\alpha p_\alpha S(\hat\rho_\alpha)$ by the entropy change caused by the measurement backaction. Nevertheless, the QC-mutual information is also non-negative-definite and turns out to be always upper bounded by $H(\{p_\alpha\})=-\sum_\alpha p_\alpha \log p_\alpha$, the Shannon entropy of the measurement outcome. Apparently, Eq.~(\ref{SLFB}) implies the possibility of violating Eq.~(\ref{WDF}). However, by carefully analyzing the energetics of measurement / information erasure, modeled as the interaction between the memory and the system / heat bath, one can show that the total amount of work required for these two processes is always lower bounded by the QC-mutual information:\cite{TS09}
\begin{equation}
W_{\rm meas} + W_{\rm eras} \ge \beta^{-1} I(\hat\rho_{S1}:{\rm M}).
\end{equation}
This result refines the famous Landauer's principle,\cite{RL61} which assumes zero energy cost for measurement (indeed valid for classical and degenerate memories) and only concerns the information erasure process.\cite{CHB82} We note that there are also some other microscopically derived bounds on the minimal work cost or entropy production required for erasing the information encoded in the memory.\textcolor{black}{\cite{JG15,FB20}}

Finally, we mention that it is possible to derive the classical version of Eq.~(\ref{SLFB}) by applying Jensen's inequality to the generalized Jarzynski equality (also known as the Sagawa-Ueda equality):\cite{TS10}
\begin{equation}
\langle  e^{-\beta (w-\Delta F_S) - i_{SM}}\rangle=1,
\label{SUE}
\end{equation} 
where $i_{SM}$ is a trajectory-level random variable whose ensemble average gives the usual mutual information. The quantum generalization was achieved in Ref.~\refcite{ZG16}, where several different types of $i_{SM}$ were found to validate Eq.~(\ref{SUE}) and one of them is consistent with the QC-mutual information (\ref{IQC}).

\subsubsection{Thermodynamic uncertainty relations}
\label{Sec:TUR}
While the history of uncertainty relation in quantum mechanics is almost as long as that of quantum mechanics, the thermodynamic uncertainty relation (TUR) for classical stochastic systems,\cite{JMH20} which takes a similar form as
\begin{equation}
\frac{{\rm Var}[j]}{\langle j\rangle^2} \langle \sigma\rangle \ge 2,
\label{TUR}
\end{equation} 
was not discovered until a couple of years ago. Here $j$ is a current observable that fluctuates at the trajectory level, ${\rm Var}[j]\equiv \langle j^2\rangle - \langle j\rangle^2$, and $\langle\sigma \rangle$ is the dimensionless (i.e., the usual entropy divided by $k_B$) entropy production. This relation (\ref{TUR}) thus reflects a trade-off between the indicator of (relative) precision and dissipation, which generally decreases and grows linearly in time, respectively. Originally found in some specific biomolecular systems in the long-time limit,\cite{ACB15} the TUR (\ref{TUR})  has been generalized to arbitrary continuous Markov processes\cite{TRG16} and finite time,\cite{JMH17} possibly with non-stationary initial states\cite{KL20} and time-dependent driving.\cite{US20} It also finds its applications to nano-scale heat engines, leading to a universal constraint on the efficiency in terms of relative power fluctuations.\cite{PP18}

While there are several ways to derive Eq.~(\ref{TUR}), one elegant approach is to appropriately parametrize the Markov process and then exploit the Cram\'er-Rao inequality.\cite{AD19,YH19} A similar approach can be applied to (Markovian) open quantum systems by exploiting the quantum Cram\'er-Rao inequality:\cite{YH20} 
\begin{equation}
\frac{\rm Var_\theta[\Theta]}{(\partial_\theta\langle \Theta \rangle_{\theta})^2} \ge \frac{1}{F_{\rm Q}(\theta)},
\label{QCR}
\end{equation}
where $\theta$ is a parameter that enters into the equation of motion (Lindblad master equation), $\Theta$ is a random variable that fluctuates over different quantum trajectories, and $F_Q(\theta)$ is the quantum Fisher information\cite{SLB94} for the entire (i.e., system plus environment) pure state, which is nevertheless calculable in terms of the parameterized master equation.\cite{SG14} In particular, focusing on the classical-like time-independent dynamics
\begin{equation}
\dot{\hat \rho}_t = - i[\hat H, \hat\rho_t] + \sum_{m\neq n}{\textcolor{black}{\mathbb{D}}}[\hat L_{mn}]\hat\rho_t,\;\;\;\;\hat H=\sum_nE_n|n\rangle\langle n|,\;\;\hat L_{mn}=\sqrt{r_{mn}}|m\rangle\langle n|,
\end{equation}
where \textcolor{black}{$\mathbb{D}[\hat L]\hat\rho\equiv \hat L\hat\rho\hat L^\dag - \{\hat L^\dag\hat L,\hat\rho\}/2$} and the transition rates $r_{mn}$ satisfy $r_{mn}>0$ $\forall m\neq n$, we can parametrize the jump operators as
\begin{equation}
\hat L_{mn}(\theta) = \sqrt{1+\theta\left(1-\sqrt{\frac{r_{nm}\pi_m}{r_{mn}\pi_n}}\right)}\hat L_{mn},
\end{equation}
with $\pi_n$'s determined from the stationary solution $\hat\rho_{\rm ss}=\sum_n \pi_n |n \rangle\langle n|$. Then choosing $\Theta$ to be an arbitrary current observable $j$, whose increment associated with jump $\hat L_{mn}$ coincides its decrement associated with the reverse jump $\hat L_{nm}$, we find that Eq.~(\ref{QCR}) at stationary and $\theta=0$ reads
\begin{equation}
\frac{{\rm Var}[j]}{\langle j \rangle^2} \ge \frac{1}{T\mathcal{X}},\;\;\;\;\mathcal{X}=\sum_{m\neq n}(\sqrt{r_{mn}\pi_n}-\sqrt{r_{nm}\pi_m})^2.
\label{jTX}
\end{equation} 
According to the inequality $(x-y)^2\le(x^2-y^2)\log(x/y)/2$, we know that $\mathcal{X}\le\frac{1}{4}\sum_{m\neq n}(r_{mn}\pi_n - r_{nm}\pi_m)\log\frac{r_{mn}\pi_n}{r_{nm}\pi_m} =\Sigma/2$ with $\Sigma$ being the entropy-production rate.\footnote{Precisely speaking, to obtain this entropy expression, we have assumed the local detailed balance condition and that the system is coupling to a single bath with a fixed temperature.} Therefore, Eq.~(\ref{jTX}) implies the TUR (\ref{TUR}). Of course, by choosing other parametrization and observables, one can obtain many other TUR-like relations for more general open quantum dynamics.\cite{YH21}


Finally, we would like to mention the so-called fluctuation-theorem TUR,\cite{TVV19} which takes the form of
\begin{equation}
\frac{{\rm Var}[j]}{\langle j\rangle} \ge \frac{2}{e^{\langle\sigma\rangle}-1},
\label{FTTUR}
\end{equation}
and is thus always looser than Eq.~(\ref{TUR}). As the name indicates, this relation can be derived solely from the detailed fluctuation theorem for any joint distributions of entropy production and current observation $j$:\cite{RGG10}
\begin{equation}
P(\sigma,j) = e^{\sigma} \tilde P(-\sigma,-j).
\label{Psj}
\end{equation}
In fact, starting from Eq.~(\ref{Psj}), one can obtain a tighter result in which the rhs of the fluctuation-theorem TUR (\ref{FTTUR}) is replaced by $2/[\cosh(2g(\langle\sigma\rangle/2))-1]$, where $g(x)$ is the inverse function of $x\tanh x$.\cite{AMT19} Note that Eq.~(\ref{FTTUR}) applies to both classical stochastic systems and open quantum systems, provided the validity of Eq.~(\ref{Psj}).


\section{Conclusions and outlook}
\label{Sec:CO}

We have surveyed various bounds appearing in nonequilibrium dynamics from a theoretical point of view.
Before ending this review, let us summarize our discussions so far and provide some future directions.

We have started from the introduction of the quantum speed limit in Sec.~\ref{Sec:SL}.
Since the original derivation based on the uncertainty relation between time and energy fluctuation by Mandelstam and Tamm, various types of quantum speed limits have been obtained from, e.g., the information-theoretical and geometric approaches.
These approaches have been applied to classical and stochastic quantum systems, which also elucidate the relation between the transition speed and other quantities like the entropy production rate.
While many speed limits are useful for few-body systems, relevant speed limits for macroscopic systems are recently obtained using, e.g.,  the local conservation law of probability.

Inequalities can also provide crucial tools to investigate the foundation of quantum statistical mechanics.
As reviewed in Sec.~\ref{Sec:QT}, various bounds concerning the energy spectra of the Hamiltonian have been found to justify equilibration and thermalization in isolated quantum many-body systems.
Although complete understanding is still lacking, timescales of equilibration have been challenged by many researchers, which leads to the discovery of several constraints about them.
We have also discussed equilibration timescales in dissipative (quantum) systems, 
particularly focusing on recent works on the discrepancy between the timescale and the inverse Liouvillian gap.

In Sec.~\ref{Sec:LRbound}, we have discussed one of the most fundamental constraints in quantum many-body systems, i.e., the Lieb-Robinson bound.
After introducing the standard linear-cone Lieb-Robinson bound in (quasi-)locally interacting systems as well as a refined one for the free-fermion case, 
we summarize the modern results on the Lieb-Robinson bound in long-range interacting systems.
In the latter case, there appears either linear, sublinear, or logarithmic cone depending on the spatial dimension $d$ and the decay exponent of the long-range interaction $\alpha$.
We have also explained how the Lieb-Robinson bound is useful to extract equilibrium properties of quantum many-body systems, such as the clustering property and the entanglement area law of the ground state.

The next section, Sec.~\ref{Sec:EG}, is devoted to the bounds on the generation of entanglement, a unique quantity in quantum physics.
We have discussed general bounds based on the information-theoretic approach and 
tighter bounds for many-body Hamiltonians.
We have also given a brief introduction to the recently developed entanglement-membrane theory and derived the bounds concerning the entanglement velocity in this framework.

In  Sec.~\ref{Sec:EB}, we have discussed several examples of error bounds for approximated quantum dynamics, including circuitization of a continuous time evolution, prethermal dynamics in Floquet quantum many-body systems, and the approximated constrained dynamics due to a large energy gap.
Finally, in Sec.~\ref{Sec:MT}, we have treated other important topics, i.e., absence of time crystals, bounds on quantum chaos and scrambling, and bounds in quantum thermodynamics.

Despite recent remarkable progress, our understanding of nonequilibrium physics is far from complete.
Indeed, there are a number of open questions on the topics we have discussed so far.
First of all, several concepts introduced in this review, such as the eigenstate thermalization hypothesis and the Maldacena-Shanker-Stanford bound, still remain conjectures.
It is a big challenge to rigorously prove these conjectures solely from the principle of quantum mechanics.
Similarly, it would be desirable if microscopic quantum theory could justify the results obtained by the entanglement membrane theory, which relies on the coarse-grained picture in a phenomenological manner.

Second, while we have explained many bounds, not all of them are necessarily tight.
Some of them require clarifications of the physical situations to achieve the equality condition.
In addition, we may tighten our bounds by taking account of additional properties of the system of interest (e.g., locality) or even find the condition of optimality of the inequality (as done in Sec.~\ref{Sec:LRbound}).
One of the most important problems is timescale of equilibration in many-body systems (see Sec.~\ref{Sec:QT}), where tight bounds seem still lacking.

Third, it would be of great importance to generalize/extend the established bounds to different types of systems beyond conventional applicability.
One of the intriguing directions is to extend the bounds to open quantum systems.
For example, Secs.~\ref{Sec:LRbound},~\ref{Sec:EG}, and~\ref{Sec:EB} only considered unitary dynamics.
While the standard Lieb-Robinson bound can be generalized to quantum systems described by the GKSL equation,\cite{DP10} 
it remains less known whether the recent  advancement of the Lieb-Robinson bound (Sec.~\ref{Sec:LRbound}) and its application (Secs.~\ref{Sec:EG} and \ref{Sec:EB}) can be straightforwardly extended to the dissipative case (see Refs.~\refcite{MJK13c,TSC15,RS19,AYG21} for some recent progress).
Furthermore, it would be interesting to investigate a more complicated but realistic dynamics, such as quantum dynamics with non-Markovian dissipation.

Fourth, it is of course essential to understand how our theoretical bounds are verified in experiments and even applied to predict fundamental limitations of controlling systems in laboratories.
For verification of constraints in quantum dynamics, a promising candidate is artificial quantum systems, e.g., ultracold atomic systems,\cite{IB12} Rydberg atoms,\cite{AB20} trapped ions,\cite{RB12} superconducting qubits,\cite{AAH12} and NV centers.\cite{VVD13}
Since these systems also provide an ideal platform for performing quantum information processing and other quantum tasks (such as quantum sensing\cite{CLD17} and metrology\cite{VG11}
),
it is important to understand the cost/limitation of quantum control that may be followed from each bound.
We note that such an attempt has been made in the context of, e.g., shortcuts to adiabaticity\cite{SC17,KF17}, optimal quantum control\cite{TC09}, and predictating the limitation of the adiabatic theorem from orthogonality catastrophe\cite{OL17,JHC21}, where the quantum speed limit offers a crucial limitation.
For verification and application of bounds in thermodynamics in small systems, 
we notice the advantage of colloidal and biological systems, as well as other nano-scale devices such as a single-electron box.

Finally, we stress that many of the bounds reviewed in this manuscript were found in the last decade, indicating that the field would be just an initial stage of growth.
Of course, studies on dynamical constraints can help us discover fundamental principles of nonequilibrium statistical mechanics, which is a traditional but incomplete subject in physics.
On another front, they can cultivate interdisciplinary research areas related to, e.g., condensed matter physics, atomic, molecular, and optical physics, high-energy physics, classical and quantum information theory, mathematics, statistics,  and even biology and social science.
Therefore, universal bounds will surely offer profound insights into our understanding of nonequilibrium science and deserve further exploration.


\section*{Acknowledgements}
We thank K. Fujimoto for helpful comments on the manuscript.
We also thank for Francesco Buscemi, Adolfo del Campo, Jyong-Hao Chen, Arnab Das, Jad C. Halimeh, Zohar Nussinov, Lorenzo Piroli for useful email correspondence.
Z. G. is supported by the Max-Planck-Harvard Research Center for Quantum Optics (MPHQ).










\section*{References}










\begin{thebibliography}{0}


\bibitem{GEU63} G. E. Uhlenbeck and G. W. Ford, {\it Lectures in Statistical Mechanics}
(American Mathematical Society, Providence, 1963).

\bibitem{WH27} W. Heisenberg, {\it Z. Phys.} {\bf 43}, 172 (1927).

\bibitem{US05} U. Seifert, {\it Phys. Rev. Lett.} {\bf 95}, 040602 (2005).

\bibitem{EHK27} E. H. Kennard, {\it Z. Phys.} {\bf 44}, 326 (1927).

\bibitem{KS10} K. Sekimoto, {\it Stochastic Energetics} (Springer, Berlin, 2010).

\bibitem{JvN55} J. von Neumann, Mathematical Foundation of Quantum Mechanics (Princeton University Press, Princeton, 1955).


\bibitem{LM45} L. Mandelstam and I. Tamm, {\it J. Phys. USSR} {\bf 122}, 249 (1945).

\bibitem{SD17} S. Deffner and S. Campbell, {\it J. Phys. A} {\bf 50}, 453001 (2017).

\bibitem{HPR29}
H. P. Robertson, {\it Phys. Rev.} {\bf 34}, 163 (1929).

\bibitem{GNF73}
G. N. Fleming, \textit{Il Nuovo Cimento A} \textbf{16}, 232 (1973)

\bibitem{KB83}
K. Bhattacharyya, {\it J. Phys. A: Math. Gen.} {\bf 16}, 2993 (1983).





\bibitem{JA90}
J. Anandan and Y. Aharonov, {\it Phys. Rev. Lett.} {\bf 65}, 1697 (1990).


\bibitem{NM98} N. Margolus and L.B. Levitin, {\it Physica D} {\bf 120}, 188 (1998).


 
 \bibitem{LBL09}
 L. B. Levitin and T. Toffoli, {\it Phys. Rev. Lett.} {\bf 103}, 160502 (2009).
 

\bibitem{PP93}
 P. Pfeifer, {\it Phys. Rev. Lett.} {\bf 70}, 3365 (1993).

\bibitem{AU92}
A. Uhlmann {\it Phys. Lett. A} {\bf 161}, 329 (1992).

\bibitem{SD13E}
S. Deffner and E. Lutz, \textit{J. Phys.
A: Math. Theor.} \textbf{46}, G5302 (2013).


 
\bibitem{PJJ10}
P. J. Jones and P. Kok, \textit{Phys. Rev. A} \textbf{82}, 022107 (2010).

\bibitem{CH67}
C. Helstrom, {\it Phys. Lett. A} {\bf 25}, 101 (1967).

\bibitem{JL19}
 J. Liu, H. Yuan, X.-M. Lu, and X. Wang, \textit{J. Phys. A: Math. Theor.} \textbf{53}, 023001 (2019).
 
 \bibitem{SLB94} S. L. Braunstein and C. M. Caves, {\it Phys. Rev. Lett.} {\bf 72}, 3439 (1994).

 \bibitem{DPP15}
 D. P. Pires, L. C. C\'eleri, and D. O. Soares-Pinto, \textit{Phys. Rev. A} \textbf{91}, 042330 (2015).

 \bibitem{DPP16}
 D. P. Pires, M. Cianciaruso, L. C. C\'eleri, G. Adesso, and D. O. Soares-Pinto, \textit{Phys. Rev. X} \textbf{6}, 021031 (2016).

\bibitem{BS18} B. Shanahan, A. Chenu, N. Margolus, and A. del Campo, {\it Phys. Rev. Lett.} {\bf 120}, 070401 (2018).

\bibitem{MO18} M. Okuyama and M. Ohzeki, {\it Phys. Rev. Lett.} {\bf 120}, 070402 (2018).

\bibitem{NS18} N. Shiraishi, K. Funo, and K. Saito, {\it Phys. Rev. Lett.} {\bf 121}, 070601 (2018).

\bibitem{US12}
U. Seifert, \textit{Rep. Prog. Phys.} \textbf{75}, 126001 (2012).

\bibitem{TH01}
T. Hatano and S.-i. Sasa, \textit{Phys. Rev. Lett.} \textbf{86}, 3463 (2001).


\bibitem{VTV20} 
V. T. Vo, T. Van Vu, and Y. Hasegawa, \textit{Phys. Rev. E} \textbf{102}, 062132 (2020).

\bibitem{TVV21}
 T. Van Vu and Y. Hasegawa, \textit{Phys. Rev. Lett.} \textbf{126}, 010601 (2021).

\bibitem{SBN20} 
S. B. Nicholson, L. P. Garcia-Pintos, A. del Campo, and J. R. Green, \textit{Nat. Phys.} \textbf{16}, 1211 (2020).

\bibitem{SI20} 
S. Ito and A. Dechant, \textit{Phys. Rev. X} \textbf{10}, 021056 (2020). 
 
\bibitem{KA22} 
K. Adachi, R. Iritani, and R. Hamazaki, \textit{Commun. Phys.} \textbf{5}, 129 (2023). 
 
\bibitem{LPG22D}
L. P. Garcia-Pintos,
``Diversity and fitness uncertainty allow for faster evolutionary rates", arXiv:2202.07533.
 
\bibitem{MMT13}
M. M. Taddei, B. M. Escher, L. Davidovich, and R. L. de Matos Filho, \textit{Phys. Rev. Lett.} \textbf{110}, 050402 (2013).

\bibitem{AdC13} 
A. del Campo, I. L. Egusquiza, M. B. Plenio, and S. F. Huelga, \textit{Phys. Rev. Lett.} \textbf{110}, 050403 (2013).
 
\bibitem{SD13} S. Deffner and E. Lutz, {\it Phys. Rev. Lett.} {\bf 111}, 010402 (2013).

\bibitem{LPG21}
L. P. Garcia-Pintos, S. Nicholson, J. R. Green, A. del Campo, and A. V. Gorshkov, \textit{Phys. Rev. X} \textbf{12}, 011038 (2022).

\bibitem{VG76}
 V. Gorini, A. Kossakowski, and E. C. G. Sudarshan, \textit{J. Math. Phys.} \textbf{17}, 821 (1976).

\bibitem{GL76}
G. Lindblad,  \textit{Commun. Math. Phys.} \textbf{48}, 119 (1976).

\bibitem{KF19}
K. Funo, N. Shiraishi, and K. Saito, \textit{New J. Phys.} \textbf{21}, 013006 (2019).

\bibitem{SD17NJP}
 S. Deffner, \textit{New J. Phys.} \textbf{19}, 103018 (2017).
 

\bibitem{MB19}
M. Bukov, D. Sels, and A. Polkovnikov, \textit{Phys. Rev. X} \textbf{9}, 011034 (2019).

\bibitem{MRL21}
M. R. Lam, N. Peter, T. Groh, W. Alt, C. Robens, D. Meschede, A. Negretti, S. Montangero, T. Calarco, and A. Alberti, \textit{Phys. Rev. X} \textbf{11}, 011035 (2021).


\bibitem{RH21} R. Hamazaki, \textit{PRX Quantum} \textbf{3}, 020319 (2022).

\bibitem{FD99}
F. Dalfovo, S. Giorgini, L. P. Pitaevskii, and S. Stringari, \textit{Rev. Mod. Phys.} \textbf{71}, 463 (1999).

\bibitem{EM27}
E. Madelung, \textit{Z. Phys.} \textbf{40}, 322 (1927).

\bibitem{TCW94}
T. C. Wallstrom, \textit{Phys. Rev. A} \textbf{49}, 1613 (1994).


\bibitem{AD18}
 A. Dechant and S.-i. Sasa, \textit{Phys. Rev. E} \textbf{97}, 062101 (2018).

\bibitem{CV09}
C. Villani, {\it Optimal transport: old and new} (Springer, 2009).

\bibitem{AS05}
A. Shimizu and T. Morimae, \textit{Phys. Rev. Lett.} \textbf{95}, 090401 (2005).

\bibitem{BY16}
B. Yadin and V. Vedral, \textit{Phys. Rev.
A} \textbf{93}, 022122 (2016).

\bibitem{BD98}
B. Derrida, \textit{Phys. Rep.} \textbf{301}, 65 (1998).



\bibitem{LDL80}
L. D. Landau and E. M. Lifshitz, {\it Statistical Physics Part I}, (Elsevier, 2013).


\bibitem{JVN29}
J. v. Neumann, \textit{Zeit. für Phys.} \textbf{57}, 30 (1929); English translation (by R. Tumulka), \textit{Eur. Phys. J. H.} \textbf{35}, 201 (2010).


\bibitem{IMG14}
I. M. Georgescu, S. Ashhab, and Franco Nori, \textit{Rev. Mod. Phys.}, \textbf{86}, 153 (2014).

\bibitem{IB08}
I. Bloch, J. Dalibard, and W. Zwerger, \textit{Rev. Mod. Phys.}, \textbf{80}, 885 (2008).

\bibitem{PB57}
P. Bocchieri and A. Loinger, \textit{Phys. Rev.} \textbf{107}, 337 (1957). 

\bibitem{AP11}
A. Polkovnikov, K. Sengupta, A. Silva, and M. Vengalattore,
\textit{Rev. Mod. Phys.} \textbf{83}, 863 (2011).

\bibitem{JE15}
J. Eisert, M. Friesdorf, and C. Gogolin,
\textit{Nat. Phys.}, \textbf{11}, 124 (2015).

\bibitem{MU20}
M. Ueda, \textit{Nat. Rev. Phys.} \textbf{2}, 669 (2020).


\bibitem{CG16}
C. Gogolin and J. Eisert,
\textit{Rep. Prog. Phys.}, \textbf{79}, 056001 (2016).

\bibitem{LD16} L. D'Alessio, Y. Kafri, A. Polkovnikov, and M. Rigol, {\it Adv. Phys.} {\bf 65}, 239 (2016).


\bibitem{TM18JP} T. Mori, T. N. Ikeda, E. Kaminishi, and M. Ueda, {\it J. Phys. B} {\bf 51}, 112001 (2018).



\bibitem{HT98}
H. Tasaki, \textit{Phys. Rev. Lett.} \textbf{80}, 1373 (1998).

\bibitem{PR08}
P. Reimann, \textit{Phys. Rev. Lett.} \textbf{101}, 190403 (2008).

\bibitem{NL09}
N. Linden, S. Popescu, A. J. Short, and A. Winter, \textit{Phys. Rev. E} \textbf{79}, 061103 (2009).

\bibitem{AJS11}
A. J. Short, \textit{New J. Phys.} \textbf{13}, 053009 (2011).

\bibitem{AJS12}
A. J. Short and T. C. Farrelly, \textit{New J. Phys.} \textbf{14}, 013063 (2012).

\bibitem{TF17}
T. Farrelly, F. G. S. L. Brand\~ao, and M. Cramer,
\textit{Phys. Rev. Lett.} \textbf{118}, 140601 (2017).

\bibitem{FGSL15}
F. G. S. L. Brand\~ao and M. Cramer,
``Equivalence of Statistical Mechanical Ensembles for Non-Critical Quantum Systems", arXiv:1502.03263.

\bibitem{HW19PRL}
H. Wilming, M. Goihl, I. Roth, and J. Eisert, \textit{Phys. Rev. Lett.} \textbf{123}, 200604 (2019).

\bibitem{JMD91}
J. M. Deutsch, \textit{Phys. Rev. A} \textbf{43}, 2046 (1991).

\bibitem{MS94}
M. Srednicki, \textit{Phys. Rev. E} \textbf{50}, 888 (1994).

\bibitem{MR08}
M. Rigol, V. Dunjko, and M. Olshanii, \textit{Nature} \textbf{452}, 854
(2008).

\bibitem{GDP15}
G. De Palma, A. Serafini, V. Giovannetti, and
M. Cramer, \textit{Phys. Rev. Lett.} \textbf{115}, 220401 (2015).

\bibitem{MS99}
M. Srednicki, \textit{J. Phys. A: Math. Gen.} \textbf{32}, 1163 (1999).

\bibitem{EK13}
E. Khatami, G. Pupillo, M. Srednicki, and M. Rigol,
\textit{Phys. Rev. Lett.} \textbf{111}, 050403 (2013).

\bibitem{RH1719}
R. Hamazaki, “Theoretical study on thermalization in
isolated quantum systems,” (2017), Master’s thesis,
arXiv:1901.01481.

\bibitem{IMK19}
I. M. Khaymovich, M. Haque, and P. A. McClarty,
\textit{Phys. Rev. Lett.} \textbf{122}, 070601 (2019).

\bibitem{FH10}
F. Haake, ``Quantum signatures of chaos", (Springer
Science \& Business Media, 2010)

\bibitem{OB84}
O. Bohigas, M. J. Giannoni, and C. Schmit,
{\it Phys. Rev. Lett.} {\bf 52}, 1 (1984).

\bibitem{SM04}
S. M\"uller, S. Heusler, P. Braun, F. Haake, and A. Altland, \textit{Phys. Rev. Lett.} \textbf{93}, 014103 (2004).

\bibitem{SM05}
S. M\"uller, S. Heusler, P. Braun, F. Haake, and A. Altland, \textit{Phys. Rev. E} \textbf{72}, 046207 (2005).

\bibitem{BB18}
B. Bertini, P. Kos, and T. Prosen, \textit{Phys. Rev. Lett.} \textbf{121}, 264101 (2018).

\bibitem{MVB77}
M. V. Berry, \textit{J. Phys. A: Math.
Gen.} \textbf{10}, 2083 (1977).

\bibitem{DN93}
D. N. Page
\textit{Phys. Rev. Lett.} \textbf{71}, 1291 (1993).

\bibitem{IA16}
I. Arad, T. Kuwahara, and Z. Landau, \textit{J. Stat. Mech.} 033301  (2016).

\bibitem{MS17}
M. Serbyn, Z. Papi\'c, and D. A. Abanin, \textit{Phys. Rev. B} \textbf{96}, 104201 (2017).

\bibitem{RH18PRL}
R. Hamazaki and M. Ueda, \textit{Phys. Rev. Lett.} \textbf{120}, 080603 (2018).

\bibitem{GB10}
G. Biroli, C. Kollath, and A. M. L\"auchli, \textit{Phys. Rev. Lett.} \textbf{105}, 250401 (2010).

\bibitem{TNI13}
T. N. Ikeda, Y. Watanabe, and M. Ueda, \textit{Phys. Rev. E} \textbf{87}, 012125 (2013).

\bibitem{EI17}
E. Iyoda, K. Kaneko, and T. Sagawa, \textit{Phys. Rev. Lett.} \textbf{119}, 100601 (2017).

\bibitem{TM16arxiv}
Takashi Mori,
``Weak eigenstate thermalization with large deviation bound",
arXiv:1609.09776

\bibitem{HT09}
H. Touchette, \textit{Phys. Rep.} \textbf{478}, 1 (2009).

\bibitem{TY18}
T. Yoshizawa, E. Iyoda, and T. Sagawa, \textit{Phys. Rev. Lett.} \textbf{120}, 200604 (2018).

\bibitem{MR07}
M. Rigol, V. Dunjko, V. Yurovsky, and M. Olshanii, \textit{Phys. Rev. Lett.} \textbf{98}, 050405 (2007).

\bibitem{MR09}
M. Rigol, \textit{Phys. Rev. Lett.} \textbf{103}, 100403 (2009).

\bibitem{FHL16}
F. H. L Essler and M. Fagotti, \textit{J. Stat. Mech.} 064002, (2016).

\bibitem{LV16}
L. Vidmar and M. Rigol, \textit{J. Stat. Mech.} 064002, (2016).

\bibitem{VO07}
V. Oganesyan and D. A. Huse, \textit{Phys. Rev. B} \textbf{75}, 155111 (2007).

\bibitem{MZ08M}
M. Žnidarič, T. Prosen, and P. Prelovšek
\textit{Phys. Rev. B} \textbf{77}, 064426 (2008).

\bibitem{AP10M}
A. Pal and D. A. Huse, \textit{Phys. Rev. B} \textbf{82}, 174411 (2010).

\bibitem{RN15}
R. Nandkishore and D. A. Huse. \textit{Annu. Rev. Condens. Matter Phys.}, \textbf{6}, 15 (2015).

\bibitem{DAA19}
D. A. Abanin, E. Altman, I. Bloch, and M. Serbyn, \textit{Rev. Mod. Phys.} \textbf{91}, 021001 (2019).

\bibitem{NS17PRL}
N. Shiraishi and T. Mori, \textit{Phys. Rev. Lett.} \textbf{119}, 030601 (2017).

\bibitem{CJT18}
C. J. Turner, A. A. Michailidis, D. A. Abanin, M. Serbyn, and Z. Papic, \textit{Nat. Phys.} \textbf{14}, 745 (2018).

\bibitem{MS21}
M. Serbyn, D. A. Abanin, and Z. Papic, \textit{Nat. Phys.} \textbf{17}, 675 (2021).

\bibitem{SP19}
S. Pai, M. Pretko, and R. M. Nandkishore, Phys. Rev. X 9,
021003 (2019).

\bibitem{PS20}
P. Sala, T. Rakovszky, R. Verresen, M. Knap, and F. Pollmann, Phys. Rev. X 10, 011047 (2020).

\bibitem{VK20}
V. Khemani, Michael Hermele, and R. Nandkishore, Phys. Rev. B 101, 174204 (2020).

\bibitem{SG10PRE}
S. Goldstein, J. L. Lebowitz, C. Mastrodonato, R. Tumulka, and N. Zanghi, \textit{Phys. Rev. E} \textbf{81}, 011109 (2010).

\bibitem{SG10Eur}
S. Goldstein, J. L. Lebowitz, R. Tumulka, and N. Zanghi, \textit{Eur. Phys. J. H.} \textbf{35}, 173 (2010).

\bibitem{PR15}
P. Reimann, \textit{Phys. Rev. Lett.} \textbf{115}, 010403 (2015).

\bibitem{SS21PRL}
S. Sugimoto, R. Hamazaki, and M. Ueda, \textit{Phys. Rev. Lett.} \textbf{126}, 120602 (2021).

\bibitem{SS21arxiv}
S. Sugimoto, R. Hamazaki, and M. Ueda, \textit{Phys. Rev. Lett.} \textbf{129}, 030602 (2022).

\bibitem{BN17L}
B. Neyenhuis, J. Zhang, P. W. Hess, J. Smith, A. C. Lee, P. Richerme, Z.-X. Gong, A. V. Gorshkov, and C. Monroe, \textit{Sci. Adv.} \textbf{3}, 1 (2017).

\bibitem{TK20PRL}
T. Kuwahara and K. Saito, Eigenstate Thermalization from Clustering Property of Correlations, Phys. Rev. Lett. 124, 200604 (2020).

\bibitem{HW18Eq}
H. Wilming, T. R. de Oliveira, A. J. Short, and J. Eisert, in \textit{Thermodynamics in the Quantum Regime–Recent Progress and Outlook}, ed. F. Binder {\it et al.} 
(Springer, Berlin, 2018), pp. 435-455.

\bibitem{TRO18}
T. R. de Oliveira, C. Charalambous, D. Jonathan, M.
Lewenstein, and A. Riera, \textit{New J. Phys.} \textbf{20}, 033032 (2018).

\bibitem{SG13PRL}
S. Goldstein, T. Hara, and H. Tasaki, \textit{Phys. Rev. Lett.} \textbf{111}, 140401 (2013).

\bibitem{SG15NJP}
S. Goldstein, T. Hara, and H. Tasaki, \textit{New J. Phys.} \textbf{17}, 045002 (2015).

\bibitem{PR16Nat}
P. Reimann, \textit{Nature Comm.} \textbf{7}, 10821 (2016).

\bibitem{ZN22}
Z. Nussinov and S. Chakrabarty, \textit{Ann.  Phys.} \textbf{168970} (2022).

\bibitem{LPG17}
L. P. Garc\'ia-Pintos, N. Linden, A. S. L. Malabarba, A. J. Short, and A. Winter, \textit{Phys. Rev. X} \textbf{7}, 031027 (2017).



\bibitem{RH20PRX}
R. Heveling, L. Knipschild, and J. Gemmer, \textit{Phys. Rev. X} 
\textbf{10}, 028001 (2020).


\bibitem{AR11o}
A. Rivas and S.F. Huelga, {\it Open Quantum Systems. An Introduction} (Springer, Heidelberg, 2011).


\bibitem{MZ15} M. Znidaric, {\it Phys. Rev. E} {\bf 92}, 042143 (2015).

\bibitem{KM16} K. Macieszczak, M. Guta, I. Lesanovsky, and J. P. Garrahan, {\it Phys. Rev. Lett} {\bf 116}, 240404 (2016).

\bibitem{DAL17} D. A. Levin and Y. Peres, {\it  Markov chains and mixing times} (American Mathematical Society, 2017).

\bibitem{MJK12}
M. J. Kastoryano, D. Reeb, and M. M. Wolf,  {\it J. Phys. A} {\bf 45}, 075307 (2012).

\bibitem{MJK13c}
M. J. Kastoryano and J. Eisert, {\it J. Math. Phys.} {\bf 54}, 102201 (2013).

\bibitem{EV20}
E. Vernier, {\it SciPost Phys.} {\bf 9}, 049 (2020).

\bibitem{TM20} T. Mori and T. Shirai, {\it Phys. Rev. Lett.} {\bf 125}, 230604 (2020).

\bibitem{TH21} T. Haga, M. Nakagawa, R. Hamazaki, and M. Ueda, {\it Phys. Rev. Lett.} {\bf 127}, 070402 (2021).


\bibitem{SY18}
S. Yao and Z. Wang, {\it Phys. Rev. Lett.} {\bf 121}, 086803 (2018).

\bibitem{FKK18}
F. K. Kunst, E. Edvardsson, J. C. Budich, and E. J. Bergholtz, {\it Phys. Rev. Lett.} {\bf 121}, 026808 (2018).

\bibitem{ZG18PRX}
Z. Gong, Y. Ashida, K. Kawabata, K. Takasan, S. Higashikawa, and M. Ueda, {\it Phys. Rev. X} {\bf 8}, 031079 (2018).

\bibitem{NO20}
N. Okuma, K. Kawabata, K. Shiozaki, and M. Sato, {\it Phys. Rev. Lett.} {\bf 124}, 086801 (2020).

\bibitem{FS19}
F. Song, S. Yao, and Z. Wang, {\it Phys. Rev. Lett.} {\bf 123}, 170401 (2019).

\bibitem{KK19s}
K. Kawabata, K. Shiozaki, M. Ueda, and M. Sato, {\it Phys. Rev. X} {\bf 9}, 041015 (2019).

\bibitem{RH20}
R. Hamazaki, K. Kawabata, N. Kura, and M. Ueda, {\it Phys. Rev. Research} {\bf 2}, 023286 (2020).

\bibitem{PD96}
P. Diaconis, {\it Proc. Natl. Acad. Sci. U.S.A} {\bf 93}, 1659 (1996).

\bibitem{TM21}
T. Mori, {\it Phys. Rev. Research} {\bf 3}, 043137 (2021). 

\bibitem{JB21}
J. Bensa and M. Znidaric,
{\it Phys. Rev. X} {\bf 11}, 031019 (2021)

\bibitem{MZ08}
M. Znidaric,  {\it Phys. Rev. A} {\bf 78}, 032324 (2008).

\bibitem{WTK20}
W.-T. Kuo, A. A. Akhtar, D. P. Arovas, and Y.-Z. You, {\it Phys. Rev. B} {\bf 101}, 224202 (2020).

\bibitem{MC22} \textcolor{black}{M. Cheneau, ``Experimental tests of Lieb–Robinson bounds'', arXiv:2206.15126.}

\bibitem{MFF15} M. Foss-Feig, Z.-X. Gong, C. W. Clark, and A. V. Gorshkov, {\it Phys. Rev. Lett.} {\bf 114}, 157201 (2015).  

\bibitem{EHL72} E. H. Lieb and D. W. Robinson, {\it Commun. Math. Phys.} {\bf 28}, 251 (1972). 

\bibitem{BN06} B. Nachtergaele and R. Sims, {\it Commun. Math. Phys.} {\bf 265}, 119 (2006). 

\bibitem{MBH10} M. B. Hastings, ``Locality in Quantum Systems'', arXiv:1008.5137.

\bibitem{BN17} B. Nachtergaele, R. Sims, and A. Young, \textcolor{black}{in {\it Mathematical Problems in Quantum
Physics}, ed. F. Bonetto {\it et al.}, Contemporary Mathematics Vol. 717 (American Mathematical Society, Providence, RI, 2018), pp. 93–115.}

\bibitem{SB06} S. Bravyi, M. B. Hastings, and F. Verstraete, {\it Phys. Rev. Lett.} {\bf 97}, 050401 (2006).  

\bibitem{IPS10} I. Pr\'emont-Schwarz, A. Hamma, I. Klich, and F. Markopoulou-Kalamara, {\it Phys. Rev. A} {\bf 81}, 040102(R) (2010).  

\bibitem{TK21b} T. Kuwahara and K. Saito, {\it Phys. Rev. Lett.} {\bf 127}, 070403 (2021).

\bibitem{JF22} J. Faupin, M. Lemm, and I. M. Sigal, \textcolor{black}{{\it Phys. Rev. Lett.} {\bf 128}, 150602 (2022).}

\bibitem{CY22} C. Yin and A. Lucas, \textcolor{black}{{\it Phys. Rev. X} {\bf 12}, 021039 (2022).} 

\bibitem{TK22} \textcolor{black}{T. Kuwahara, T. V. Vu, and K. Saito, ``Optimal light cone and digital quantum simulation of interacting bosons", arXiv:2206.14736.}

\bibitem{PC11} P. Calabrese, F. H. L. Essler, and M. Fagotti, {\it Phys. Rev. Lett.} {\bf 106}, 227203 (2011).

\bibitem{MC12} M. Cheneau, P. Barmettler, D. Poletti, M. Endres, P. Schau\ss, T. Fukuhara, C. Gross, I. Bloch, C. Kollath, and S. Kuhr, {\it Nature} {\bf 481}, 484 (2012). 

\bibitem{PJ14} P. Jurcevic, B. P. Lanyon, P. Hauke, C. Hempel, P. Zoller, R. Blatt, and C. F. Roos, {\it Nature} {\bf 511}, 202 (2014). 

\bibitem{ZG19} Z. Gong, N. Kura, M. Sato, and M. Ueda, ``Lieb-Robinson Bounds on Entanglement Gaps from Symmetry-Protected Topology'', arXiv:1904.12464. 

\bibitem{RB97} R. Bhatia, {\it Matrix Analysis} (Springer, New York, 1997).

\bibitem{YA20} Y. Ashida, Z. Gong, and M. Ueda, {\it Adv. Phys.} {\bf 69}, 249 (2020).

\bibitem{ZW20} Z. Wang and K. R.A. Hazzard, {\it Phys. Rev. X Quantum} {\bf 1}, 010303 (2020).

\bibitem{WK59} W. Kohn, {\it Phys. Rev.} {\bf 115}, 809 (1959).

\bibitem{JDC64} J. D. Cloizeaux, {\it Phys. Rev.} {\bf 135}, A685 (1964).

\bibitem{CB07} C. Brouder, G. Panati, M. Calandra, C. Mourougane, and N. Marzari, {\it Phys. Rev. Lett.} {\bf 98}, 046402 (2007).

\bibitem{BY13} B. Yan, S. A. Moses, B. Gadway, J. P. Covey, K. R. A. Hazzard, A. M. Rey, D. S. Jin, and J. Ye, {\it Nature} {\bf 501}, 521 (2013).    

\bibitem{PR14} P. Richerme, Z.-X. Gong, A. Lee, C. Senko, J. Smith, M. Foss-Feig, S. Michalakis, A. V. Gorshkov, and C. Monroe, {\it Nature} {\bf 511}, 198 (2014).  

\bibitem{SC17t} S. Choi, J. Choi, R. Landig, G. Kucsko, H. Zhou, J. Isoya, F. Jelezko, S. Onoda, H. Sumiya, V. Khemani, C. von Keyserlingk, N. Y. Yao, E. Demler, and M. D. Lukin, {\it Nature} {\bf 543}, 221 (2017).   

\bibitem{ZE17} Z. Eldredge, Z.-X. Gong, J. T. Young, A. H. Moosavian, M. Foss-Feig, and A. V. Gorshkov, {\it Phys. Rev. Lett.} {\bf 119}, 170503 (2017)

\bibitem{MCT19} M. C. Tran, A. Y. Guo, Y. Su, J. R. Garrison, Z. Eldredge, M. Foss-Feig, A. M. Childs, and A. V. Gorshkov, {\it Phys. Rev. X} {\bf 9}, 031006 (2019).

\bibitem{MBH06} M. B. Hastings and T. Koma, {\it Commun. Math. Phys.} {\bf 265}, 781 (2006). 

\bibitem{ZXG14} Z.-X. Gong, M. Foss-Feig, S. Michalakis, A. V. Gorshkov, {\it Phys. Rev. Lett.} {\bf 113}, 030602 (2014).

\bibitem{TM17} T. Matsuta, T. Koma, and S. Nakamura, {\it Ann. Henri Poincar\'e} {\bf 18}, 519 (2017).

\bibitem{DVE20} D. V Else, F. Machado, C. Nayak, and N. Y. Yao, {\it Phys. Rev. A} {\bf 101}, 022333 (2020).

\bibitem{CFC19} C.-F. Chen and A. Lucas, {\it Phys. Rev. Lett.} {\bf 123}, 250605 (2019).

\bibitem{TK20}  T. Kuwahara and K. Saito, {\it Phys. Rev. X} {\bf 10}, 031010 (2020).

\bibitem{MCT20} M. C. Tran, C.-F. Chen, A. Ehrenberg, A. Y. Guo, A. Deshpande, Y. Hong, Z.-X. Gong, A. V. Gorshkov, and A. Lucas, {\it Phys. Rev. X} {\bf 10}, 031009 (2020). 

\bibitem{MCT21} M. C. Tran, A. Y. Guo, A. Deshpande, A. Lucas, and A. V. Gorshkov, {\it Phys. Rev. X} {\bf 11}, 031016 (2021).  

\bibitem{MCT21b} M. C. Tran, A. Y. Guo, C. L. Baldwin, A. Ehrenberg, A. V. Gorshkov, and A. Lucas, {\it Phys. Rev. Lett.} {\bf 127}, 160401 (2021).   

\bibitem{MBH21} M. B. Hastings, ``Gapped Quantum Systems: From Higher Dimensional Lieb-Schultz-Mattis to the Quantum Hall Effect", arXiv:2111.01854.

\bibitem{MBH04a} M. B. Hastings, {\it Phys. Rev. Lett.} {\bf 93}, 140402 (2004).

\bibitem{DV16} D. Vodola, L. Lepori, E. Ercolessi, and G. Pupillo, {\it New J. Phys.} {\bf 18}, 015001 (2016).

\bibitem{MBH04b} M. B. Hastings, {\it Phys. Rev. Lett.} {\bf 93}, 126402 (2004).  

\bibitem{SHS17} S. Hern\'andez-Santana, C. Gogolin, J. I. Cirac, and A. Ac\'in, {\it Phys. Rev. Lett.} {\bf 119}, 110601 (2017).  

\bibitem{MK14} M. Kliesch, C. Gogolin, M. J. Kastoryano, A. Riera, and J. Eisert, {\it Phys. Rev. X} {\bf 4}, 031019 (2014).  

\bibitem{TK20b} T. Kuwahara, K. Kato, and F. G. S. L. Brand\~ao, {\it Phys. Rev. Lett.} {\bf 124}, 220601 (2020). 

\bibitem{TK22PRX} \textcolor{black}{T. Kuwahara and K. Saito, {\it Phys. Rev. X} {\bf 12}, 021022 (2022).}

\bibitem{AMA22} \textcolor{black}{\'A. M. Alhambra, ``Quantum many-body systems in thermal equilibrium", arXiv:2204.08349.}

\bibitem{HW18} H. Watanabe, {\it Phys. Rev. B} {\bf 98}, 155137 (2018).

\bibitem{ZW21} Z. Wang, M. Foss-Feig, and K. R. A. Hazzard, {\it Phys. Rev. Research} {\bf 3}, L032047 (2021).

\bibitem{EHL61} E. H. Lieb, T. Schultz, and D. J. Mattis, {\it Ann. Phys. (N.Y.)} {\bf 16}, 407 (1961).

\bibitem{MO00} M. Oshikawa, {\it Phys. Rev. Lett.} {\bf 84}, 1535 (2000).

\bibitem{MBH04} M. B. Hastings, {\it Phys. Rev. B} {\bf 69}, 104431 (2004). 

\bibitem{QN85} Q. Niu, D. J. Thouless, and Y.-S. Wu, {\it Phys. Rev. B} {\bf 31}, 3372 (1985).  

\bibitem{KK19} K. Kudo, H. Watanabe, T. Kariyado, and Y. Hatsugai, {\it Phys. Rev. Lett.} {\bf 122}, 146601 (2019).  

\bibitem{JE10} J. Eisert, M. Cramer, and M. B. Plenio, {\it Rev. Mod. Phys.} {\bf 82}, 277 (2010).  

\bibitem{MBH07} M. B. Hastings, {\it J. Stat. Mech.} (2007) P08024.

\bibitem{IA13} I. Arad, A. Kitaev, Z. Landau, and U. Vazirani, ``An area law and sub-exponential algorithm for 1D systems", arXiv:1301.1162. 

\bibitem{TK20c} T. Kuwahara and K. Saito, {\it Nat. Commun.} {\bf 11}, 4478 (2020).

\bibitem{FGSLB13} F. G. S. L. Brand\~ao, M. Horodecki, {\it Nat. Phys.} {\bf 9}, 721 (2013).

\bibitem{FGSLB15} F. G. S. L. Brand\~ao, M. Horodecki, {\it Commun. Math. Phys.} {\bf 333}, 761 (2015).

\bibitem{JC18} J. Cho, {\it Phys. Rev. X} {\bf 8}, 031009 (2018).

\bibitem{MBP05} M. B. Plenio, J. Eisert, J. Drei{\ss}ig, and M. Cramer, {\it Phys. Rev. Lett.} {\bf 94}, 060503 (2005).

\bibitem{MC06} M. Cramer, J. Eisert, M. B. Plenio, and J. Drei{\ss}ig, {\it Phys. Rev. A} {\bf 73}, 012309 (2006).

\bibitem{AA21} A. Anshu, I. Arad, and D. Gosset, \textcolor{black}{in {\it Proceedings of the 54th Annual ACM SIGACT Symposium on Theory of Computing} (Association for Computing Machinery, New York, NY, 2022), pp.12–18.}

\bibitem{ZB19} B. Zeng, X. Chen, D.-L. Zhou, and X.-G. Wen, {\it Quantum information meets quantum matter} (Springer, Berlin, 2019).

\bibitem{KVA13} K. V. Acoleyen, M. Mari\"en, and F. Verstraete, {\it Phys. Rev. Lett.} {\bf 111}, 170501 (2013).

\bibitem{MBH05} M. B. Hastings and X.-G. Wen, {\it Phys. Rev. B} {\bf 72}, 045141 (2005).

\bibitem{TJO07} T. J. Osborne, {\it Phys. Rev. A} {\bf 75}, 032321 (2007).

\bibitem{SB10} S. Bravyi, M. B. Hastings, and S. Michalakis, {\it J. Math. Phys.} {\bf 51}, 093512 (2010).

\bibitem{ZXG17} Z.-X. Gong, M. Foss-Feig, F. G. S. L. Brand\~ao, and A. V. Gorshkov, {\it Phys. Rev. Lett.} {\bf 119}, 050501 (2017).  

\bibitem{WD01} W. D\"ur, G. Vidal, J. I. Cirac, N. Linden, and S. Popescu, {\it Phys. Rev. Lett.} {\bf 87}, 137901 (2001).  

\bibitem{SB07} S. Bravyi, {\it Phys. Rev. A} {\bf 76}, 052319 (2007).  

\bibitem{NL09b} N. Linden, J. A. Smolin, and A. Winter, {\it Phys. Rev. Lett.} {\bf 103}, 030501 (2009).  

\bibitem{CHB03} C. H. Bennett, A. W. Harrow, D. W. Leung, and J. A. Smolin, {\it IEEE Trans. Inf. Theory} {\bf 49}, 1895 (2003).

\bibitem{EHL13} E. H. Lieb and A. Vershynina, {\it Quantum Inf. Comput.} {\bf 13}, 0986 (2013).  

\bibitem{KMRA14} K. M. R. Audenaert, {\it J. Math. Phys.} {\bf 55}, 112202 (2014).

\bibitem{MM16} M. Mari\"en, K. M. R. Audenaert, K. Van Acoleyen, and F. Verstraete, {\it Commun. Math. Phys.} {\bf 346}, 35 (2016). 

\bibitem{AC04} A. Childs, D. Leung, and G. Vidal, {\it IEEE Trans. Inf. Theory} {\bf 50}, 1189 (2004).  

\bibitem{BS04} B. Schumacher and R. F. Werner, ``Reversible quantum cellular automata", arXiv:quant-ph/0405174.

\bibitem{PA19} P. Arrighi, {\it Nat. Comput.} {\bf 18}, 885 (2019).

\bibitem{TF20} T. Farrelly, {\it Quantum} {\bf 4}, 368 (2020).

\bibitem{JIC17} J. I. Cirac, D. Perez-Garcia, N. Schuch, and F. Verstraete, {\it J. Stat. Mech.} (2017) 083105.

\bibitem{DG12} D. Gross, V. Nesme, H. Vogts, and R. F. Werner, {\it Commun. Math. Phys.} {\bf 310}, 419 (2012).

\bibitem{BRD18} B. R. Duschatko, P. T. Dumitrescu, and A. C. Potter, {\it Phys. Rev. B} {\bf 98}, 054309 (2018).

\bibitem{DR20} D. Ranard, M. Walter, and F. Witteveen, arXiv:2012.00741.

\bibitem{ZG21} Z. Gong, L. Piroli, and J. I. Cirac, {\it Phys. Rev. Lett.} {\bf 126}, 160601 (2021).

\bibitem{HCP16} H. C. Po, L. Fidkowski, T. Morimoto, A. C. Potter, and A. Vishwanath, {\it Phys. Rev. X} {\bf 6}, 041070 (2016).

\bibitem{TZ17} T. Zhou and D. J. Luitz, {\it Phys. Rev. B} {\bf 95}, 094206 (2017).

\bibitem{LP20} L. Piroli and J. I. Cirac, {\it Phys. Rev. Lett.} {\bf 125}, 190402 (2020).

\bibitem{IP09} I. Peschel and V. Eisler, {\it J. Phys. A} {\bf 42}, 504003 (2009).

\bibitem{KF20} K. Fujimoto, R. Hamazaki, and Y. Kawaguchi, {\it Phys. Rev. Lett.} {\bf 124}, 210604 (2020). 

\bibitem{TJ20} T. Jin, A. Krajenbrink, and D. Bernard, {\it Phys. Rev. Lett.} {\bf 125}, 040603 (2020).

\bibitem{KF21} K. Fujimoto, R. Hamazaki, and Y. Kawaguchi, {\it Phys. Rev. Lett.} {\bf 127}, 090601 (2021).

\bibitem{CJ18} C. Jonay, D. A. Huse, and A. Nahum, ``Coarse-grained dynamics of operator and state entanglement'', arXiv:1803.00089.

\bibitem{TZ20} T. Zhou  and A. Nahum, {\it Phys. Rev. X} {\bf 10}, 031066 (2020).

\bibitem{AN18} A. Nahum, S. Vijay, and J. Haah, {\it Phys. Rev. X} {\bf 8}, 021014 (2018).

\bibitem{AN18b} A. Nahum, J. Ruhman, and D. A. Huse, {\it Phys. Rev. B} {\bf 98}, 035118 (2018).

\bibitem{CWvK18} C. W. von Keyserlingk, T. Rakovszky, F. Pollmann, and S. L. Sondhi, {\it Phys. Rev. X} {\bf 8}, 021013 (2018).

\bibitem{BB20} B. Bertini and L. Piroli, {\it Phys. Rev. B} {\bf 102}, 064305 (2020).

\bibitem{TZ19} T. Zhou and A. Nahum, {\it Phys. Rev. B} {\bf 99}, 174205 (2019).

\bibitem{ZG21b} Z. Gong, A. Nahum, and L. Piroli, \textcolor{black}{{\it Phys. Rev. Lett.} {\bf 128}, 080602 (2022).}

\bibitem{PH16}  P. Hosur, X. Qi, D. A. Roberts, and B. Yoshida, {\it J. High Energy Phys. 02} (2016) 004. 


\bibitem{HFT59} H. F. Trotter, {\it Proc. Am. Math. Soc.} {\bf 10}, 545 (1959).

\bibitem{MS76} M. Suzuki, {\it Commun. Math. Phys.} {\bf 51}, 183 (1976).

\bibitem{SL96} S. Lloyd, {\it Science} {\bf 273}, 1073 (1996).

\bibitem{AMC19} A. M. Childs and Y. Su, {\it Phys. Rev. Lett.} {\bf 123}, 050503 (2019).

\bibitem{MS91} M. Suzuki, {\it J. Math. Phys.} {\bf 32}, 400 (1991).

\bibitem{AMC21} A. M. Childs, Y. Su, M. C. Tran, N. Wiebe, and S. Zhu, {\it Phys. Rev. X} {\bf 11}, 011020 (2021).

\bibitem{DW14} D. Wecker, B. Bauer, B. K. Clark, M. B. Hastings, and M. Troyer, {\it Phys. Rev. A} {\bf 90}, 022305 (2014).

\bibitem{RB18} R. Babbush, N. Wiebe, J. McClean, J. McClain, H. Neven, and G. Kin-Lic Chan, {\it Phys. Rev. X} {\bf 8}, 011044 (2018).


\bibitem{GHL19} G. H. Low and I. L. Chang, {\it Quantum} {\bf 3}, 163 (2019).

\bibitem{DWB07} D. W. Berry, G. Ahokas, R. Cleve, and B. C. Sanders, {\it Commun. Math. Phys.} {\bf 270}, 359 (2007).

\bibitem{TJO06} T. J. Osborne, {\it Phys. Rev. Lett.} {\bf 97}, 157202 (2006).

\bibitem{JE06} J. Eisert and T. J. Osborne, {\it Phys. Rev. Lett.} {\bf 97}, 150404 (2006).

\bibitem{AMA21} \'A. M. Alhambra and J. I. Cirac, {\it Phys. Rev. X Quantum} {\bf 2}, 040331 (2021).



\bibitem{MB16} M. Bukov, L. D'Alessio, and A. Polkovnikov, {\it Ann. Phys.} {\bf 64:2}, 139 (2015).

\bibitem{AE17} A. Eckardt, {\it Rev. Mod. Phys.} {\bf 89}, 011004 (2017).

\bibitem{TO19} T. Oka and S. Kitamura,
{\it Annu. Rev. Condens. Matter Phys.} {\bf 10}, 387 (2019).

\bibitem{AH22}
A. Haldar and A. Das, {\it J. Phys.: Condens. Matter} {\bf 34}, 234001 (2022).

\bibitem{TK16} T. Kuwahara, T. Mori, and K. Saito, {\it Ann. Phys.} {\bf 367}, 96 (2016).

\bibitem{AL14} A. Lazarides, A. Das, and R. Moessner, {\it Phys. Rev. E} {\bf 90}, 012110 (2014).

\bibitem{HK14} H. Kim, T. N. Ikeda, and D. A. Huse, {\it Phys. Rev. E} {\bf 90}, 052105 (2014).

\bibitem{LDA14} L. D’Alessio, and M. Rigol, {\it Phys. Rev. X} {\bf 4}, 041048 (2014).


\bibitem{DAA15} D. A. Abanin, W. De Roeck, and F. Huveneers, {\it Phys. Rev. Lett.} {\bf 115}, 256803 (2015).

\bibitem{TM16} T. Mori, T. Kuwahara, and K. Saito, {\it Phys. Rev. Lett.} {\bf 116}, 120401 (2016).

\bibitem{DA17PRB} D. Abanin, W. De Roeck, W. W. Ho, and F. Huveneers, {\it Phys. Rev. B} {\bf 95}, 014112 (2017). 

\bibitem{DA17} D. Abanin, W. De Roeck, W. W. Ho, and F. Huveneers, {\it Commun. Math. Phys.} {\bf 354}, 809 (2017). 

\bibitem{WWH18} W. W. Ho, I. Protopopov, and D. A. Abanin, {\it Phys. Rev. Lett.} {\bf 120}, 200601 (2018).

\bibitem{AR20} A. Rubio-Abadal, M. Ippoliti, S. Hollerith, D. Wei, J. Rui, S. L. Sondhi, V. Khemani, C. Gross, and I. Bloch, {\it Phys. Rev. X} {\bf 10}, 021044 (2020).

\bibitem{FM20} F. Machado, D. V. Else, G. D. Kahanamoku-Meyer, C. Nayak, and N. Y. Yao, {\it Phys. Rev. X} {\bf 10}, 011043 (2020).

\bibitem{DVE17}
D. V. Else, B. Bauer, and C. Nayak,
 {\it Phys. Rev. X} {\bf 7}, 011026 (2017).

\bibitem{TM18} T. Mori, {\it Phys. Rev. B} {\bf 98}, 104303 (2018).


\bibitem{AP21} A. Pizzi, A. Nunnenkamp , and J. Knolle, {\it Phys. Rev. Lett.} {\bf 127}, 140602 (2021).

\bibitem{AP21PRB} A. Pizzi, A. Nunnenkamp , and J. Knolle, {\it Phys. Rev. B} {\bf 104}, 094308 (2021).

\bibitem{BY21} B. Ye, F. Machado, and N. Y. Yao, {\it Phys. Rev. Lett.} {\bf 127}, 140603 (2021).

\bibitem{AH18} A. Haldar, R. Moessner, and A. Das, {\it Phys. Rev. B} {\bf 97}, 245122 (2018).

\bibitem{AH21} A. Haldar, D. Sen, R. Moessner, and A. Das, {\it Phys. Rev. X} {\bf 11}, 021008 (2021).




\bibitem{HB17} H. Bernien, S. Schwartz, A. Keesling, H. Levine, A. Omran, H. Pichler, S. Choi, A. S. Zibrov, M. Endres, M. Greiner, V. Vuleti\'{c}, and M. D. Lukin, {\it Nature} {\bf 551}, 579 (2017).

\bibitem{ZG20} Z. Gong, N. Yoshioka, N. Shibata, and R. Hamazaki, {\it Phys. Rev. Lett.} {\bf 124}, 210606 (2020).

\bibitem{ZG20b} Z. Gong, N. Yoshioka, N. Shibata, and R. Hamazaki, {\it Phys. Rev. A} {\bf 101}, 052122 (2020).

\bibitem{JJS11} J. J. Sakurai and J. Napolitano, {\it Modern Quantum Mechanics} (Addison-Wesley, Boston, 2011).

\bibitem{EB07} E. Brion, L. H. Pedersen, and K. M{\o}lmer, {\it J. Phys. A} {\bf 40}, 1033 (2007).

\bibitem{JRS66} J. R. Schrieffer and P. A. Wolff, {\it Phys. Rev.} {\bf 149}, 491 (1966).

\bibitem{SB11} S. Bravyi, D. P. DiVincenzo, and D. Loss, {\it Ann. Phys.} {\bf 326}, 2793 (2011).

\bibitem{TG06} T. Gorin, T. Prosen, T. H. Seligman, and M. \v{Z}nidari\v{c}, {\it Phys. Rep.} {\bf 435}, 33 (2006).

\bibitem{ND96} N. Datta, J. Fr\"ohlich, L. Rey-Bellet, and R. Fern\'andez, {\it Helv. Phys. Acta} {\bf 69}, 752 (1996).

\bibitem{SB17} S. Bachmann,W. De Roeck, and M. Fraas, {\it Phys. Rev. Lett.} {\bf 119}, 060201 (2017).

\bibitem{JCH20} J. C. Halimeh and P. Hauke, Phys. Rev. Lett. 125, 030503 (2020).

\bibitem{JCH21} J. C. Halimeh, H. Lang, J. Mildenberger, Z. Jiang, and P. Hauke, {\it PRX Quantum} {\bf 2}, 040311 (2021).

\bibitem{JCH21b} J. C. Halimeh, L. Homeier, C. Schweizer, M. Aidelsburger, P. Hauke, and F. Grusdt, {\it Phys. Rev. Res.} {\bf 4}, 033120 (2022).

\bibitem{JCH22} J. C. Halimeh, H. Lang, and P. Hauke, {\it New J. Phys.} {\bf 24}, 033015 (2022).





\bibitem{FW12} F. Wilczek, {\it Phys. Rev. Lett.} {\bf 109}, 160401 (2012).

\bibitem{HW15} H. Watanabe and M. Oshikawa, {\it Phys. Rev. Lett.} {\bf 114}, 251603 (2015).

\bibitem{VK19} V. Khemani, R. Moessner, S. L. Sondhi, ``A Brief History of Time Crystals'', arXiv:1910.10745.

\bibitem{HW20} H. Watanabe, M. Oshikawa, and T. Koma, {\it J. Stat. Phys.} {\bf 178}, 926 (2020). 

\bibitem{YH19TC} Y. Huang, ``Absence of temporal order in states with spatial correlation decay'', arXiv:1912.01210.

\bibitem{KS17} K. Sacha and J. Zakrzewski, {\it Rep. Prog. Phys.} {\bf 81}, 016401 (2017).

\bibitem{DVE20}
D. V. Else, C. Monroe, C. Nayak, and N. Y. Yao,
{\it Annu. Rev. Condens. Matter Phys.} {\bf 11}, 467 (2020).

\bibitem{DVE16} D. V. Else, B. Bauer, and C. Nayak, {\it Phys. Rev. Lett.} {\bf 117}, 090402 (2016).

\bibitem{NYY17}
N. Y. Yao, A. C. Potter, I. D. Potirniche, and A. Vishwanath,
{\it Phys. Rev. Lett.} {\bf 118}, 030401 (2017).

\bibitem{JZ17}
J. Zhang, P. W. Hess, A. Kyprianidis, P. Becker, A. Lee, J. Smith, G. Pagano, I. D. Potirniche, A. C. Potter, A. Vishwanath, N. Y. Yao, and C. Monroe,
{\it Nature} {\bf 543}, 217 (2017).


\bibitem{JR18} J. Rovny, R. L. Blum, and S. E. Barrett, {\it Phys. Rev. Lett.} {\bf 120}, 180603 (2018).

\bibitem{HK21} H. Keßler, P. Kongkhambut, C. Georges, L. Mathey, J. G. Cosme, and A. Hemmerich, {\it Phys. Rev. Lett.} {\bf 127}, 043602 (2021).

\bibitem{AK21}
A. Kyprianidis, F. Machado, W. Morong, P. Becker, K. S. Collins, D. V. Else, L. Feng, P. W. Hess, C. Nayak, G. Pagano, N. Y. Yao, and C. Monroe,
{\it Science} {\bf 372.6547}, 1192 (2021).

\bibitem{ZG18}
Z. Gong, R. Hamazaki, and M. Ueda,
 {\it Phys. Rev. Lett.} {\bf 120}, 040404 (2018).

\bibitem{BZ19}
B. Zhu, J. Marino, N. Y Yao, M, D Lukin, and E. A. Demler,
 {\it New J. Phys.} {\bf 21}, 073028 (2019).



\bibitem{FI18}
F. Iemini, A. Russomanno, J. Keeling, M. Schir\`o, M. Dalmonte, and R. Fazio
 {\it Phys. Rev. Lett.} {\bf 121}, 035301 (2018).
 
\bibitem{FMG19}
 F. M. Gambetta, F. Carollo, M. Marcuzzi, J. P. Garrahan, and I. Lesanovsky,
  {\it Phys. Rev. Lett.} {\bf 122}, 015701 (2019).
  
\bibitem{BB19}
 B. Bu\v{c}a, J. Tindall, and D. Jaksch, {\it Nat. Commun.} {\bf 10}, 1730 (2019).  

\bibitem{VKK19}
V. K. Kozin and O. Kyriienko
  {\it Phys. Rev. Lett.} {\bf 123}, 210602 (2019).




\bibitem{AIL69} A. I. Larkin and Y. N. Ovchinnikov, {\it Sov. Phys. JETP} {\bf 28(6)}, 1200 (1969).  

\bibitem{AK14} S. Kitaev, in {\it talk given at Fundamental Physics Prize Symposium} (2014). 

\bibitem{AK15} S. Kitaev, in {\it KITP strings seminar and Entanglement} (2015). 

\bibitem{SHS15} S. H. Shenker and D. Stanford, {\it J. High Energy Phys. 05} (2015) 132. 

\bibitem{RH18} R. Hamazaki, K. Fujimoto, and M. Ueda, ``Operator Noncommutativity and Irreversibility in Quantum Chaos'', arXiv:1807.02356.

\bibitem{AP10} A. Polkovnikov, {\it Ann. Phys.} {\bf 325}, 8 (2010).

\bibitem{EBR17} E. B. Rozenbaum, S. Ganeshan, and V. Galitski, {\it Phys. Rev. Lett.} {\bf 118}, 086801 (2017).

\bibitem{JK18} J. Kurchan, {\it J. Stat. Phys.} {\bf 171}, 965 (2018). 

\bibitem{JSC18} J. S. Cotler, D. Ding, and G. R. Penington, {\it Ann. Phys.} {\bf 396}, 318 (2018).

\bibitem{RAJ18} R. A. Jalabert, I. Garc\'ia-Mata, and D. A. Wisniacki, {\it Phys. Rev. E} {\bf 98}, 062218 (2018).

\bibitem{TS19} T. Scaffidi  and E. Altman, {\it Phys. Rev. B} {\bf 100}, 155128 (2019).

\bibitem{TX20} 
T. Xu, T. Scaffidi, and X. Cao, {\it Phys. Rev. Lett.} {\bf 124}, 140602 (2020).

\bibitem{KH20} 
K. Hashimoto, K. Huh, K. Kim and R. Watanabe, {\it J. High Energy Phys. 11} (2020) 068. 

\bibitem{WK21} W. Kirkby, D. H. J. O'Dell, and J. Mumford, {\it Phys. Rev. A} {\bf 104},  043308 (2021).

\bibitem{JM16} J. Maldacena, S. H. Shenker, and D. Stanford, {\it J. High Energy Phys. 08} (2016) 106. 

\bibitem{SHS14}  S. H. Shenker and D. Stanford, {\it J. High Energy Phys. 03} (2014) 067. 


\bibitem{NT18} N. Tsuji, T. Shitara, and M. Ueda, {\it Phys. Rev. E} {\bf 98}, 012216 (2018).

\bibitem{SS93} S. Sachdev and J. Ye, {\it Phys. Rev.  Lett.} {\bf 79}, 3339 (1993).

\bibitem{JM16b} J. Maldacena and D. Stanford, {\it Phys. Rev. D} {\bf 94}, 106002 (2016).

\bibitem{JP16}  
J. Polchinski and V. Rosenhaus, {\it J. High Energy Phys. 04} (2016) 001. 

\bibitem{JSC17} 
J. S. Cotler, G. Gur-Ari, M. Hanada, J. Polchinski, P. Saad, S. H. Shenker, D. Stanford, A. Streicher, M. Tezuka, {\it J. High Energy Phys. 05} (2017) 118. 

\bibitem{YG17} 
Y. Gu, X. L. Qi, and D. Stanford, {\it J. High Energy Phys. 05} (2017) 125. 

\bibitem{SS15} S. Sachdev, {\it Phys. Rev. X} {\bf 5}, 041025 (2015).


\bibitem{IK17} I. Kukuljan, S. Grozdanov, and T. Prosen, {\it Phys. Rev. B} {\bf 96}, 060301(R) (2017).

\bibitem{TK21} T. Kuwahara and K. Saito, {\it Phys. Rev. Lett.} {\bf 126}, 030604 (2021).

\bibitem{JG16} J. Goold, M. Huber, A. Riera, L. del Rio, and P. Skrzypczyk, {\it J. Phys. A} {\bf 49}, 143001 (2016).

\bibitem{AEA04} A. E. Allahverdyan, R. Balian, and Th. M. Nieuwenhuizen, {\it Europhys. Lett.} {\bf 67}, 565 (2004).

\bibitem{RA13} R. Alicki and M. Fannes, {\it Phys. Rev. E} {\bf 87}, 042123 (2013).

\bibitem{WP78} W. Pusz and S. L. Woronowicz, {\it Commun. Math. Phys.} {\bf 58}, 273 (1978).

\bibitem{AL78} A. Lenard, {\it J. Stat. Phys.} {\bf 19}, 575 (1978).

\bibitem{KVH13} K. V. Hovhannisyan, M. Perarnau-Llobet, M. Huber, and A. Ac\'in, {\it Phys. Rev. Lett.} {\bf 111}, 240401 (2013).

\bibitem{MF14} M. Frey, K. Funo, and M. Hotta, {\it Phys. Rev. E} {\bf 90}, 012127 (2014).

\bibitem{AMA19} \'A. M. Alhambra, G. Styliaris, N. A. Rodr\'iguez-Briones, J. Sikora, and E. Mart\'in-Mart\'inez, {\it Phys. Rev. Lett.} {\bf 123}, 190601 (2019).

\bibitem{YM22} \textcolor{black}{Y. Mitsuhashi, K. Kaneko, and T. Sagawa, {\it Phys. Rev. X} {\bf 12}, 021013 (2022).}

\bibitem{NYH16} \textcolor{black}{N. Y. Halpern, P. Faist, J. Oppenheim, and A. Winter, {\it Nat. Commun.} {\bf 7}, 12051 (2016).}


\bibitem{MPL15} M. Perarnau-Llobet, K. V. Hovhannisyan, M. Huber, P. Skrzypczyk, N. Brunner, and A. Ac\'in, {\it Phys. Rev. X} {\bf 5}, 041011 (2015).



\bibitem{MC11} M. Campisi, P. H\"anggi, and P. Talkner, {\it Rev. Mod. Phys.} {\bf 83}, 771 (2011).

\bibitem{SV16} S. Vinjanampathy and J. Anders, {\it Contemp. Phys.} {\bf 57}, 545 (2016). 

\bibitem{MH13} M. Horodecki and J. Oppenheim, {\it Nat. Commun.} {\bf 4}, 2059 (2013). 

\bibitem{FB15} F. Brand\~ao, M. Horodecki, N. Ng, J. Oppenheim, and S. Wehner, {\it Proc. Natl. Acad. Sci. U.S.A.} {\bf 112}, 3275 (2015). 

\bibitem{EC19} E. Chitambar and G. Gour, {\it Rev. Mod. Phys.} {\bf 91}, 025001 (2019).

\bibitem{CJ97} C. Jarzynski, {\it Phys. Rev. Lett.} {\bf 78}, 2690 (1997).

\bibitem{HT00} H. Tasaki, ``Jarzynski Relations for Quantum Systems and Some Applications", arXiv:cond-mat/0009244.

\bibitem{JK00} J. Kurchan, ``A quantum fluctuation theorem", arXiv:cond-mat/0007360.

\bibitem{KM09} K. Maruyama, F. Nori, and V. Vedral, {\it Rev. Mod. Phys.} {\bf 81}, 1 (2009).

\bibitem{LS29} L. Szilard, {\it Z. Phys.} {\bf 53}, 840 (1929).

\bibitem{JMRP15} J. M. R. Parrondo, J. M. Horowitz, and T. Sagawa, {\it Nat. Phys.} {\bf 11}, 131 (2015).

\bibitem{TS08} T. Sagawa and M. Ueda, {\it Phys. Rev. Lett.} {\bf 100}, 080403 (2008).

\bibitem{TS09} T. Sagawa and M. Ueda, {\it Phys. Rev. Lett.} {\bf 102}, 250602 (2009).

\bibitem{RL61} R. Landauer, {\it IBM J. Res. Dev.} {\bf 5}, 183 (1961).

\bibitem{CHB82} C. H. Bennett, {\it Int. J. Theor. Phys.} {\bf 21}, 905 (1982).

\bibitem{JG15} J. Goold, M. Paternostro, and K. Modi, {\it Phys. Rev. Lett.} {\bf 114}, 060602 (2015).

\bibitem{FB20} F. Buscemi, D. Fujiwara, N. Mitsui, and M. Rotondo, {\it Phys. Rev. A} {\bf 102}, 032210 (2020).

\bibitem{TS10} T. Sagawa and M. Ueda, {\it Phys. Rev. Lett.} {\bf 104}, 090602 (2010).

\bibitem{ZG16} Z. Gong, Y. Ashida, and M. Ueda, {\it Phys. Rev. A} {\bf 94}, 012107 (2016).


\bibitem{JMH20} J. M. Horowitz and T. R. Gingrich , {\it Nat. Phys.} {\bf 16}, 15 (2020). 

\bibitem{ACB15} A. C. Barato and U. Seifert, {\it Phys. Rev. Lett.} {\bf 114}, 158101 (2015). 

\bibitem{TRG16} T. R. Gingrich, J. M. Horowitz, N. Perunov, and J. England, {\it Phys. Rev. Lett.} {\bf 116}, 120601 (2016). 

\bibitem{KL20} K. Liu, Z. Gong, and M. Ueda, {\it Phys. Rev. Lett.} {\bf 125}, 140602 (2020).

\bibitem{US20} T. Koyuk and U. Seifert, {\it Phys. Rev. Lett.} {\bf 125}, 260604 (2020).  

\bibitem{PP18} P. Pietzonka and U. Seifert, {\it Phys. Rev. Lett.} {\bf 120}, 190602 (2018).

\bibitem{JMH17} J. M. Horowitz and T. R. Gingrich, {\it Phys. Rev. E} {\bf 96}, 020103(R) (2017).

\bibitem{AD19} A. Dechant, {\it J. Phys. A} {\bf 52}, 035001 (2019).

\bibitem{YH19} Y. Hasegawa and T. V. Vu, {\it Phys. Rev. E} {\bf 99}, 062126 (2019).

\bibitem{YH20} Y. Hasegawa, {\it Phys. Rev. Lett.} {\bf 125}, 050601 (2020).


\bibitem{SG14} S. Gammelmark and K. M{\o}lmer, {\it Phys. Rev. Lett.} {\bf 112}, 170401 (2014).

\bibitem{YH21} Y. Hasegawa, {\it Phys. Rev. Lett.} {\bf 126}, 010602 (2021).

\bibitem{TVV19} Y. Hasegawa and T. V. Vu, {\it Phys. Rev. Lett.} {\bf 123}, 110602 (2019).

\bibitem{RGG10} R. Garc\'ia-Garc\'ia, D. Dom\'inguez, V. Lecomte, and A. B. Kolton, {\it Phys. Rev. E} {\bf 82}, 030104(R) (2010).

\bibitem{AMT19} A. M. Timpanaro, G. Guarnieri, J. Goold, and G. T. Landi , {\it Phys. Rev. Lett.} {\bf 123}, 090604 (2019).


\bibitem{DP10} D. Poulin, {\it Phys. Rev. Lett.} {\bf 104}, 190401 (2010).


\bibitem{TSC15} T. S. Cubitt, A. Lucia, S. Michalakis, and D. Perez-Garcia, {\it Commun. Math. Phys.} {\bf 337}, 1275 (2015).

\bibitem{RS19} R. Sweke, J. Eisert, and M. Kastner, {\it J. Phys. A: Math. Theor.} {\bf 52}, 424003 (2019).

\bibitem{AYG21} A. Y. Guo, S. Lieu, M. C. Tran, and A. V. Gorshkov, ``Clustering of steady-state correlations in open systems with long-range interactions", arXiv:2110.15368.

\bibitem{IB12} \textcolor{black}{I. Bloch, J. Dalibard, and S. Nascimb\`ene, {\it Nat. Phys.} {\bf 8}, 267 (2012).}

\bibitem{AB20} \textcolor{black}{A. Browaeys and T. Lahaye, {\it Nat. Phys.} {\bf 16}, 132 (2020).}

\bibitem{RB12} \textcolor{black}{R. Blatt and C. F. Roos, {\it Nat. Phys.} {\bf 8}, 277 (2012).}

\bibitem{AAH12} \textcolor{black}{A. A. Houck, H. E. T\"ureci, and J. Koch, {\it Nat. Phys.} {\bf 8}, 292 (2012).}

\bibitem{VVD13} \textcolor{black}{V. V. Dobrovitski, G.D. Fuchs, A. L. Falk, C. Santori, and D. D. Awschalom, {\it Annu. Rev. Condens. Matter Phys.} {\bf 4}, 23 (2013).}

\bibitem{CLD17} C. L. Degen, F. Reinhard, and P. Cappellaro, {\it Rev. Mod. Phys.} {\bf 89}, 035002 (2017).

\bibitem{VG11} V. Giovannetti, S. Lloyd, and L. Maccone, {\it Nat. Photonics} {\bf 5}, 222 (2011).


\bibitem{SC17} S. Campbell and S. Deffner, {\it Phys. Rev. Lett.} {\bf 118}, 100601  (2017).

\bibitem{KF17} K. Funo, J. Zhang, C. Chatou, K. Kim, M. Ueda, and A. del Campo, {\it Phys. Rev. Lett.} {\bf 118}, 100602 (2017).


\bibitem{TC09}
T. Caneva, M. Murphy, T. Calarco, R. Fazio, S. Montangero, V. Giovannetti, and G. E. Santoro, {\it Phys. Rev. Lett.} {\bf 103}, 240501 (2009).

\bibitem{OL17}
O. Lychkovskiy, O. Gamayun, and V. Cheianov, {\it Phys. Rev. Lett.} {\bf 119}, 200401  (2017).

\bibitem{JHC21}
J.-H. Chen and V. Cheianov, ``Bounds on quantum adiabaticity in driven many-body systems from generalized orthogonality catastrophe and quantum speed limit," arXiv:2112.06900.












\end{thebibliography}
\end{document}